# Infrared Optical Spectroscopy of Molten Fluorides: Methods, Electronic and Vibrational Data, Structural Interpretation, and Relevance to Radiative Heat Transfer

Will B. Derdeyn[1], Sara Mastromarino[2], Ruchi Gakhar[3], Mark H. Anderson[1,4], Mikhail A. Kats[5,6,7], Raluca O. Scarlat[2*]

## Affiliations

[1]Department of Engineering Physics, University of Wisconsin-Madison, Madison, Wisconsin 53706, USA
[2]Department of Nuclear Engineering, University of California, Berkeley, California 94720, USA
[3]Pyrochemistry and Molten Salt Systems Department, Idaho National Laboratory, Idaho Falls, Idaho 83415, USA
[4]Department of Mechanical Engineering, University of Wisconsin-Madison, Madison, Wisconsin 53706, USA
[5]Department of Electrical and Computer Engineering, University of Wisconsin-Madison, Madison, Wisconsin 53706, USA
[6]Department of Materials Science and Engineering, University of Wisconsin-Madison, Madison, Wisconsin 53706, USA
[7]Department of Physics, University of Wisconsin-Madison, Madison, Wisconsin 53706, USA
*Correspondence to: scarlat@berkeley.edu

## Abstract

To help address the need for predicting radiative heat transfer (RHT) behavior of molten salts, we conducted a comprehensive review of methods and data from infrared optical spectroscopic measurements on molten fluoride salts. Transmittance, reflectance, and trans-reflectance experimental methods are discussed, along with the corresponding data reduction methodology and the limitations of each technique. Optical spectroscopy is a convenient indirect probe for changes in structural parameters with temperature and composition. Electronic and vibrational absorption data for transition-metal, lanthanide, and actinide solutes and vibrational absorption data for alkali and alkaline earth fluoride solvents are compiled, and the corresponding structural interpretation is discussed and compared with other experimental and theoretical work. We find that solvent and solute vibrational absorption can be significant in the mid-infrared, resulting in near-infrared edges of significance to RHT. Extrapolation and averaging of existing edge data leads to estimated gray absorption coefficient values at 700 ˚C of 493 $m^{-1}$ for FLiBe and 148 $m^{-1}$ for FLiNaK, both within the range of 1 – 6000 $m^{-1}$ identified to be of engineering relevance for radiative heat transfer analysis.

## ACRONYMS

ARE Aircraft Reactor Experiment

MSRE Molten Salt Reactor Experiment

MSBR Molten Salt Breeder Reactor

MSR molten salt reactor

FHR fluoride-salt-cooled, high-temperature reactor

CN coordination number

EM electromagnetic

KK Kramers-Kronig relations

CO classical oscillator method

UV ultraviolet

NIR near infrared

MIR mid infrared

FIR far infrared

NMR nuclear magnetic resonance

ORNL Oak Ridge National Laboratory

RHT radiative heat transfer

SINAP Shanghai Institute of Applied Physics

FTS Fourier transform spectrometer

MD molecular dynamics

AIMD ab-initio molecular dynamics

XRD x-ray diffraction

EXAFS extended x-ray absorption fine structure

HT-NMR high temperature nuclear magnetic resonance

HT-PFG-NMR high temperature pulsed field gradient nuclear magnetic resonance

QC-DFT quantum-chemical density-functional-theory

## NOMENCLATURE

$\lambda$ = wavelength

$\upsilon$ = spectroscopic wavenumber

$\omega$ = angular frequency

$T$ = temperature

$I$ = light intensity

$R$ = reflectance

$\mathcal{T}$ = transmittance

$Abs$ = absorptivity

$\epsilon$ = emissivity

$\tilde{r}$ = complex EM-field reflectance

$r$ = magnitude of EM-field reflectance

$\tilde{n}_i$ = complex refractive index of medium $i$

$n$ = real part of the complex refractive index, i.e., the refractive index

$k$ = imaginary part of the complex refractive index, i.e., the extinction coefficient

$\kappa_\lambda$ = spectral linear absorption coefficient

$\kappa$ = gray linear absorption coefficient

$s$ = distance traveled by light

$A$ = absorbance

$\varepsilon$ = molar absorptivity

$c$ = molar concentration

$\bar{\theta}$ = frequency-dependent phase shift of reflected EM wave

$\tilde{\epsilon}_r$= complex relative permittivity

$\epsilon_r{}'$ = real component of the complex permittivity

$\epsilon_r{}''$ = imaginary component of complex permittivity

$\epsilon_{r,\infty}$ = high-frequency limit of permittivity

$\nu_j$ = wavenumber of the $j$th resonance

$4\pi\rho_j$ = strength of the $j$th resonance

$\gamma_j$ = damping factor of the $j$th resonance

$\mu_r$ = magnetic permeability

$\theta$ = angle of light propagation

$h$ = salt height (related to the trans-reflectance method)

## 1. INTRODUCTION

Molten-fluoride salts are currently under investigation in support of several proposed molten salt nuclear reactor (MSR) designs [1]. Historically, molten fluorides saw considerable use in the Aircraft Reactor Experiment (ARE) and Molten Salt Reactor Experiment (MSRE), two test reactors built and operated at Oak Ridge National Laboratory (ORNL) during the 1950s and 60s. The ARE used NaF-$ZrF_4$- $UF_4$ (53-41-6 mol%) fuel salt at steady state temperatures up to 860 ˚C and ran at up to 2.5 MWth for a total of 241 hours [2]. Following successful operation of the ARE, the MSRE ran at up to 8 MWth for over 15,000 hours and used $LiF$-$BeF_2$-$ZrF_4$-$UF_4$ (65-29-5-1 mol%) fuel salt at steady state temperatures up to 650 ˚C, along with 2LiF-$BeF_2$ (FLiBe) as secondary coolant [3][4]. The Molten Salt Breeder Reactor (MSBR), the larger scale follow-up to the MSRE, proposed to run at 1000 MWe and use $LiF$-$BeF_2$-$ThF_4$-$UF_4$ (72-16-12-0.4 mol%) fuel salt, along with $NaF$-$NaBF_4$ (8-92 mol%) secondary coolant [5]. Unfortunately, the MSBR program was cut short before the reactor could be built. However, there has been renewed interest in the commercialization of MSRs and the sub-class fluoride-salt-cooled, high-temperature reactors (FHRs) [6][7] and the use of fluorides for fusion reactors and concentrated solar power [8].

MSRs and FHRs are typically designed to operate between 500 and 700 ˚C; hence, reactor designers must consider RHT for prediction of steady state and transient reactor performance. Participating medium effects, wherein the fluid absorbs and re-emits radiant energy in exchange with itself and any surrounding surfaces, often govern the relative contribution of RHT to the total heat transfer in the reactor. These effects have been investigated through several computational fluid dynamics simulations [9][10] and RHT models [11][12]. The degree to which overall heat transfer is enhanced was found highly sensitive to the gray linear absorption coefficient, $\kappa$, which determines the amount of light absorbed and re-emitted along a given direction per unit length ($m^{-1}$ units). RHT is significant when the nondimensional product of pipe diameter and $\kappa$ is in the range of 0.1 to 60 [10]. For example, for a one-centimeter heat-exchanger tube diameter, $\kappa$ of 10 to 6000 $m^{-1}$ would lead to significant RHT effects; for a ten-centimeter heat-exchanger tube diameter, $\kappa$ of 1 to 600 $m^{-1}$ would lead to significant RHT effects.

Molten salt heat transfer models that capture RHT typically use gray $\kappa$, meaning that $\kappa$ is assumed to be constant over all wavelengths, to avoid the high computational cost of spectrally resolved calculations [13]. In reality, the linear absorption coefficient has a spectral dependence, $\kappa_\lambda(\lambda)$, which introduces the challenge of picking a single or banded gray $\kappa$ that adequately captures the RHT behavior. There is limited $\kappa_\lambda$ data for proposed molten-fluoride coolants and fuel (e.g., FLiBe, FLiNaK) or for solute species that could be introduced via corrosion or fuel leakage or are part of the fuel in the case of MSRs, thus it is difficult to ascertain if the use of gray $\kappa$ is acceptable for RHT-modeling in fluoride salts.

At the expected temperature ranges of MSRs and FHRs, absorption in the near-infrared (NIR) and mid-infrared (MIR) ranges will be most important, as shown in Figure 1. We predict RHT-relevance by plotting the Planck distribution in wavelength (Figure 1), here calculated for a temperature of 700 ˚C and showing a peak at 3 μm. The distribution increases in intensity and shifts to shorter wavelengths as temperature increases.

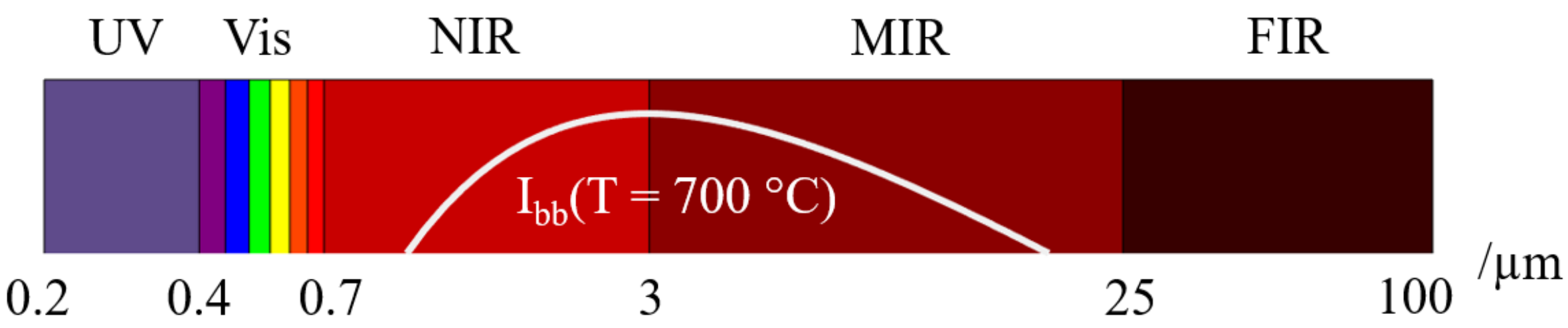

**Figure 1. Portion of electromagnetic spectrum covered by studies in this review. The following regions are defined: ultra-violet (UV), visible (Vis), near-infrared (NIR), mid-infrared (MIR), and far-infrared (FIR). Planck distribution in wavelength at T = 700°C is plotted in white on a logarithmic wavelength scale. The lower limit of the y-axis scale has been chosen such that the displayed portion of the Planck distribution makes up 97% of the total blackbody power at T = 700 °C.**

We refer to measurement of absorption spectra within the overall range depicted in Figure 1 as optical absorption spectroscopy. In the ultra-violet (UV), visible, and NIR, absorption mostly occurs due to excitation of valence electrons, referred to as electronic absorption. In the NIR, MIR and far-infrared (FIR), absorption may also occur due to excitation of molecular vibrations, referred to as vibrational absorption. The various available electronic and vibrational energy levels will result in a unique spectrum for a given solute-solvent combination. Spectroscopists typically compare results with complementary methods to infer structural information, such as the coordination number (CN) [14]. Despite offering this useful information, optical absorption spectroscopy has been somewhat overlooked in molten-salt chemistry in recent years. Rollet and Salanne comprehensively reviewed molten halide structure, focusing on several computational and experimental methods, but rarely mention optical absorption spectroscopy (though we note that Raman spectroscopy receives considerable attention) [15].

We determined that a review of molten-fluoride optical absorption spectroscopy could encourage investigation by providing a database of methods and results and offering a discussion of structural interpretations. Several excellent reviews exist for the broader field of molten-salt optical spectroscopy [16][17][18], mostly involving nitrates and chlorides. These articles are useful for general best-practices and theoretical background of spectroscopic methods, but there is limited information regarding fluorides. The closest to a molten-fluoride review is Toth's ORNL report, 'Containers for Molten Fluoride Spectroscopy' [19]. Toth concluded that for 'optically pure' fluorides, $SiO_2$ would be an acceptable container, but that a metal cell with diamond windows would be the most robust. The report rarely mentions reflectance setups, and there is no summary or discussion of absorption data.

As such, we conducted a comprehensive search of molten-fluoride methods and data. After providing a brief background on the relevant optics concepts, we present and discuss the spectroscopy methods, dividing them by transmittance (3.1) and reflectance techniques (3.2). Detailed information on the methods is compiled in Table 1 (transmittance) and Table 3 (reflectance), and we give each setup an index (T# for transmittance, R# for reflectance) that we use to refer to throughout the text. We then present the data as a collection of digitized spectra, given in Figure 4, and give each plot an index (A#) that we refer to throughout the text. Each spectra is given an index (V# for vibrational, E# for electronic), and we summarize in Table 4 (solvent vibrational edge), Table 5 (solvent and solute vibrational resonance), Table 6 (lighter solute electronic), and Table 7 (heavier solute electronic), listing the various peak locations and intensities and deduced speciation. Next, we discuss the structural interpretation of the data and compare with more recent data from complementary methods, dividing into vibrational (5.1) and electronic (5.2) absorption. Lastly, we summarize our discussion of structural interpretation from the available spectra of solvents and solutes and point out consequences of the NIR data on RHT, identifying the relatively small set of compositions that have been studied, and highlighting key gaps in speciation in molten fluorides that could potentially be closed with further measurements. Due to the experimental difficulty of optical absorption spectroscopy measurements on high-temperature melts, the in-depth description and discussion of the experimental set-ups describe the state of the art of what has been achieved thus far, and highlight the more promising directions for experimental set-ups.

## 2. OPTICS BACKGROUND AND TERMINOLOGY

### 2.1. Measurable Parameters

We define the following measurable optical parameters: transmittance, $\mathcal{T}(\lambda,T) = I_t(\lambda,T)/I_i(\lambda,T)$, reflectance, $R(\lambda,T) = I_r(\lambda,T)/I_i(\lambda,T)$, absorptivity, $Abs(\lambda,T) = I_a(\lambda,T)/I_i(\lambda,T)$, and emissivity, $\epsilon(\lambda,T) = I_e(\lambda,T)/I_{bb}(\lambda,T)$ of a molten salt sample at temperature $T$. $I$ is the light intensity. The subscript $i$ refers to light incident on the sample, the subscript $t$ refers to light transmitted by the sample, the subscript $r$ refers to light reflected by the sample, the subscript $a$ refers to light absorbed by the sample, the subscript $e$ refers to light emitted by the sample, and the subscript $bb$ refers to a blackbody at temperature $T$. In this manuscript we assume non-scattering surfaces and media because, in the studies we found, salt-gas and salt-window interfaces could be assumed to be optically flat, and the presence of suspended particles was avoided.

For a semi-transparent sample, $1 = R(\lambda,T) + \mathcal{T}(\lambda,T) + Abs(\lambda,T)$, and for an opaque sample, $1 = R(\lambda,T) + Abs(\lambda,T)$. For a reciprocal sample at thermal equilibrium, $\epsilon(\lambda,T) = Abs(\lambda,T)$, and through substitution the previous equations become $1 = R(\lambda,T) + \mathcal{T}(\lambda,T) + \epsilon(\lambda,T)$ and $1 = R(\lambda,T) + \epsilon(\lambda,T)$.

### 2.2. Optical Properties of Materials

We define the complex refractive index of a given medium, $\tilde{n}$.

$$\tilde{n} = n + ik \quad (1)$$

The real component, $n$, sometimes simply referred to as the refractive index, determines the speed at which electromagnetic (EM) waves propagate through the material. The imaginary component, $k$, also known as the extinction coefficient, determines the rate at which the EM fields lose strength due to absorption. The spectral linear absorption coefficient, which captures the loss of optical intensity as light propagates, is $\kappa_\lambda = 4\pi k/\lambda$. We note that our notation for $\kappa_\lambda$ follows that of RHT literature, rather than in optics (where it would be the extinction coefficient).

At the interface between two media with complex refractive indices, $\tilde{n}_1$ and $\tilde{n}_2$, the reflectance at normal incidence is given by:

$$R = \left|\frac{\tilde{n}_1 - \tilde{n}_2}{\tilde{n}_1 + \tilde{n}_2}\right|^2 \quad (2)$$

Within a given medium, the transmittance of light across a distance $s$ is given by:

$$\mathcal{T} = e^{-\kappa_\lambda s} \quad (3)$$

Equations 2 and 3 are the basis of methods for obtaining $\kappa_\lambda$ from measurement of $R$ or $\mathcal{T}$ of a molten fluoride sample.

## 3. EXPERIMENTAL METHODS

Here, we compile and summarize the experimental setups for molten fluoride transmittance and reflectance measurements. We discuss the advantages and disadvantages of the various setups and provide recommendations for future designs. Molten fluoride transmittance and reflectance setups are relatively distinct in design; hence, they are treated separately in sections 3.1 (transmittance method) and 3.2 (reflectance method).

### 3.1. Transmittance

#### 3.1.1. Background

First, we outline two techniques that are used to extract $\kappa_\lambda$ or the molar absorptivity, $\varepsilon$, via measurement of transmittance. We describe Beer's Law in section 3.1.1.1, typically used to obtain $\varepsilon$ of a solute species. In section 3.1.1.2, we describe a method wherein transmittance measurements are performed at two or more path lengths, typically used to obtain $\kappa_\lambda$ of a solvent species.

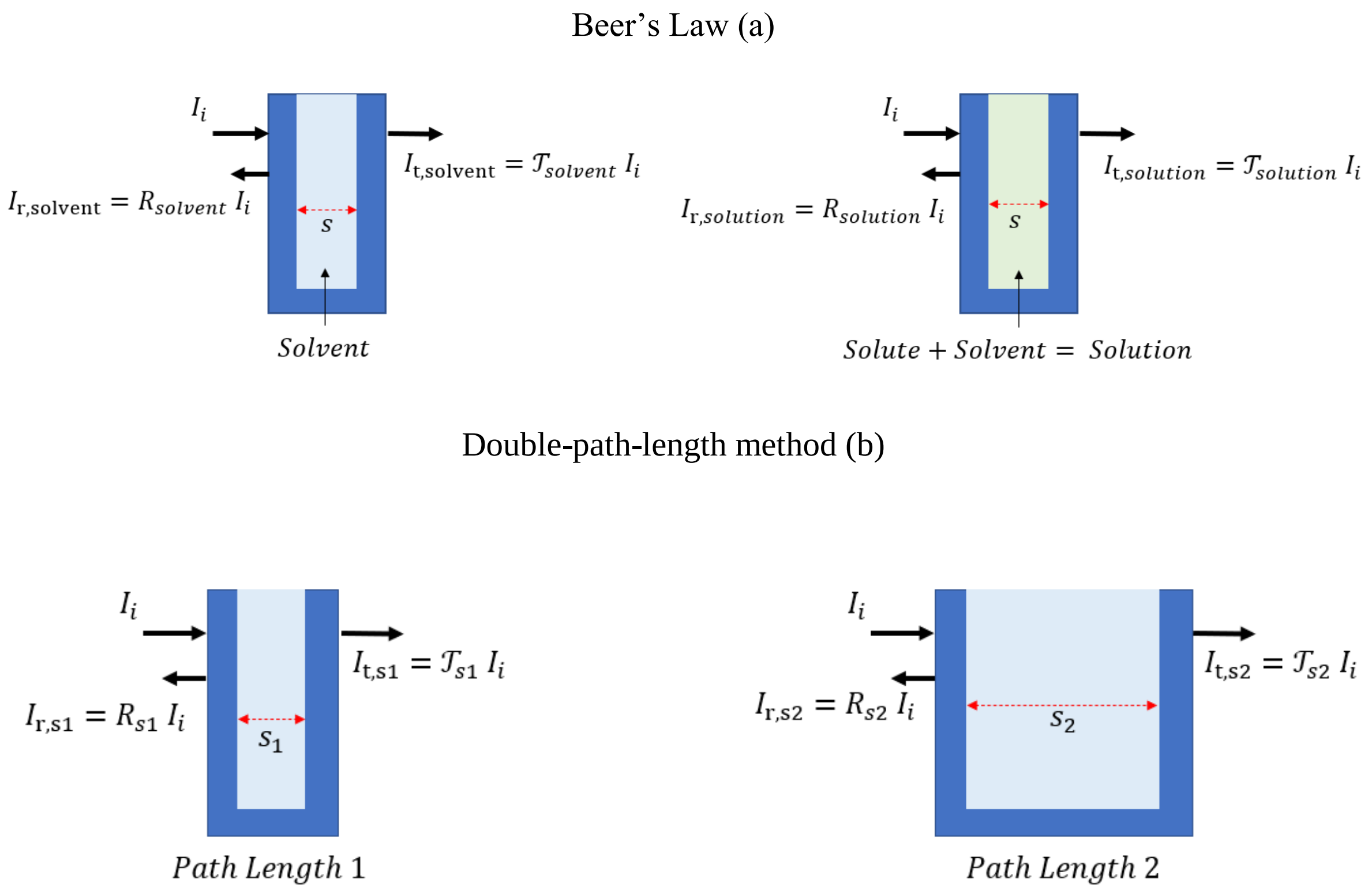

**Figure 2. Schematics for (a) Beer's Law and (b) double-path-length transmittance measurement methods.**

#### 3.1.1.1. Beer's Law

We give Beer's Law in equation 4.

$$A_i = \varepsilon_i c_i s \quad (4)$$

Here, $A$ is the absorbance, $\varepsilon$ is the molar absorptivity (units of $M^{-1}cm^{-1}$), $c$ is the concentration (units of M), and $s$ is the distance traveled by light, a.k.a. the path length (units of cm). The subscript $i$ refers to species $i$. Beer's law shows that in the dilute concentration limit, the absorbance of a species will increase linearly with its concentration; however, deviations from linearity occur at higher concentrations and for highly absorbing species [20]. The spectral linear absorption coefficient, $\kappa_\lambda$, is related to molar absorptivity, $\varepsilon$, by: $\kappa_\lambda = \varepsilon c \ln(10)$. Absorbance, $A$, is related to transmittance, $\mathcal{T}$, by: $A = -\log_{10}\mathcal{T}$. The overall absorbance of a solution consisting of a solvent and a single solute is given by: $A_{solution} = A_{solvent} + A_{solute}$. Here, $A_{solution}$ is the absorbance of the solution consisting of a single solute within a solvent, $A_{solvent}$ is the absorbance of just the solvent, and $A_{solute}$ is the absorbance of just the solute. Hence, to obtain the molar absorptivity of a given solute, $\varepsilon_{solute}$, one must measure $A_{solvent}$, $A_{solute}$, and the solute concentration $c_{solute}$. Typically, this procedure is performed over a range of concentrations to show that the measurements are within the dilute limit.

Almost all the molten-fluoride transmittance (T1-13, T15-17) and several reflectance (R5, R7) studies we review used the Beer's law method.

3.1.1.2. Double-Path-Length Method

The standard approach for measuring $\kappa_\lambda$ of a solvent species is to perform two transmittance measurements with different path lengths [21][22], a method we refer to as the double-path-length method. The ratios of the transmittances are used to calculate $\kappa_\lambda$ via equations 5 and 6.

$$\frac{\mathcal{T}_{s1}}{\mathcal{T}_{s2}} = \frac{I_{t,s1}}{I_{t,s2}} \quad (5)$$

$$\kappa_\lambda = \frac{1}{s2 - s1} \ln\left[\frac{\mathcal{T}_{s1}}{\mathcal{T}_{s2}}\right] \quad (6)$$

Where, $s1$ and $s2$ are the path lengths of the two measurements. The path lengths can either be measured individually or the path length difference can be measured.

Only one molten-fluoride transmittance setup (T14) has used the double-path-length method, but we also include a molten-glass reflectance setup (R6) that would be compatible with molten fluorides. We note that Coyle adapted a chloride and nitrate setup for fluorides by replacing the concentric quartz cuvettes with Inconel tubes capped by diamond windows, however the setup has not been tested with fluorides yet [13].

### 3.1.2. Compilation of transmittance setups

Table 1 lists the transmittance setups covered in this review. For clarity, the setups have been defined by each unique combination of cell, furnace and optical path. In the first column, we include the setup index (T#), primary author and reference, year of publication, and data-processing method (i.e., Beer's Law or double-path-length method). The second column provides schematics and equipment details (i.e., furnace, cell, optical path, cover gas). The third column lists parameters such as wavelength range, maximum $\kappa_\lambda$ measured, maximum temperature, path length, optical window used, amount of salt required, and reported uncertainty. The next columns provide details on the studies performed, such as the salt compositions used, and provides references to relevant plots (A#) and spectra indices (E#/V#).

**Table 1. Experimental setups for measurement of $\kappa_\lambda$ via transmittance. If unspecified, sample size was estimated using dimensions from the schematics and assuming the optical path to be completely filled with salt.**

| Experimental Set-up | Schematic and Equipment Details | Parameters | Studies Performed | | |
|---|---|---|---|---|---|
| | | | Salt Composition | | Data |
| | | | Solute | Solvent | |

<table>
<tr><td>T1<br><br>Young [23]<br><br>1959<br><br>Beer's Law</td><td>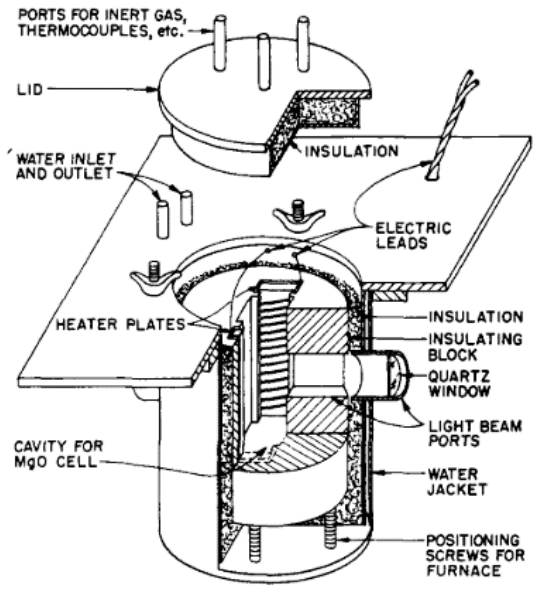

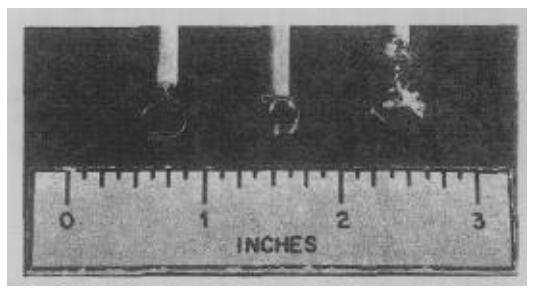

<br>Furnace: Pt-wrapped quartz plates in Lavite block, contained in Cu assembly<br>Cell: Pt pendent drop (shown in bottom image)<br>Optics: Cary 14M grating spectrometer, sample compartment<br>Cover gas: Ar</td><td>Wavelength: 0.3 – 1.4 μm<br><br>$\kappa_\lambda$:NR<br><br>Path Length: ~8 – 10 mm<br><br>Temperature: 780 ˚C<br><br>Window: None<br><br>Uncertainty: NR<br><br>Sample Size: ~0.7 g</td><td>$CoF_2$<br>$CrF_3$<br>$PrF_3$<br>$UF_4$<br>$UO_2F_2$</td><td>FLiNaK</td><td>A2|E1<br>A3|E2<br>A4|E30<br>A5|E31<br>A6|E32<br>[24]</td></tr>
<tr><td>T2<br><br>Young [23]<br><br>1959<br><br>Beer's Law</td><td>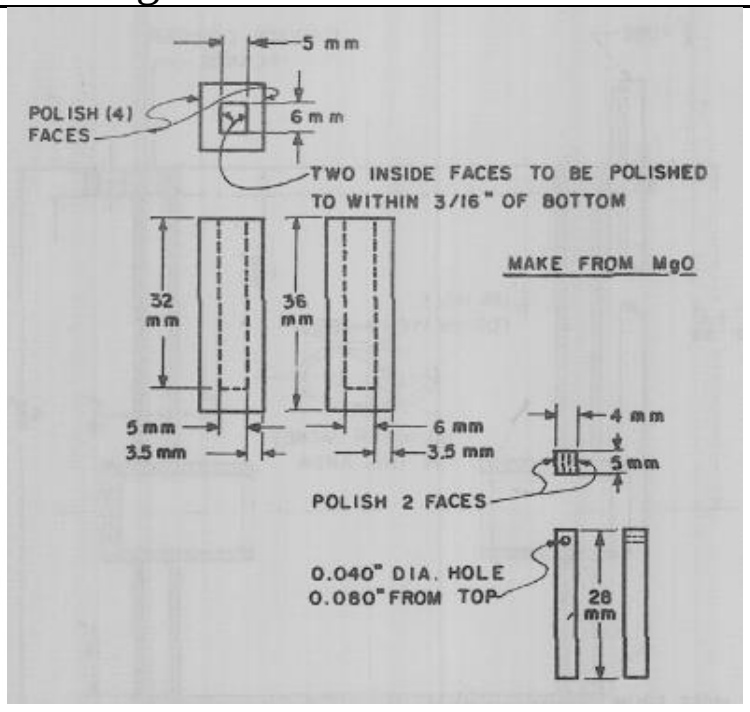

<br>Furnace: T1<br>Cell: MgO cuvette (shown in schematic)<br>Optics: Cary 14M grating spectrometer, sample compartment<br>Cover gas: Ar</td><td>Wavelength: 0.3 – 0.7 μm<br><br>$\kappa_\lambda$: NR<br><br>Path Length: 5 mm<br><br>Temperature: 780 °C<br><br>Window: MgO<br><br>Uncertainty: NR<br><br>Sample size: ~1.4 g</td><td>$NiF_2$</td><td>FLiNaK</td><td>A1|E3<br>[24]</td></tr>
</table>

| | | | | | |
|---|---|---|---|---|---|
| T3<br><br>Young [25]<br><br>1964<br><br>Beer's Law | 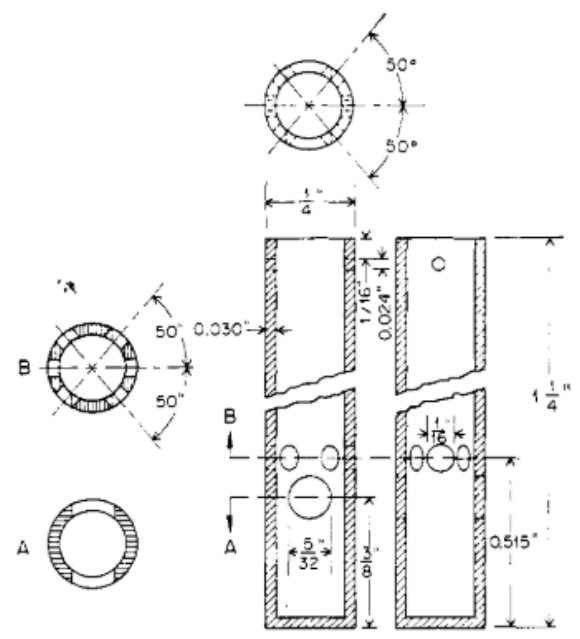<br>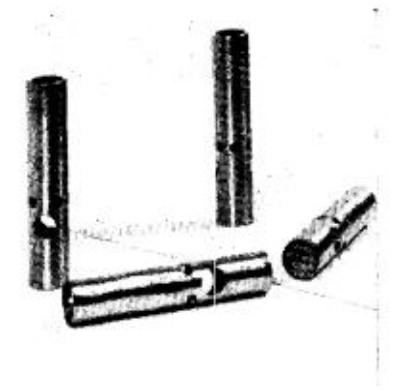<br>Furnace: T1<br>Cell: Windowless Cu or Pt tube (cross sectional schematic above, and image below)<br>Optics: Cary 14M grating spectrometer, sample compartment<br>Cover gas: Ar | Wavelength: 0.3 – 2 μm<br><br>$\kappa_\lambda$: NR<br><br>Path Length: ~8 mm<br><br>Temperature: 525 ˚C<br><br>Window: None<br><br>Uncertainty: 10%<br><br>Sample size: ~1.5 g | $NiF_2$ | FLiNaK | A7\|E4<br>[26] |
| T4<br><br>Young [27][28]<br><br>1967<br><br>Beer's Law | 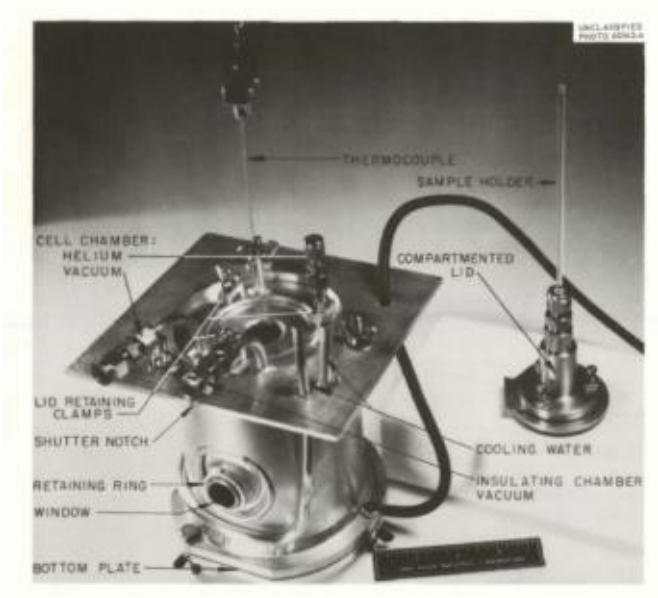<br><br>Furnace: Vacuum-tight inverted Ni tee inside of Ni block with 12 heater rods, contained within sealed assembly (shown in image)<br>Cell: T3, 0.25" OD Cu and graphite (w/o keeper holes)<br>Optics: Cary 14M grating spectrometer, sample compartment<br>Cover gas: He | Wavelength: 0.2 – 2.3 μm<br><br>$\kappa_\lambda$: NR<br><br>Path Length: ~8 mm<br><br>Temperature: 550 ˚C<br><br>Window: none<br><br>Uncertainty: 10%<br><br>Sample size: <1.5 g | $UF_4$<br>$UF_3$<br><br>$NiF_2$<br>$FeF_2$<br>$CrF_2$<br>$CrF_3$ | FLiBe, FLiNaK<br><br>FLiBe | A8\|E33<br>A9\|E33<br>A10\|E34-35<br>A11\|E34-35<br>[27]<br><br>A13\|E5<br>A14\|E6<br>A15\|E7<br>A16\|E8<br>[29] |
| T5<br><br>Young [30]<br><br>1969<br><br>Beer's Law | 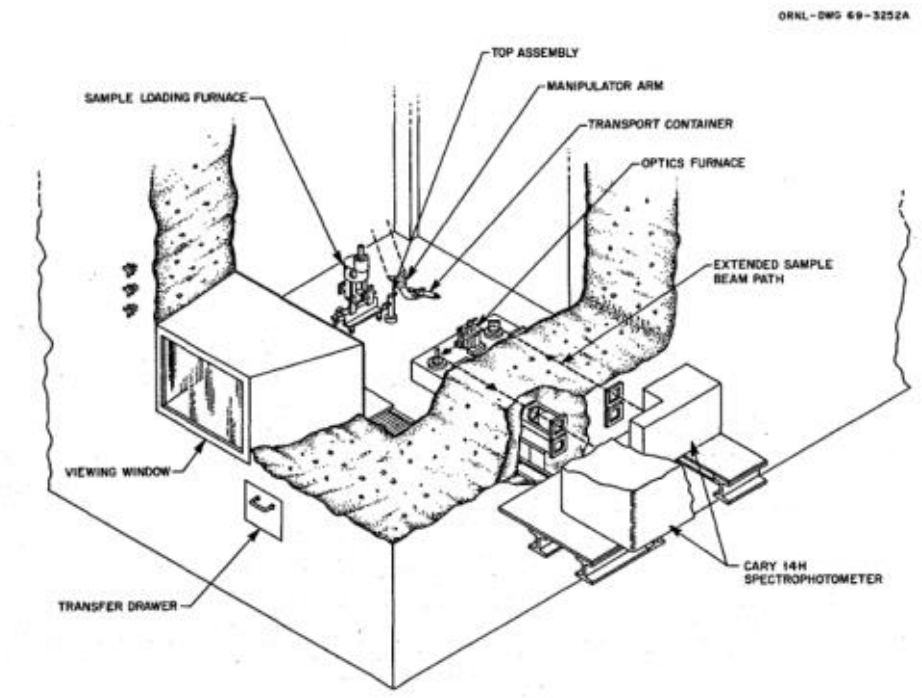<br><br>Furnace: T4<br>Cell: T3, Cu (size not specified)<br>Optics: Cary 14H grating spectrometer inside of hot cell with optical feedthroughts (Shown in schematic)<br>Cover gas: He | Wavelength: 0.2 – 2.1 μm<br><br>$\kappa_\lambda$: NR<br><br>Path Length: NR<br><br>Temperature: 575 ˚C<br><br>Window: None<br><br>Uncertainty: 10% | $PaF_4$<br><br><br>$PuF_3$ | Solvent: $LiF$-$BeF_2$-$ThF_4$ (72-16-12 mol%) | A17\|E37<br>A18\|E37<br>[31]<br><br><br>A19\|E42<br>[32] |
| T6<br><br>Young [33]<br><br>1972<br><br>Beer's Law | 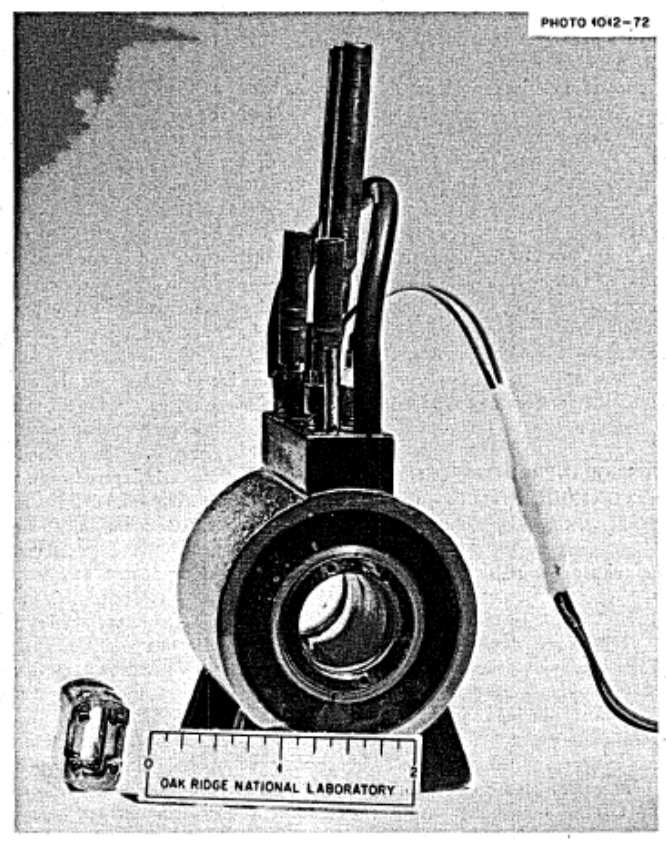<br><br>Fig. 7.6. Photograph of LaF-windowed cell and infrared furnace.<br><br>Furnace: Heating-wire-wrapped BN tube contained within sealed metal tube (shown in right of image)<br>Cell: Ni body with Au-sealed $LaF_3$ windows (shown in left of image)<br>Optics: no details given<br>Cover gas: Ar or vacuum | Wavelength: 2.6 – 3.8<br><br>$\kappa_\lambda$: NR<br><br>Path Length: NR<br><br>Temperature: 450 ˚C<br><br>Window: $LaF_3$<br><br>Uncertainty: NR<br><br>Sample size : NR | $D_3BO_3$ | $NaBF_4$ | [34] |

| | | | | |
|---|---|---|---|---|
| T7<br><br>Whiting [35]<br><br>1973<br><br>Beer's Law<br><br>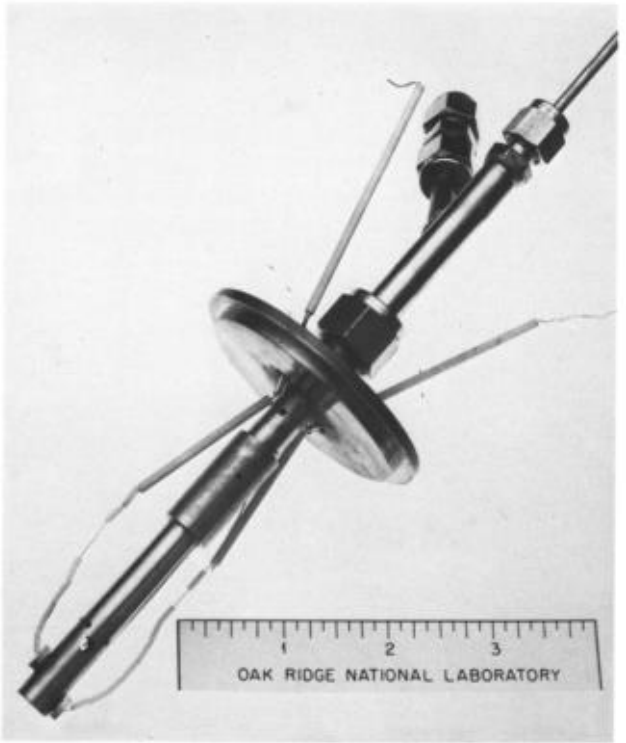<br>Furnace: T4<br>Cell: Modified 0.375" OD hard carbon, graphite, or Ni windowless (T3) with electrodes inserted through BN plugs at bottom or lowered directly through beam path<br>Optics: Cary 14M grating spectrometer, sample compartment<br>Cover gas: He | Wavelength: 0.2 – 2 μm<br><br>$\kappa_\lambda$: NR<br><br>Path Length: NR<br><br>Temperature: 540 ˚C<br><br>Window: None<br><br>Uncertainty: NR<br><br>Sample size: ~4 g | $UF_4$<br>$UF_3$<br><br>$MnF_3$<br>$MnF_2$<br>$CuF_2$ | LiF-$BeF_2$-$ZrF_4$ (66:29:5 mol%)<br><br>FLiNaK | A12\|E36<br>[36]<br><br>A23\|E14<br>A24\|E15<br>A25\|E16<br>[35] |
| T8<br><br>Bamberger [37]<br><br>1968<br><br>Beer's Law<br><br>Furnace: T4<br>Cell: silica tube<br>Optics: Cary 14M grating spectrometer, sample compartment<br>Cover gas: $SiF_4$ | Wavelength: 0.64, 1.09 μm<br><br>$\kappa_\lambda$: NR<br><br>Path Length: NR<br><br>Temperature: 698 °C<br><br>Window: Silica<br><br>Uncertainty: 5%<br><br>Sample size: NR | $UF_4$ | FLiBe | E38-41<br>[37] |

| | | | | | |
|---|---|---|---|---|---|
| T9<br><br>Bamberger [38]<br><br>1974<br><br>Beer's Law | Furnace : T4<br>Cell : 10 mm ID silica tube<br>Optics: Cary 14M grating spectrometer, sample compartment<br>Cover gas: NR | Wavelength: 0.25 – 0.7 µm<br><br>$\kappa_\lambda$: NR<br><br>Path length: 10 mm<br><br>Temperature: 600 °C<br><br>Window: Silica<br><br>Uncertainty: NR<br><br>Sample size: ~3 g | $Te_2$<br><br>$LiTe_3$<br><br>$Li_2Te$ + oxidants | FLiBe | A20\|E9-10 [38] |
| T10<br><br>Whiting [39]<br><br>1972<br><br>Beer's Law | Furnace: T4<br>Cell: 10 mm path length rectangular quartz cell<br>Optics: Cary 14M grating spectrometer, sample compartment<br>Cover gas: He | Wavelength: 0.2 – 2 µm<br><br>$\kappa_\lambda$: NR<br><br>Path length: 10 mm<br><br>Temperature: 500 °C<br><br>Window: Quartz<br><br>Uncertainty: NR<br><br>Sample size: 4 g | $NaO_2$<br>$K_2CrO_4$<br>$KNO_2$ | FLiNaK | A21\|E11<br>A22\|E12<br>E13<br>[40][39] |
| T11<br><br>Bates [41]<br><br>1972<br><br>Beer's Law | Furnace: T4<br>Cell: 2mm path length silica cell<br>Optics: Perkin-Elmer model 621 grating spectrometer, sample compartment<br>Cover gas: NR | Wavelength: ~3.72 µm<br><br>$\kappa_\lambda$: NR<br><br>Path length: 2 mm<br><br>Temperature: 425 °C<br><br>Window: Silica<br><br>Uncertainty: NR<br><br>Sample size: NR | $D_3BO_3$ | $NaBF_4$ | A26\|V69 [41] |

| | | | | |
|---|---|---|---|---|
| T12<br><br>Toth<br>[42]<br><br>1969<br><br>Beer's Law<br><br>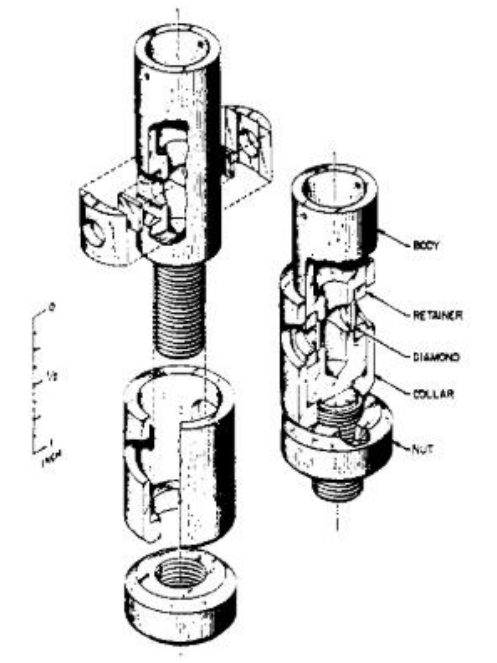<br><br>Furnace: T4<br>Cell: Type ATJ graphite body with un-sealed diamond windows<br>Optics: Cary 14H grating spectrometer, sample compartment<br>Cover gas: He | Wavelength:<br>0.2 – 2.5 μm<br><br>$\kappa_\lambda$: NR<br><br>Path Length:<br>6.35 mm<br><br>Temperature:<br>690 ˚C<br><br>Window:<br>Diamond<br>Type IIa<br><br>Uncertainty:<br>NR<br><br>Sample size:<br>~2 g | $NiF_2$<br>$UF_4$<br><br>$UF_4$<br><br>$UF_4$<br>$UF_3$ | FLiBe<br><br>FLiNaK<br>FLiBe<br>LiF-$BeF_2$<br>(48-52 mol%)<br><br>FLiBe | A27\|E17<br>A28\|E43<br>[42]<br><br>A29\|E44<br>A30\|E45-46<br>A31\|E47-48<br>[14]<br><br>[43] |
| T13<br><br>Toth<br>[44]<br>[45]<br><br>1973<br><br>Beer's Law<br><br>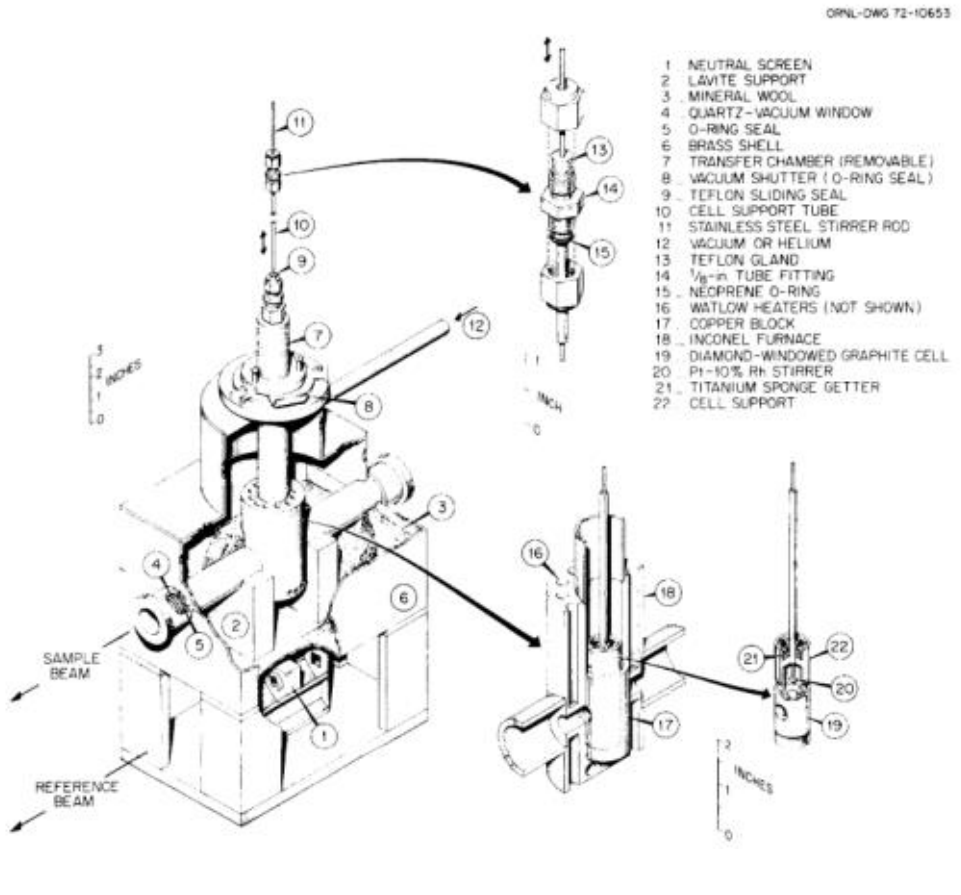<br><br>Furnace: Cylindrical inconel block with ring of heater rods, contained within sealed Cu assembly (schematic shown above from [45])<br>Cell: T12, modifed to include either Pt stirring loop [44] or Pt-10%Rh sparger tube [46] could be inserted into cell<br>Optics: Cary 14H grating spectrometer, sample compartment<br>Cover gas: He | Wavelength:<br>0.3 – 2.5 μm<br><br>$\kappa_\lambda$: NR<br><br>Path Length:<br>6.35 mm<br><br>Temperature:<br>800 ˚C<br><br>Window:<br>Diamond<br>Type Iia<br><br>Uncertainty:<br>NR<br><br>Sample size:<br>0.6 g | $UF_4$<br>$UF_3$<br><br>$NbF_4$<br>$K_2NbF_7$<br><br>$UF_4$<br>$UF_3$<br>HF<br>$H_2$<br><br>$LiTe_2$<br>$LiTe_3$ | FLiBe<br>LiF-$BeF_2$<br>(48-52 mol%)<br><br>FLiBe<br>LiF-$BeF_2$<br>(48-52 mol%)<br><br>FLiBe<br>LiF-$BeF_2$<br>(48-52 mol%)<br>LiF-$BeF_2$<br>( 57-43 mol%)<br>LiF-$BeF_2$-$ThF_4$<br>(72-16-12 mol%)<br><br>FLiBe | A32\|E49<br>A33\|E49<br>A34\|E50-52<br>[44]<br><br>A35\|E18<br>[47]<br><br>E53<br>[46]<br><br>E19<br>[48] |

| | | | | | |
|---|---|---|---|---|---|
| T14<br><br>Varlamov [49]<br><br>1989<br><br>Double-path-length method | 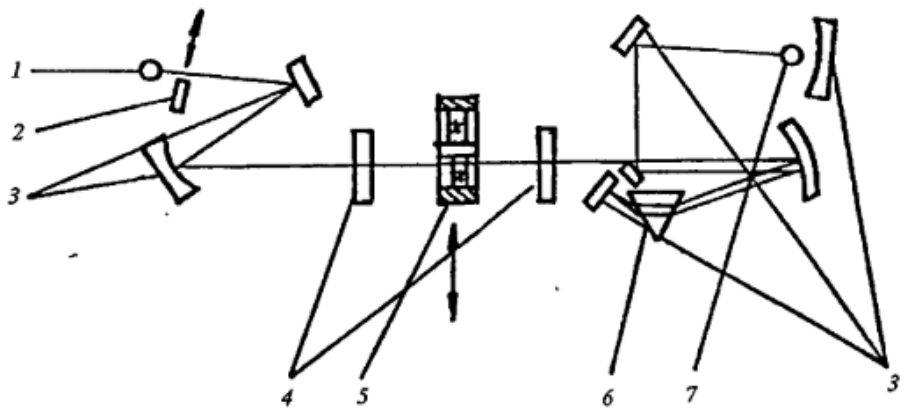<br><br>Furnace: no details given<br>Cell: windowless Pt plates or sapphire windows<br>Optics:<br>(1) Radiation source; (2) modulator; (3) rotating and focusing mirrors; (4) high temperature adapter windows; (5) holder with samples; (6) monochromator prism; (7) radiation detector<br>Cover gas: Ar or He | Wavelength:<br>4 – 14 μm<br><br>$\kappa_\lambda$:<br>10 – 2000 $m^{-1}$<br><br>Path Length:<br>NR<br><br>Temperature:<br>1527 K<br><br>Window:<br>sapphire<br><br>Uncertainty:<br>10%<br>(100 – 1000 $m^{-1}$)<br>20%<br>(10 $m^{-1}$ > )<br><br>Sample size:<br>NR | LiF<br><br>$CaF_2$<br><br>$BaF_2$ | | A36\|V1-6<br>A37\|V7-15<br>A38\|V16-23<br>[49] |
| T15<br><br>S. Liu [50]<br><br>2017<br><br>Beer's Law | 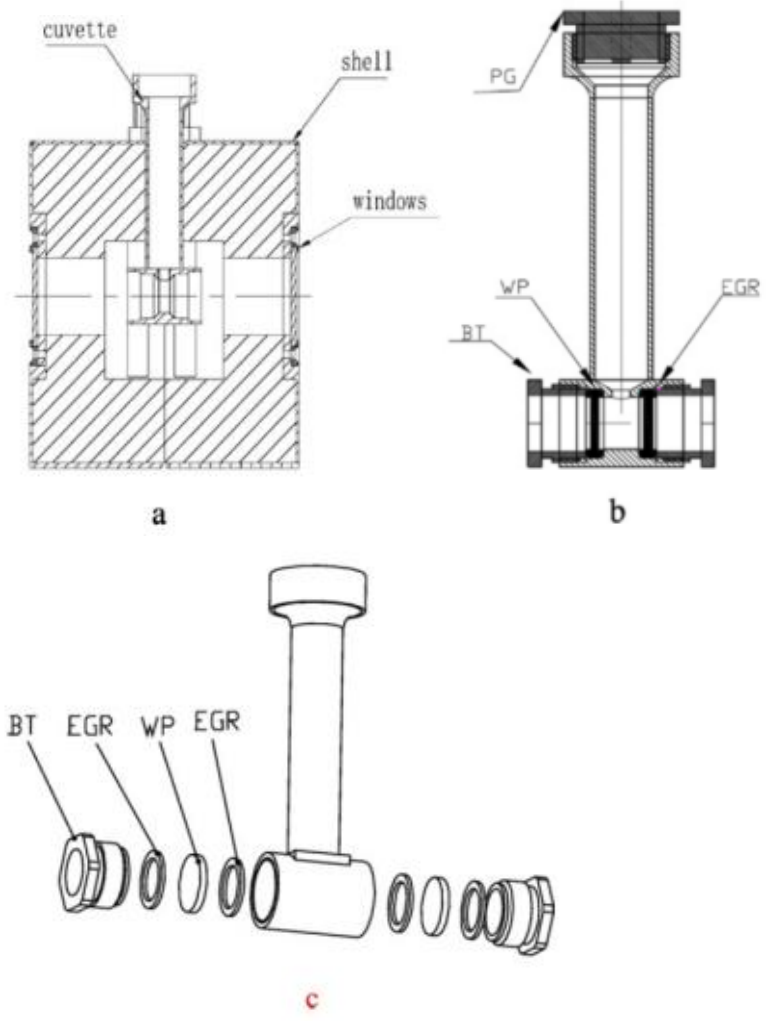<br><br>Furnace: Aluminum oxide oven with beam path drilled out. Schematic shown above (a)<br>Cell: Hastelloy C inverted tee with graphite-sealed windows. Schematics shown above (b,c) – BT: bolt, WP: window, EGR: graphite o-ring<br>Optics: Perkin Elmer spectrometer (unspecified type) sample compartment<br>Cover gas: Ar | Wavelength:<br>0.83 – 20 μm<br><br>$\kappa_\lambda$: NR<br><br>Path Length:<br>10 mm<br><br>Temperature:<br>600 ˚C<br><br>Window:<br>Diamond, SiC<br><br>Uncertainty:<br>NR<br><br>Sample size:<br>~2.3 g | LiOH | FLiBe | A39\|V24, V70<br>A40\|V24, V70<br>A41\|V71<br>[50] |

<table>
<tr>
<td>T16<br><br>H. Liu<br>[51]<br><br>2018<br><br>Beer's Law</td>
<td>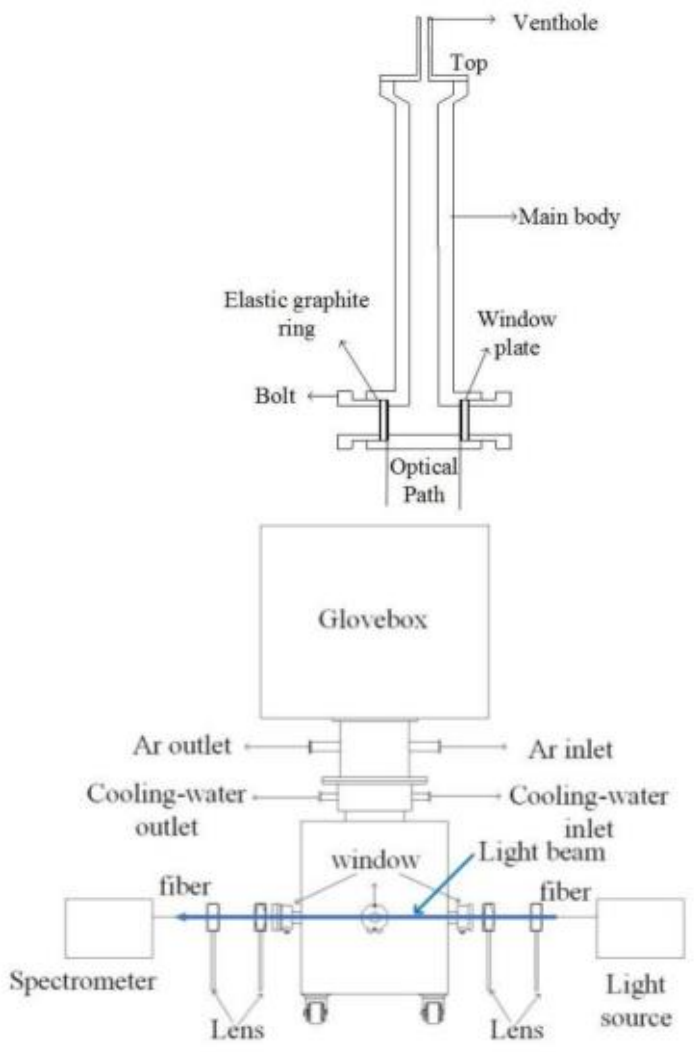

Furnace: Aluminum oxide oven with beam ports, directly coupled to glovebox with cell hoist assembly. Shown in schematic below<br>Cell: T15, modified to include hole for gas sparging tube. Shown in schematic above<br>Optics: QE Pro 65 grating spectrometer (Ocean Optics Co.)<br>Deuterium-Tungsten lamp, DT-2000 (Hangzhou SPL Photonics Co.), focusing optics, and fiber optic cable. Shown in schematic below<br>Cover gas: Ar</td>
<td>Wavelength: 0.2 – 1.0 μm<br><br>$\kappa_\lambda$: NR<br><br>Path Length: 10 mm<br><br>Temperature: 800 ˚C<br><br>Window: Diamond, SiC, Quartz, Sapphire, $CaF_2$<br><br>Uncertainty: NR<br><br>Sample size: ~2.3 g</td>
<td>$UF_4$</td>
<td>FLiBe</td>
<td>A42|E54<br>[51]</td>
</tr>
<tr>
<td>T17<br><br>Y. Liu<br>[52]<br><br>2020<br><br>Beer's Law</td>
<td>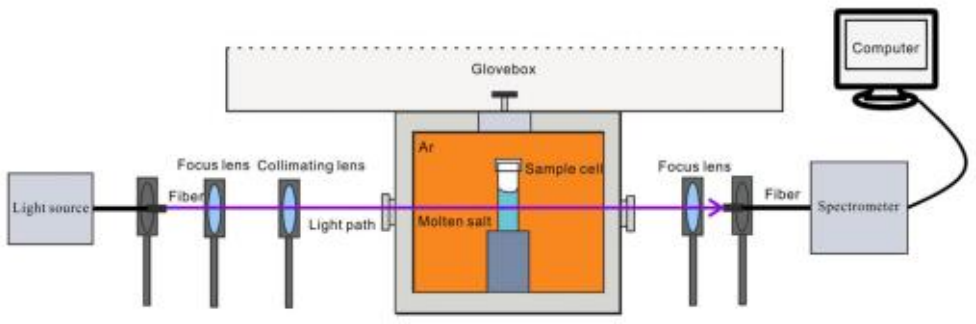

Furnace: T16<br>Cell: quartz cuvette<br>Optics: T16, with Deuterium-Tungsten lamp, iDH2000 (ideaoptics)<br>Cover gas: Ar</td>
<td>Wavelength: 0.2 – 0.9 μm<br><br>$\kappa_\lambda$: 250 $m^{-1}$<br><br>Path Length: 10 mm<br><br>Temperature: 600 ˚C<br><br>Window: Quartz<br><br>Uncertainty: NR<br><br>Sample size: 3 g</td>
<td>$CrF_2$<br>$CrF_3$</td>
<td>FLiBe<br><br>FLiNaK<br><br>LiF-$BeF_2$ (69-31 mol%)</td>
<td>A43|E20-21<br>A44|E21-23<br>A45|E24-25<br>[52]</td>
</tr>
</table>

NR indicates the parameter was not reported

## 3.1.3. Discussion

### 3.1.3.1. Furnaces

Resistively heated furnaces have been ubiquitously used to reach the desired measurement temperatures for molten fluoride transmittance setups. The first ORNL furnace consisted of a Lavite block with beam ports and a sample cavity heated by Pt-wound fused-silica plates (T1). The furnace was not vacuum-tight, and it was found that significant out-gassing would occur from the insulation or quartz heaters, which was detrimental to air-sensitive samples. Later ORNL furnaces (T4, T13) used metal blocks (typically Ni) with a cavity drilled out for the optical cell and light path, and several smaller holes drilled out for heater rods (i.e., cartridge heaters). The metal block was then situated within a larger, vacuum-tight metal enclosure with suitable optical windows and feedthroughs for the optical cell and heater. These vacuum-tight furnaces exhibited excellent spatial temperature uniformity, minimized contamination due to insulating materials and heater out-gassing, and could be held under vacuum or flushed with inert gas during operation. This was particularly important because the ORNL optical cells were not vacuum tight. Sealed quartz tubes or metal containers were used to transport samples from a glovebox to the furnace. After transport, vacuum was typically pulled on the furnace, followed by flushing of Ar or He.

Varlamov also used a vacuum-tight furnace (T14), though did not describe the heating element. Reports of SINAP setups (T15-17) did not give details on the heating element construction; thus, it is likely that commercial ceramic (aluminum oxide) ovens were used and modified to allow for beam ports and a hoist assembly for the optical cell. In the more recent studies (T16, T17), the oven was placed inside of a glovebox well with optical windows, eliminating the need to design a vacuum-tight furnace or cell.

#### 3.1.3.2. Cells

*Diamond windows*

The overall uncertainty of $\kappa_\lambda$ derived from transmittance measurements is primarily determined by the uncertainty in the path-length measurement. Several optical cells were used by ORNL researchers, with increasing accuracy and repeatability of path length measurement with successive iterations. The first cell developed was the Pt pendent drop (T2), consisting of an open horizontal tube which would be dipped into the salt, for which the authors stated that the results should be considered qualitative because the path length was not well-repeatable between reference and sample measurements. The windowless cell (T3-5, T7), consisting of a vertical capsule with small holes drilled through horizontally, enabled much more repeatable measurements, resulting in an overall uncertainty of 10%. For the diamond-windowed graphite cell (T12-13), uncertainty was not reported; however, it can be assumed to have similar uncertainty as the silica cell setup of Bamberger (T8), which was reported to be 5%. Indeed, Toth reported $UF_4$ (in FLiBe) features in the 2 μm region (Plot A28, E43) which had been previously unresolvable with the windowless cell (Plot A10, E33). This improved resolution enabled Toth to perform remarkable studies leading to better understanding of the coordination behavior of $U^{4+}$ complexes and detailed equilibria data for the $U^{3+}/U^{4+}$ redox couple (further discussed in 5.2.3.2).

Varlamov used windowless Pt plates or sapphire windows to contain liquid fluoride samples, but did not give further details or schematics (T14). SINAP researchers adapted Toth's diamond-windowed graphite cell, using instead a Hastelloy body sealed by elastic graphite gaskets (T15-16). Single-crystal SiC windows were used, in addition to diamond. The authors did not discuss their reasoning as to choosing Hastelloy versus graphite. Hastelloy has superior mechanical properties at high temperature, so it may have been chosen for better durability. Graphite is more chemically inert, but for a purified salt sample, Hastelloy (or other Ni-based alloys) would likely be acceptable. It might have been chosen for greater ease in achieving a vacuum-tight seal. Toth reported that molten fluorides do not wet to graphite or diamond; thus, a tight seal was not necessary to prevent leakage. Although not used for transmittance studies, Andersen's trans-reflectance cell design (R5) could be adapted to build a cell with a Ni or Au-lined Ni body with Au-sealed diamond windows (further discussed in 3.2.3.2).

*Oxide windows*

An intriguing possibility is the use of oxide (e.g., silica, MgO, sapphire, etc.) windows, which are more economical than diamond and have good durability at high temperature. ORNL researchers used silica or quartz cells in contact with telluride compounds in FLiBe (T9), several oxygen-containing species in FLiNaK (T10), and $D_3BO_3$ in $NaBF_4$ (T11). Young used a MgO cell for $NiF_2$ in FLiNaK (T1), but reported incompatibility if $UF_4$ or $ZrF_4$ were present. Bamberger (T8) provided evidence for the stability of silica against attack by $UF_4$ in FLiBe by introducing highly toxic $SiF_4$ cover gas to limit this reaction (eq. 7).

$$UF_4(d) + SiO_2(s) \leftrightarrow UO_2(s) + SiF_4(g) \quad (7)$$

Toth analyzed similar reactions of LiF and $BeF_2$ with various oxide windows ($SiO_2$, $Al_2O_3$, and MgO) and found that most reactions were unfavorable . They hypothesized that dissolved oxides could attack silica to form silicates via the following reactions:

$$Li_2O(d) + SiO_2(s) \leftrightarrow Li_2SiO_3(s) \quad (8)$$

$$Li_2O(d) + 2SiO_2(s) \leftrightarrow Li_2Si_2O_5(s) \quad (9)$$

Experimentally, Toth reported a successful 5-hour corrosion test of FLiBe in quartz at 500 ˚C In conclusion, Toth stated that 'optically pure' (i.e. 'freed of alkali oxide, HF, $H_2O$, and then filtered to remove suspended particles') fluoride melts 'behave surprisingly well in quartz under an atmosphere of helium' [19]. Liu et al provide some validation of this statement in their measurements of $CrF_2$ and $CrF_3$ in FLiBe and FLiNaK using quartz cuvettes (T17). The measurements were performed using pre-vacuum-dried, hydro-fluorinated solvents [53] (no mention of filtration) inside of a purified Ar glovebox ($H_2O$ and $O_2$ below 2 ppm). In summary, the use of oxide windows, in particular $SiO_2$, is certainly possible, but more work is needed to understand their limitations.

*Spectro-electrochemistry*

Spectro-electrochemistry measurements have also been conducted, using the graphite, hard (glassy) carbon, and Ni windowless cells (T7). These materials are convenient because of their chemical inertness towards the salt, and electrical conductivity allowing use of the cell body as one of the electrodes. In some cases, Pt electrodes were inserted via a BN plug at the bottom of the cell. The working electrode typically consisted of a Pt wire inserted into the cell from above and hanging in the beam path. Spectro-electrochemistry was not performed with the diamond-windowed graphite cell, but a gas sparging tube was inserted for an experiment with $H_2$ and HF bubbling (T13).

#### 3.1.3.3. Spectrometer and external optics

Almost all setups were designed to fit the furnace within a spectrometer sample compartment, requiring little to no modification of the optical path. A setup of particular interest which required substantial engineering of the optical path was the hot-cell-embedded Cary 14H grating spectrometer (T5). Varlamov's setup (T14) used a single-beam prism spectrometer, and an optical path schematic was given, but it is not clear if the furnace fit in the sample compartment or required external optics. The first SINAP setup (T15) fit in the sample compartment of a Perkin-Elmer spectrometer (unspecified type), but the second two (T16-17) involved the use of the fiber-optic cables, focusing optics, and modular light sources and grating spectrometers. The use of external optics is attractive because it allows for more flexibility in furnace and cell design.

High-temperature thermal radiation can disrupt visible-NIR measurements by saturating the detector, but the degree of severity is determined by the type of spectrometer and details of the optical path (e.g., numerical aperture, etc.). For a typical grating spectrometer, thermal radiation from the sample reaches the detector, but only at the narrow wavelength band selected by the grating. This effect can be accounted for by measuring a spectrum without the source, and then subtracting this intensity from the signal with the source on, as described by Liu [51]. This procedure is of course dependent on the grating being located between the sample and detector. For further discussion please see [16]. For a Fourier-transform spectrometer (FTS), the saturation effect can be much more severe. FTS are more relevant in the discussion of reflection setups, which operate in the MIR, and will be discussed in section 0.

### 3.1.4. Summary

For the purposes of quantitative visible/NIR absorption measurements, we recommend the transmittance method due to simple, repeatable path-length measurement and the relative ease of coupling a spectrometer to a transmittance cell. Three material combinations for transmittance cells have been well-demonstrated: graphite body with unsealed diamond windows, Hastelloy C body with graphite-sealed diamond and single-crystal SiC windows, and Au-lined Ni body with Au-sealed diamond windows. Silica or MgO cuvettes have been used in several studies, but their chemical incompatibility and failure modes are not fully understood.

Looking ahead toward future measurements, it would be of interest to develop a molten fluoride setup capable of measuring multiple path lengths. As mentioned in section 3.1.1.2, construction of such a setup (concentric, movable Inconel tubes with graphite-sealed diamond windows) has been completed, though testing with molten fluorides is yet to be done [13]. An alternative design is suggested in [16], consisting of an array of heater rods lining a machined stainless-steel block, surrounded by insulation and contained within a water-cooled brass housing. Several different sample cavities could be used, capable of holding quartz cuvettes of 2, 5 or 10 cm path length. This furnace was not included in the review because it was not used to conduct fluoride studies; however, a similar furnace could be built and used with fluoride-compatible cuvettes. In addition to increasing the range of sensitivity for measurement and detection of $\kappa_\lambda$ and ε of solute species, it would enable the use of the double-path-length method, enabling measurement of transparent/semi-transparent range $\kappa_\lambda$ and optical properties of solvents.

## 3.2. Reflectance & Trans-reflectance

### 3.2.1. Background

Here we outline techniques used to extract $\kappa_\lambda$ (and in some cases $\tilde{n}$) via reflectance measurements. In section 3.2.1.1 and Figure 3a, we describe optical fitting methods to extract $\tilde{n}$ in the opaque region (typically MIR/FIR). In section 3.2.1.2 and Figure 3b, we describe a framework for extracting $\kappa_\lambda$ in the semi-transparent region (typically NIR), which we refer to as the trans-reflectance method.

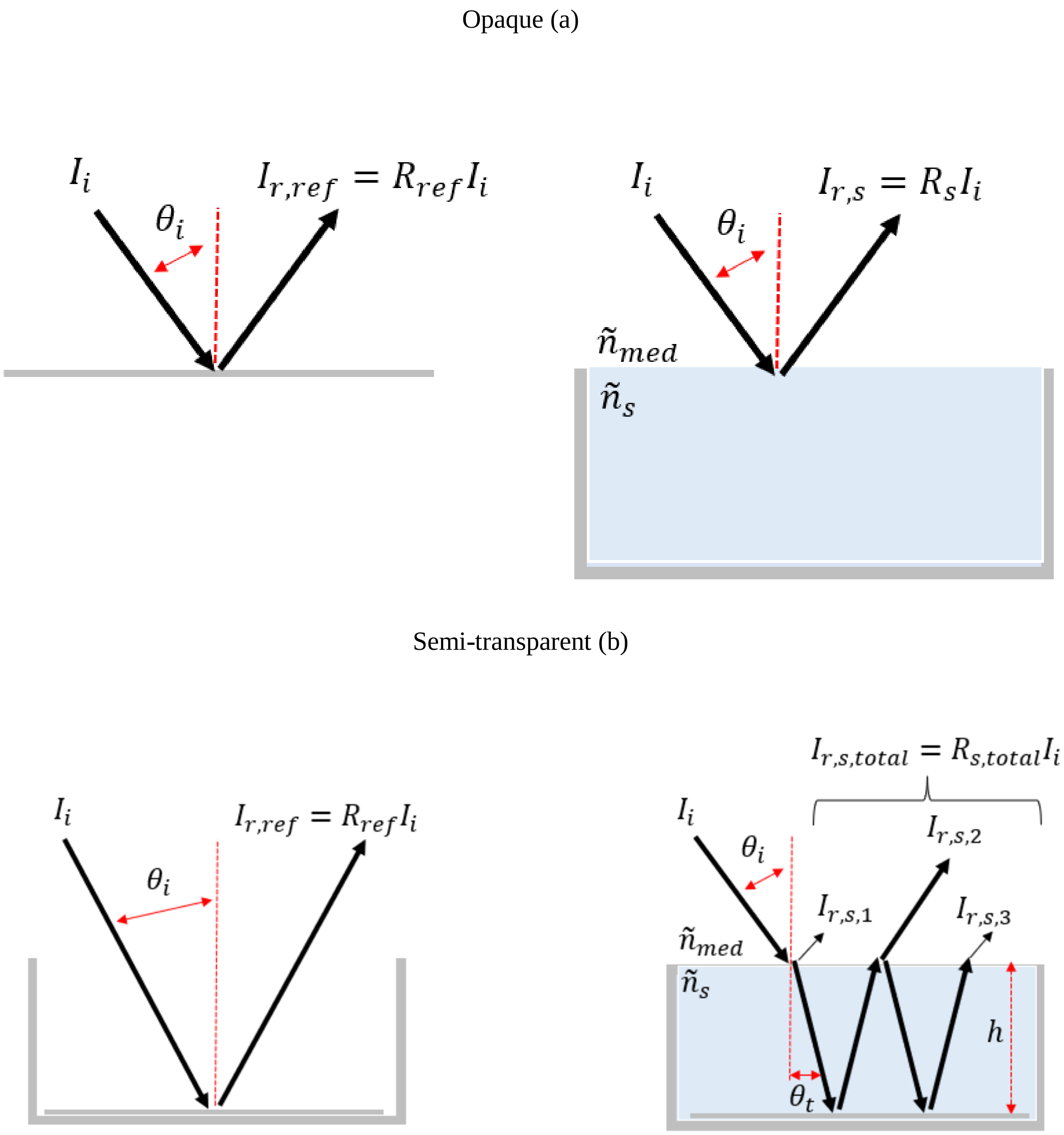

**Figure 3. Schematics of reflectance for (a) opaque and (b) semi-transparent samples.**

3.2.1.1. Opaque region

The following equation gives the sample reflectance, $R_s$:

$$R_s = R_{\text{ref}} \cdot \frac{I_{r,s}}{I_{r,ref}} \quad (10)$$

Here, $R_{\text{ref}}$ is the known reflectance of a reference sample, $I_{r,s}$ is the light intensity reflected from the sample, and $I_{r,ref}$ is the light intensity reflected from the reference sample.

An alternative method is to measure the sample emission to obtain, $\epsilon_s$, and apply Kirchoff's law: $R_s = 1 - \epsilon_s$. We note that this application of Kirchoff's law is only valid for an opaque sample. The following equation gives $\epsilon_s$:

$$\epsilon_s = \epsilon_{\text{ref}} \cdot \frac{I_{\text{e},s}}{I_{\text{e},ref}} \qquad (11)$$

We note that equation 11 is only valid if background emission is negligible [54]. The emission method is preferred for samples with high emissivity, whereas the reflectance method is preferred for samples with low emissivity. For samples within an intermediate range (~0.2-0.8), we recommend both methods as a comparison. Note that for complete agreement, the numerical aperture and the angle of measurement should be the same.

If $R_s$ is measured at near-normal incidence and $\tilde{n}_{med}$ is known (as is generally the case in air or inert gas media), then $\tilde{n}_s$ is the only unknown in equation 2 (here we substitute $\tilde{n}_s$ and $\tilde{n}_{med}$ for $\tilde{n}_1$ and $\tilde{n}_2$); however, both real and imaginary components of $\tilde{n}_s$ must be considered. Therefore, two degrees of freedom remain to be solved for, rendering a single measurement insufficient. Here we outline optical fitting methods used by molten-fluoride spectroscopists to extract $\tilde{n}_s$ from a single, near-normal reflectance measurement.

## Direct Kramers-Kronig

The first method involves the direct application of the Kramers-Kronig (KK) relations, which we refer to as the direct KK method [55]. The first step is to define the complex EM field reflectance, $\tilde{r}$, which depends on the magnitude, $r$, and phase shift, $\bar{\theta}$, as shown in the following equation.

$$\tilde{r} = re^{i\bar{\theta}} \qquad (12)$$

The measured reflectance is related by $R = |\tilde{r}|^2 = r^2$. Following the formulation by Bowlden and Wilmshurst [55], $r$ and $\bar{\theta}$ are defined as functions of $\omega$, the angular frequency. $\omega$ is related to wavelength as: $\omega = 2\pi c/\lambda$. Eventually, the following expression is reached, relating $r$ to $\bar{\theta}$.

$$\bar{\theta}(\omega) = -\frac{1}{\pi}\int_0^\infty \frac{d[\ln r(\omega')]}{d\omega'} \ln\left|\frac{\omega' - \omega}{\omega' + \omega}\right| d\omega' \qquad (13)$$

$r(\omega)$ can be obtained via measurement of $R$; thus, $\bar{\theta}(\omega)$ can be obtained through computation of the integral in equation 13. $n$ and $k$ are obtained from the following equations.

$$n = \frac{1 - r^2}{1 - 2r\cos\bar{\theta} + r^2} \qquad (14)$$

$$k = \frac{-2r\cos\bar{\theta}}{1 - 2r\cos\bar{\theta} + r^2} \qquad (15)$$

In practice, the integration limits of 0 and $\infty$ are not reachable, which induces some error in this approach. Bowlden and Wilmshurst quantified this error by applying the KK method to synthetic $\tilde{n}$ data, for the cases of an oscillator with one or two resonances. They found that if the resonances were fully captured by the integration bounds, errors near the resonance would be surprisingly small. They attributed this due to a canceling of error terms for $\bar{\theta}$ (equation 13) that occurs when $\omega'/\omega_{min} \approx \omega_{max}/\omega'$. Outside of the resonance region, they observed larger error and obviously unphysical behavior, such as negative values for $k$ (seen in Figure 5c and Figure 7a), but this was deemed acceptable for absorption spectroscopy, in which the peak location and intensity are of greater interest. They did not evaluate error in $n$, but the error in the product $(nk)$ at the resonance tended to be larger than that for $k$ alone.

We note that several reflectance studies (using R1 and R2) in this review used the KK method.

## Oscillator Model

To reduce the errors of the KK method, additional constraints must be placed on $\tilde{n}$. Here we present a widely used KK-consistent model wherein it is assumed that charges within the material behave as driven, damped, harmonic oscillators in response to an oscillating electromagnetic wave. For each resonance, a set of parameters (see Table 2) are defined that govern the spectral behavior of $\tilde{\epsilon}_r$, the relative electric permittivity. $\tilde{\epsilon}_r$ is a complex quantity, and the equations for the real and imaginary parts are given in equations 17 and 18. Following the notation of Jasperse [56] and Mead [57], the spectroscopic wavenumber, $\nu = 1/\lambda$, is used here as the spectral variable.

$$\tilde{\epsilon}_r = \epsilon_r' + i\epsilon_r'' \tag{16}$$

$$\epsilon_r' = \epsilon_{r,\infty} + \sum_j \frac{4\pi\rho_j {\nu_j}^2({\nu_j}^2 - \nu^2)}{({\nu_j}^2 - \nu^2)^2 + (\gamma_j \nu)^2} \tag{17}$$

$$\epsilon_r'' = \sum_j \frac{4\pi\rho_j {\nu_j}^2(\gamma_j \nu)}{({\nu_j}^2 - \nu^2)^2 + (\gamma_j \nu)^2} \tag{18}$$

**Table 2. List of fitting parameters for oscillator model**

| | |
|---|---|
| $\nu_j$ | Frequency of $jth$ resonance |
| $4\pi\rho_j$ | Strength of $jth$ resonance |
| $\gamma_j$ | Damping factor of $jth$ resonance |
| $\epsilon_{r,\infty}$ | Permittivity in the high-frequency limit |

Under the assumption that the relative magnetic permeability ($\mu_r$) is equal to 1, as is the case for all known materials across the wavelengths of interest in this review, the following equation may be used to obtain $\tilde{n}$.

$$\tilde{n} = \sqrt{\tilde{\epsilon_r}} \quad (19)$$

Then, $\tilde{n}$ is used to calculate $R$, compared with the experimental $R$, and iterated until an acceptable fit is reached.

We note that a single molten-fluoride reflectance study (using R4) used the oscillator method. However, Jasperse [56] used the oscillator method to extract temperature-dependent (nearly up to melting point) $n$ and $k$ of crystalline LiF.

3.2.1.2. Semi-transparent region

Use of a reflectance setup in the semi-transparent region of the salt leads to a method known as trans-reflectance. This method has found use in molten-salt optical spectroscopy because it does not require an optical window to be in direct contact with the salt sample. A representative schematic of the trans-reflectance method is given in Figure 4b. One possible setup consists of a crucible with a mirror placed at the bottom, referred to as the immersed mirror method. Another option is to suspend the mirror in the salt, referred to as the suspended mirror method. The salt layer above the mirror has a thickness of $h$ and the path length, $s$, is determined in equation 20 via $h$ and the transmitted angle, $\theta_t$. $\theta_t$ is calculated using Snell's law (eq. 21), where $\tilde{n}_{med}$ is the complex refractive index of the surrounding medium, $\theta_i$ is the angle of incidence, and $\tilde{n}_s$ is the complex refractive index of the salt.

$$s = \frac{h}{cos\theta_t} \quad (20)$$

$$\tilde{n}_{med} sin\theta_i = \tilde{n}_s sin\theta_t \quad (21)$$

Here, we present the formulation by Barker (R3) used in an immersed mirror setup. The reflectance of the salt-vapor interface is $R_{s,front}$ and the transmittance is $1 - R_{s,front}$. The reflectance of the salt-mirror interface, or back surface, is assumed to be equal to $R_{ref}$, the reflectance of the reference mirror in air or inert cover gas. The transmittance of the salt layer is $\mathcal{T}_s = \exp(-\kappa_\lambda s)$. The total reflected light intensity, $I_{r,s,total}$, can then be divided into several components. The first, $I_{r,1}$, is reflected by the front surface. The second, $I_{r,2}$, transmits through the front surface, reflects from the back surface, then transmits again through the front surface. $R_{ref}$ is typically near 1; hence, multiple reflections may occur, depending on the magnitudes of $R_{s,front}$ and $\mathcal{T}_s$.

$$I_{r,s,total} = R_{s,total} I_{\mathrm{i}} = I_{r,1} + I_{r,2} + I_{r,3} + \cdots + I_{r,\mathrm{n}} + \cdots = \sum_{i=1}^{\infty} I_{r,\mathrm{i}} \quad (22)$$

$$R_{s,total} = R_{s,front} + R_{ref}\left(1 - R_{s,front}\right)^2 \mathcal{T}_s^2 \left(1 + R_{s,front} R_{ref} \mathcal{T}_s^2\right) \quad (23)$$

The equation for $R_{s,total}$ (eq. 21) is given by Barker [58], and only accounts for $I_{r,\mathrm{i}}$ up to $\mathrm{i} = 3$. This was deemed acceptable because higher order terms would include $R_{s,front}$ raised to the second power or higher, and $R_{s,front}$ had been determined to be on the order of a few percent. In Barker's experiment, $R_{s,front}$ was measured independently by replacing the mirror with a sandblasted substrate, which effectively reduced $R_{ref}$ to near zero.

Khokhryakov (R7) presents a simplified immersed-mirror approach wherein $R_{s,front}$ is assumed to be 0, resulting in $R_{s,total} = R_{ref}\mathcal{T}_s^2$ [59]. Beer's Law is then employed to obtain $\mathcal{T}_{solute}$ (reported as $A_{solute}$).

Zhang (R6) presents yet another approach using the suspended mirror method [60]. By hanging a mirror within the melt at a slight angle, they were able to spatially separate the various reflected components, and selectively measure $I_{r,2} = R_{ref}(1 - R_{s,front})^2 \mathcal{T}_s^2 I_\mathrm{i}$. The mirror could then be lowered or raised to perform subsequent measurements at different path lengths, enabling the use of the double-path-length method.

### 3.2.2. Compilation of reflectance setups

Table 3 lists the reflectance setups covered in this review. For clarity, the setups have been defined by each unique combination of cell, furnace, and optical path. In the first column, we give the setup index (R#), primary author and reference, year of publication, and data-processing method (i.e., Direct KK, oscillator model, immersed mirror, suspended mirror). The second column provides schematics and equipment details (i.e., furnace, cell, optical path, cover gas). The third column gives parameters such as wavelength range, maximum or range of $\kappa_\lambda$ measured, maximum temperature, optical window used, amount of salt required, and reported uncertainty. The next columns give details on the studies performed, such as the salt compositions used, and provides references to relevant plots (A#) and spectra indices (E#/V#).

**Table 3. Experimental reflectance and trans-reflectance setups for measurement of $\kappa_\lambda$ via extraction of optical properties using optical theory or via various trans-reflectance techniques. If unspecified, sample size was estimated using dimensions of crucible and assuming to be 2/3 full and using an approximate salt liquid density of 2 g/cm$^3$.**

| Experimental Set-up | Schematic and Equipment Details | Parameters | Studies Performed | | |
|---|---|---|---|---|---|
| | | | Salt Composition | | Data |
| | | | Solute | Solvent | |

<table>
<tr><td>R1<br><br>Wilmshurst [61] [62]<br><br>1963<br><br>Direct KK</td><td>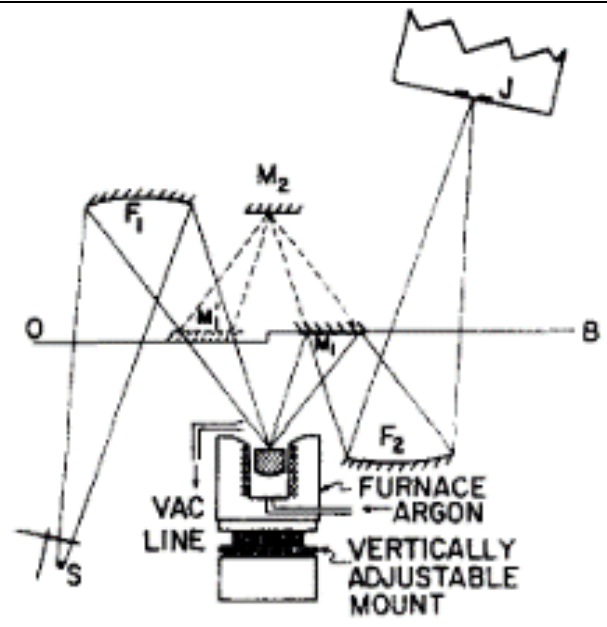

Furnace: no details given<br>Cell: Ni crucible<br>Optics: Perkin-Elmer Model 12C spectrometer (NaCl, KBr, CsI prisms)<br>Schematic of optical path [62]:<br>S – light source; OB – optical bench;<br>$F_1$, $F_2$ – spherical mirrors;<br>$M_1$, $M_2$ – mirrors; $M_1$' – standard mirror;<br>J – monochromator slit<br>Cover gas: Argon</td><td>Wavelength: 10 – 50 μm<br><br>Max $\kappa_\lambda$: 7.6E+05 $m^{-1}$<br><br>Max Temperature: 850 °C<br><br>Window: None<br><br>Uncertainty: 2% repeatability [a]<br><br>Sample size: NR</td><td colspan="2">LiF-KF<br>Range of compositions<br><br>LiF-KF-$ZrF_4$<br>(37.5-37.5-25 mol%)<br><br>NaF-KF-$ZrF_4$<br>Range of compositions</td><td>A46|V35-39<br>A47|V37, V40<br>A48|V41-45<br>[63]</td></tr>
<tr><td>R2<br><br>Fordyce [64] [65]<br><br>1965<br><br>Direct KK</td><td>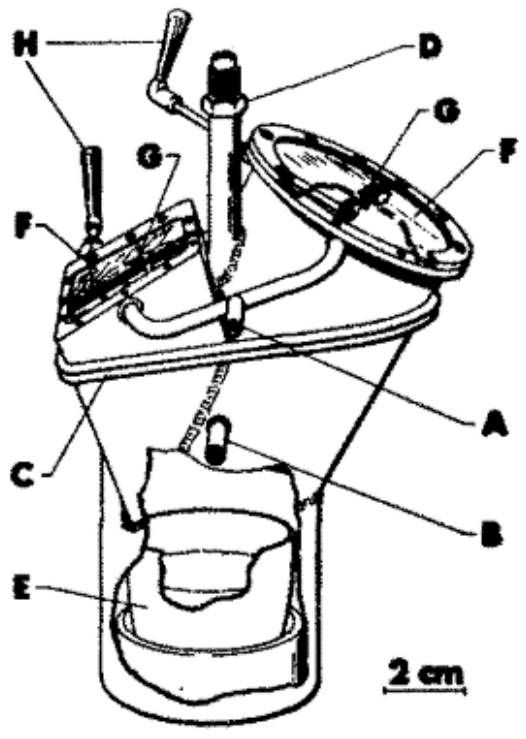

Furnace: R1<br>Cell: Ni or Pt crucible held inside of sealed, windowed inconel container<br>Schematic of optical cell:<br>A – Ar inlet; B – Ar outlet; C – cooling coil; D- thermocouple sleeve; E – Ni crucible; F – windows; G – teflon wipers; H – wiper manipulator<br>Optics: Perkin-Elmer 221 double-beam spectrometer (CsBr prism) for λ < 40 μm<br>Perkin-Elmer 12C (CsI prism) for λ > 38 μm<br>Same optical path as R1<br>Cover gas: Argon</td><td>Wavelength: 10 – 50 μm<br><br>Max $\kappa_\lambda$: 5.6E+05 $m^{-1}$<br><br>Max Temperature: 750 °C<br><br>Window: KBr [b], CsI [b]<br><br>Uncertainty: 5% precision 10% (λ > 20 μm) due to $H_2O$ vapor fine absorption<br><br>Sample size : ~50 g</td><td>$TaF_5$<br><br>$NbF_5$<br><br>$TaF_5$ + $H_2O$</td><td>LiF-KF LiF-NaF equimolar<br><br>LiF-KF equimolar<br><br>LiF-KF equimolar</td><td>A49|V46-48<br>A50|V49-50<br>A51|V47-48<br>A52|V50<br>[64]<br><br>A53|V51-52<br>A54|V52<br>[65]<br><br>A55|V53-54<br>A56|V55-57<br>[66]</td></tr>
</table>

| | | | | |
|---|---|---|---|---|
| R3<br><br>Barker [58]<br><br>1973<br><br>Immersed mirror | (a)<br>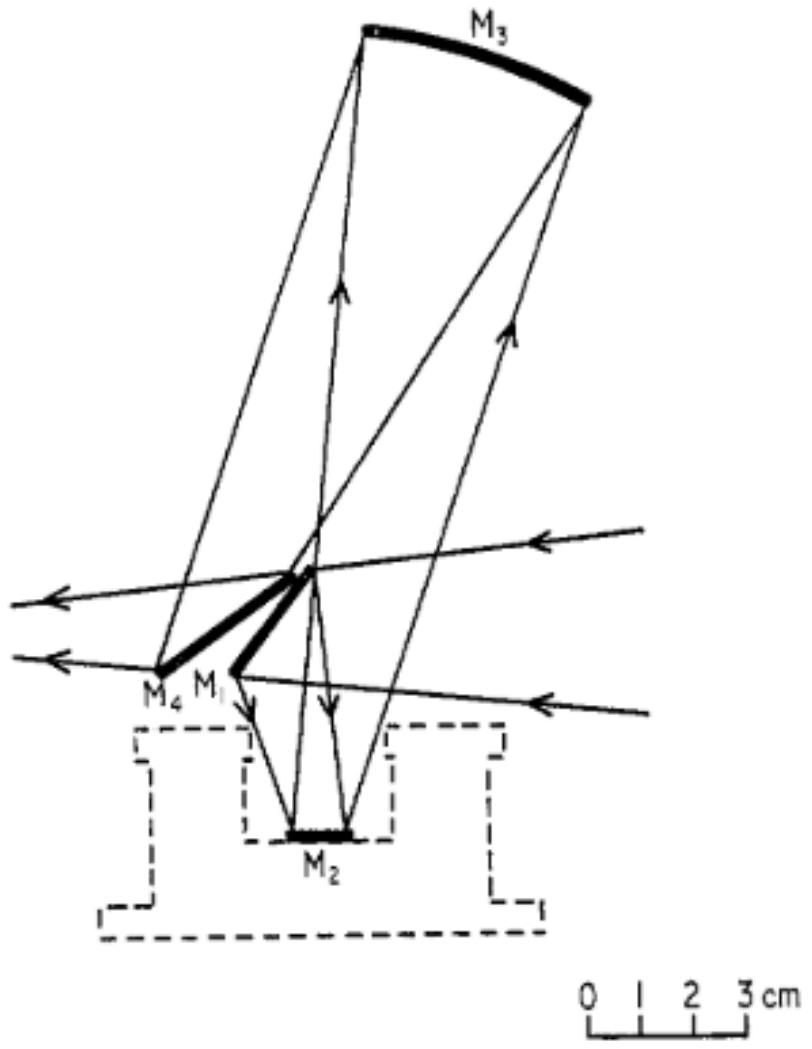<br>(b)<br>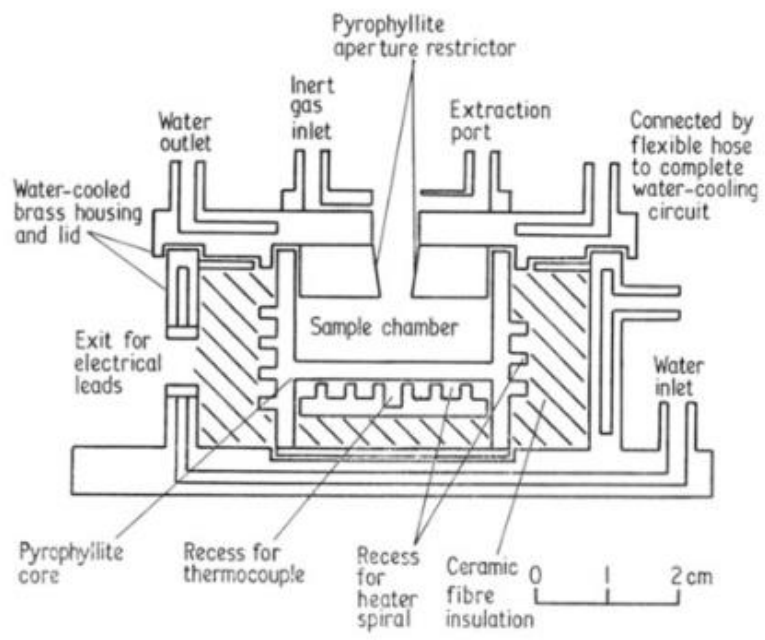<br>Furnace and cell: Pt crucible surrounded by Kanthal-wire-wrapped pyrophillite core inside of windowless brass housing. Schematic shown above (b)<br>Optics: Reflectance accessory inside of Perkin-Elmer model 457 grating spectrometer sample compartment. Schematic of optical path shown above (a)<br>Cover gas: inert (species not specified) | Wavelength:<br>8 – 40 μm<br><br>$\kappa_\lambda$:<br>10 – 1000 $m^{-1}$<br><br>Path Length:<br>2 – 4 mm<br><br>Maximum Temperature:<br>842 ˚C<br><br>Window:<br>None<br><br>Uncertainty:<br>5-10%<br><br>Sample size:<br>~1-2 g | **LiF**<br>NaCl<br>KBr | A57\|V25-29<br>[67] |

| | | | | | |
|---|---|---|---|---|---|
| R4<br><br>Mead<br>[57]<br><br>1973<br><br>Oscillator model | Furnace and cell: R3<br>Optics: Perkin-Elmer model 457 grating spectrometer for λ < 40 μm<br><br>Fourier Transform Spectrometer (FTS) for λ > 40<br>National Physical Laboratory Michelsen type interferometer (Grubb-Parsons Ltd)<br>Melinex beamsplitters<br>quartz-windowed Golay detector<br><br>Cover gas: inert (species not specified) | Wavelength:<br>10 – 100 μm<br><br>Max $\kappa_\lambda$:<br>7.0E+05 $m^{-1}$<br><br>Maximum Temperature:<br>1023 ˚C<br><br>Window:<br>none<br><br>Uncertainty:<br>2% fit error [a]<br>2% repeatability [a]<br>2% background emission [a]<br>0.5% spectrometer noise [a]<br><br>Sample size:<br>~0.1-6 g | LiF<br><br>CsF | | A58\|V58<br>A59\|V30-33<br>[57]<br><br>A60\|V59-61<br>[68] |
| R5<br><br>Andersen<br>[69]<br><br>1999<br><br>Immersed mirror / Beer's Law | 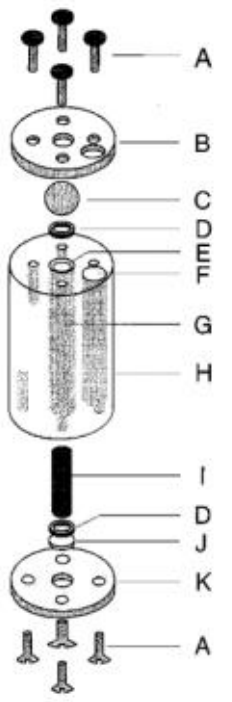<br>Furnace: Aluminum-bronze core with four heater rods<br>Cell: Au-lined Ni cylinder with Au-sealed diamond windows<br>Schematic of optical cell shown above<br>A: screws; B: top lid; C: Ni ball; D: Au gasket; E: Au lining; F: thermocouple well; G: sample chamber; H: Ni body; I: Au piston; J: diamond window; K: bottom lid<br>Optics: Bomen DA3.026 FTS<br>Globar source, KBr beamsplitter, MCT detector<br>Cover gas: Ar | Wavelength:<br>7 – 22 μm<br><br>$\kappa_\lambda$: NR<br><br>Path Length:<br>~0 - 2 mm<br><br>Temperature:<br>750 ˚C<br><br>Window:<br>Diamond<br><br>Uncertainty:<br>NR<br><br>Sample size:<br>~80 mg | $K_2NbF_7$<br>$Na_2O$ | FLiNaK | A61\|V65-68<br>[69] |

<table>
<tr>
<td rowspan="1">R6<br><br>Zhang [60]<br><br>2001<br><br>Suspended mirror / double-path-length method</td>
<td>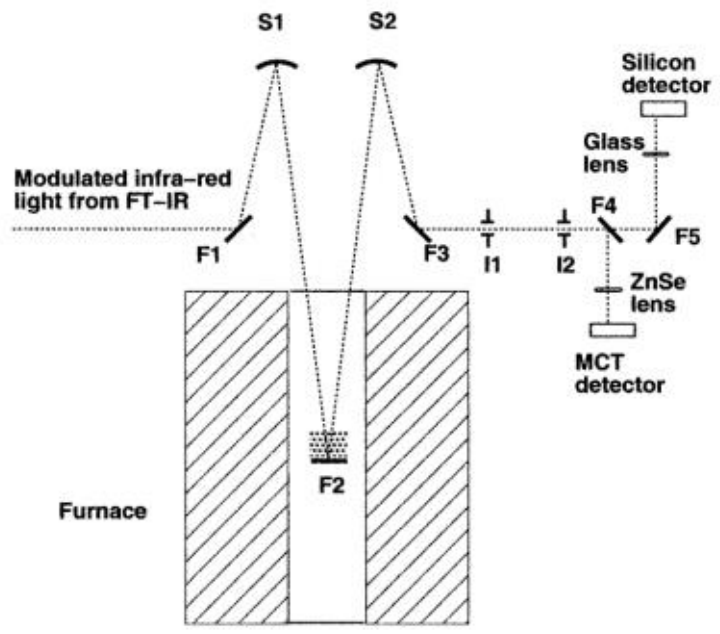

Furnace: Details not given.<br>Cell: Pt crucible with Pt suspended mirror<br>Optics: Mattson Galaxy FTS<br>Tungsten filament source, MCT or Si detector<br>Schematic of optical path<br>F1-5: flat mirrors; S1-2: spherical mirrors; I1-2: irises (aperture)<br>Cover gas: NS</td>
<td>Wavelength:<br>0.8 – 3.33 μm<br><br>$\kappa_\lambda$:<br>10 – 250 $m^{-1}$<br><br>Path Length:<br>4 – 10 mm<br><br>Temperature:<br>1300 ˚C<br><br>Window:<br>None<br><br>Uncertainty:<br><10%<br><br>Sample size:<br>NR</td>
<td colspan="2">Si, Ca, Al, B oxide glass</td>
<td>[60]</td>
</tr>
<tr>
<td rowspan="4">R7<br><br>Khokhryakov [70]<br><br>2014<br><br>Immersed mirror / Beer's Law</td>
<td rowspan="4">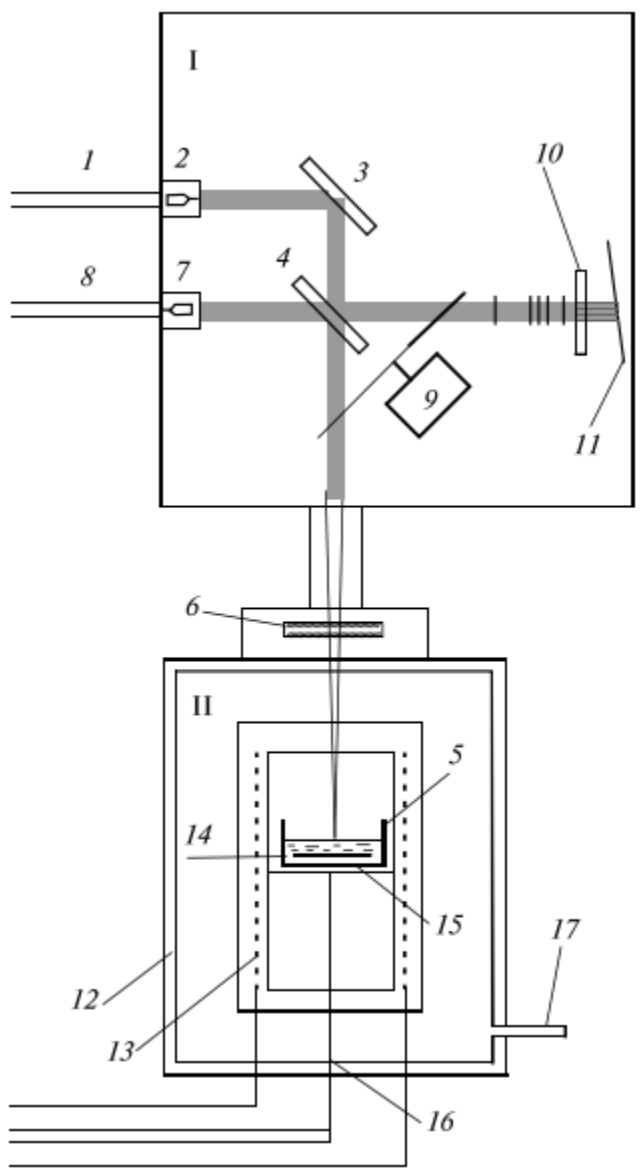

Furnace: Mo-wrapped alumina rod inside of sealed assembly<br>Cell: Pt-Rh crucible with immersed Pt-Rh mirror<br>Optics: UV-Vis S150-2048/250 grating spectrometer, 10140-1108 CCD matrix detector<br>S150-250-IR spectrometer, InGaAs linear photodiode matrix (G9201-256S)<br>Schematic of optical path shown abvove<br>1: optical fiber transmitting light from source; 2: collimator; 3: mirror; 4: 50-50 beamsplitter; 5: melt sample; 6: sapphire window; 7: collimator; 8: optical fiber transmitting light to detector; 9: shutter; 10: Pt-Rh reference sample; 11: optical trap; 12: water-cooled cell body; 13:</td>
<td rowspan="4">Wavelength:<br>0.25 – 1.67<br><br>$\kappa_\lambda$: NR<br><br>Path Length:<br>~2.0 mm<br><br>Temperature:<br>1377 ˚C<br><br>Window:<br>Sapphire[b]<br><br>Uncertainty:<br>NR<br><br>Sample size:<br>~0.6 g</td>
<td>$CeF_3$<br>$Ce_2O_3$</td>
<td>LiF-NaF<br>LiF-NaF-$ZrF_4$</td>
<td>A62|E55-56<br>[71][c]</td>
</tr>
<tr>
<td>$CoF_2$<br>$FeF_3$</td>
<td>$CaF_2$<br>$BeF_2$<br>NaF</td>
<td>A63|E26<br>A64|E27<br>A65|E28<br>A66|E29<br>[70]</td>
</tr>
<tr>
<td>$PrF_3$<br>$Pr_2O_3$</td>
<td>FLiNaK<br>LiF<br>NaF<br>KF<br>CsF</td>
<td>A67|E57-60<br>A68|E61<br>A69|E62<br>A70|E63<br>[72]</td>
</tr>
<tr>
<td>$GdF_3$</td>
<td>LiF<br>NaF<br>KF<br>CsF<br>NaF-CsF</td>
<td>A71|E64-69<br>[73]</td>
</tr>
</table>

| | | | | |
|---|---|---|---|---|
| heater; 14: Pt-Rh container; 15: Pt-Rh mirror; 16: Pt-Rh thermocouple; 17: gas port<br>Cover gas: He | | | | |

[a]Uncertainty values are provided are with respect to *R*
[b]Window not in contact with liquid salt
[c]Older version of setup used [74]

### 3.2.3. Discussion

#### 3.2.3.1. Furnaces

As with the transmittance method, the reviewed molten-fluoride reflectance setups have all used resistively heated furnaces. Wilmshurst (R1) and Zhang (R6) give schematic views showing uncovered, cylindrical ovens holding a crucible filled with molten salt (or glass), but provide no details of construction; thus, it is likely that commercial ceramic ovens were used with minimal modification. Wilmshurst reported that Ar was flushed across the melt surface. Barker designed a smaller furnace that consisted of a Kanthal-wire-wrapped pyrophyllite core, surrounded by insulation and contained within a water-cooled brass assembly (R3). A beam slit was introduced above and inert gas was flushed across. Andersen reports the use of an aluminum-bronze metal core with four heater rods (R5), without further details or a schematic. Khokhryakov's furnace (R7) consisted of a Mo-wrapped alumina rod contained within a vacuum-tight enclosure and used He flush gas.

#### 3.2.3.2. Cells

In general, reflectance cells are much simpler than transmittance cells and, in some cases, consist of simply a crucible. For the trans-reflectance method, a mirror must be immersed or suspended. Pt (R3-4 and R6), Au (R5), Pt-Rh (R7), and Ir (R7 [71]) have been demonstrated as suitable crucible and salt-immersed mirror materials. Defocusing issues can arise due to meniscus formation in smaller crucibles (Barker used 25 mm diameter, but recommended 40 mm to completely remove this effect [58]). Khokhryakov (R7) reports the use of 20 mm diameter without mention of defocusing; however, they reported crucible warping due to melt surface tension at very high temperature (~1400 ˚C). Wilmshurst and Zhang did not report crucible diameter.

Two of the more complex cell designs are described by Fordyce (R2) and Andersen (R5). Based on Fordyce's schematic (R2), their crucible appears to be about 40 mm diameter. Their cell was designed for use with high vapor pressure melts, which required holding the crucible in a sealed Inconel container with optical windows. The optical windows were flushed with Ar gas and included movable Teflon™ wipers to remove deposits. Andersen designed another unique reflectance cell (R5) consisting of a Au-lined

Ni cylinder sealed with Au gaskets against a Ni ball at one end and against a diamond window at the other. The cell was held inverted, such that the salt would rest directly on the diamond window. For short path-length measurements, an Au piston was inserted to press the salt onto the window. This is the sole trans-reflectance design in which a salt-facing window was used; thus, it is relevant also for transmittance cell design.

When performing trans-reflectance measurements, another challenge for smaller crucibles is the path-length measurement. Typically, it is difficult to access the salt during optical measurements and insertion of a cold depth probe would remove frozen salt after the measurement. Barker laboriously measured the salt depth using a Pt electrically heated contact probe combined with visual access and a vertically moving stage to determine the locations of the crucible bottom and salt surface, reporting an uncertainty of 5-10%. Depth could also be calculated by using the measured sample mass and cup dimensions, along with the salt's temperature-dependent density. Use of a larger crucible by Zhang enabled the suspended mirror approach (R6), which consisted of a Pt mirror suspended by three Pt rods. The mirror was aligned to the optical path, and the furnace was mounted below on a translation stage. After melting the sample, the furnace would be raised until the mirror became immersed, and measurements would take place at multiple path lengths, enabling the double path length method to be used. Uncertainty was reported to be less than 10%.

### 3.2.3.3. Spectrometer and external optics

All the reflectance setups required reflective external optics to direct light onto the melt and collect the reflected light. The simplest case was the reflectance accessory designed by Barker (R3), which fit in a Perkin-Elmer model 457 grating spectrometer sample compartment. Likewise, Andersen (R5) reports that their furnace fit inside of a Bomen FTS sample compartment, but does not give details regarding reflective optics. Due to their larger furnaces, Wilmshurst (R1) and Zhang (R6) used considerably external optics consisting mostly of reflective optics (preferred for MIR measurements). Likewise, Khokhryakov's large furnace required significant external optics (R7); however, their setup consisted of fiber optics and lenses.

A challenge faced by high-temperature reflection measurements in the MIR is detector saturation due to sample thermal emission. As mentioned in 3.1.3.3, the problem is present for both FTS and grating spectrometers, but is more severe for the former. Zhang (R6) briefly described the problem and outlined a mitigation strategy: increasing the power of the light source and reducing the solid angle of the light leaving the sample to keep the detector in the un-saturated regime. This saturation issue and a similar attenuation strategy is discussed in detail in [75]. Another example of optics design aimed at maximization of source intensity and throughput can be found in Appendix A of [76]. In addition to these mitigation strategies, the direct emission can be used to improve measurement reliability by performing a complementary emissivity measurement, as shown by Mead (R4).

### 3.2.4. Summary

Reflectance setups can be more complex than transmittance because they require at a minimum a reflective optical accessory (R3) and at most extensive optical components external to the spectrometer (R1, R6). That said, fluoride salts are highly absorptive in the MIR through the FIR, making reflectance measurements better suited for quantitative determination of $\kappa_\lambda$ in this region. This is generally done via $R$ measurement over a wide spectral range, followed by application of optical fitting (direct KK or oscillator model) to determine $n$ and $k$. A challenge related to MIR measurements are that optical fibers are less commercially available, requiring bulkier reflective optics. FTSs are typically used in this region and tend to be more expensive than grating spectrometers and are more susceptible to saturation. For qualitative visible-NIR measurements, the trans-reflectance method, as demonstrated by Khokhryakov (R6), has advantages over transmittance due to simpler optical cell design, typically consisting of a Pt crucible with a Pt (or Pt-Rh) immersed mirror.

If the challenge of building a MIR reflection setup is taken on, the results could be quite fruitful. We recommend the suspended mirror approach (R6), enabling quantitative trans-reflectance measurements in the visible-NIR. Following that, reflectance measurements could be made in the MIR-FIR along with optical fitting to obtain $n$ and $k$. Such a wide spectral range can be relatively easily accomplished via a FTS using appropriate combinations of sources, beamsplitters, and detectors. Of course, great care must be taken to maximize the power and throughput of the source in combination with attenuation, collimation, and solid-angle reduction to avoid saturation due to direct emission by the sample. Through optical fitting via an oscillator model or KK, this setup would enable complete optical property characterization. To improve the optical fitting accuracy, the setup could be modified to allow for measurements at varying incident angles and varying polarization. Addition of an external MIR detector would allow for complementary emission measurements, increasing the measurement reliability.

## 4. EXPERIMENTAL DATA

Here, we provide the molten-fluoride optical absorption data compiled from our search. We present spectral data in Figure 4, which consists of 71 plots generated from data digitized from graphs using the Origin™ 2016 v.93E and 2020 software. We provide the data in digital format with the supplemental materials. The molecular structure of the salt dictates the observed optical resonances. The wavelength of the resonances, in turn influences their impact on RHT. The two types of resonances that are found in the wavelength range relevant to RHT are electronic and vibrational. We summarize solvent vibrational edge absorption data in Table 4 and solvent and solute vibrational resonance absorption data in Table 5, listing the peak locations and intensities. We summarize solute electronic absorption data in Table 6 (lighter elements) and Table 7 (heavier elements).

## 4.1. Compilation of Spectra

**Figure 4. Compilation of optical spectra for molten fluoride salts in the visible and infrared range, following the order in which the corresponding setups are introduced in Table 1 and Table 3. Each plot is identified at the top left corner by plot index | setup index | spectra index; details on the spectra can be found by looking up the spectra index (e.g., E#, V#) in Table 4, Table 5, Table 6, and Table 7; details on the setup used can be found by looking up the setup index (e.g., T#, R#) in Table 1 and Table 3. Vertical-axis variables: A absorbance, $\kappa_\lambda$ spectral linear absorption coefficient, ε molar absorptivity, transmittance, emissivity, and *R* reflectance. Horizontal axis: wavelength. The µm unit was selected because of its relevance to RHT analysis. When possible, we report the reference spectra used (except for ε): 'NS' neutral density screen reference, 'AIR' air reference, 'SUB' solvent-subtracted. Data files for all plots are provided as supplemental material.**

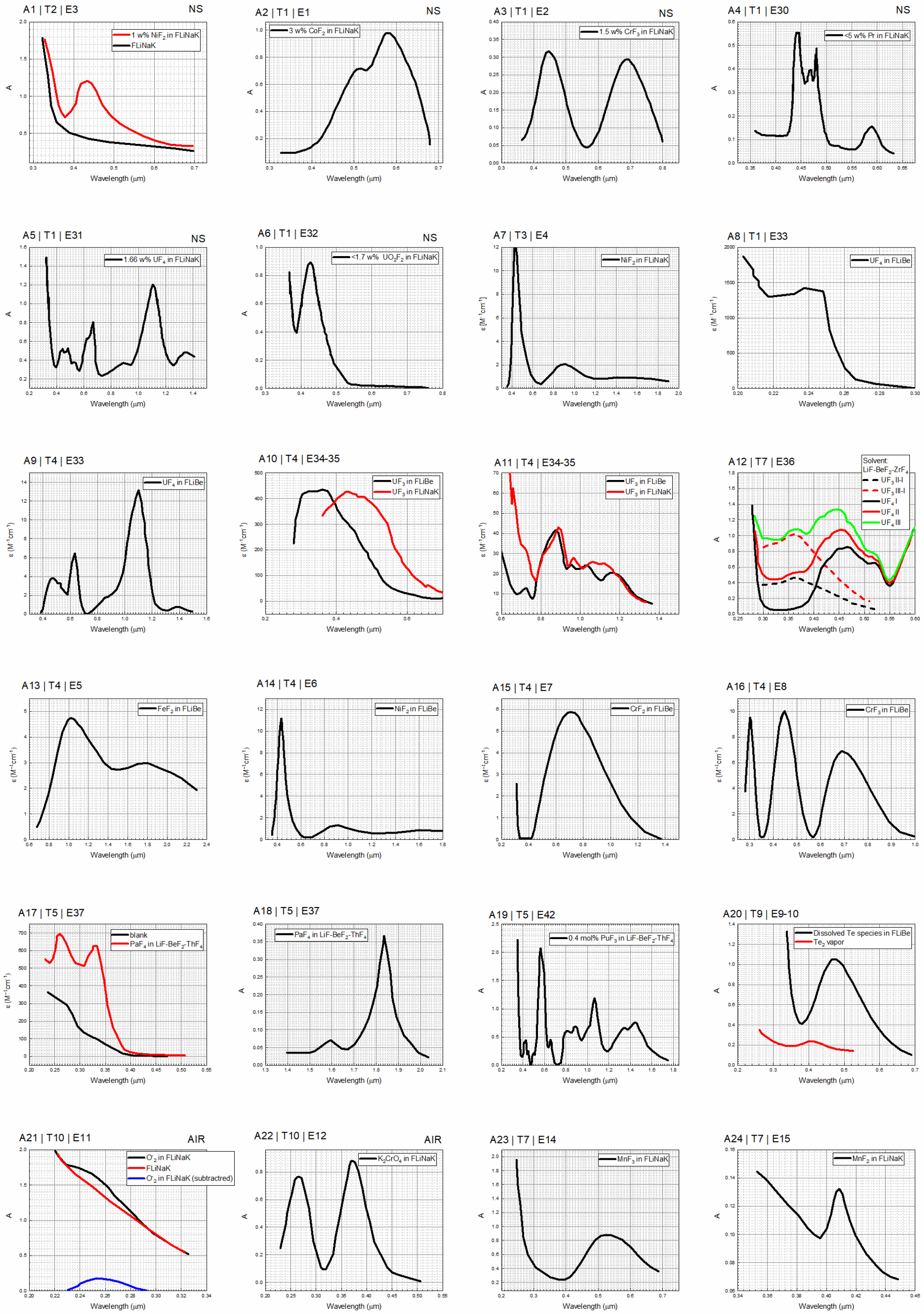
A1 | T2 | E3
NS
1 w% NiF2 in FLiNaK
FLiNaK
A2 | T1 | E1
NS
3 w% CoF2 in FLiNaK
A3 | T1 | E2
NS
1.5 w% CrF3 in FLiNaK
A4 | T1 | E30
NS
<5 w% Pr in FLiNaK
A5 | T1 | E31
NS
1.66 w% UF4 in FLiNaK
A6 | T1 | E32
NS
<1.7 w% UO2F2 in FLiNaK
A7 | T3 | E4
NiF2 in FLiNaK
A8 | T1 | E33
UF4 in FLiBe
A9 | T4 | E33
UF4 in FLiBe
A10 | T4 | E34-35
UF3 in FLiBe
UF3 in FLiNaK
A11 | T4 | E34-35
UF3 in FLiBe
UF3 in FLiNaK
A12 | T7 | E36
Solvent:
LiF-BeF2-ZrF4
UF3 II-I
UF3 III-I
UF4 I
UF4 II
UF4 III
A13 | T4 | E5
FeF2 in FLiBe
A14 | T4 | E6
NiF2 in FLiBe
A15 | T4 | E7
CrF2 in FLiBe
A16 | T4 | E8
CrF3 in FLiBe
A17 | T5 | E37
blank
PaF4 in LiF-BeF2-ThF4
A18 | T5 | E37
PaF4 in LiF-BeF2-ThF4
A19 | T5 | E42
0.4 mol% PuF3 in LiF-BeF2-ThF4
A20 | T9 | E9-10
Dissolved Te species in FLiBe
Te2 vapor
A21 | T10 | E11
AIR
O'2 in FLiNaK
FLiNaK
O'2 in FLiNaK (subtracted)
A22 | T10 | E12
AIR
K2CrO4 in FLiNaK
A23 | T7 | E14
MnF3 in FLiNaK
A24 | T7 | E15
MnF2 in FLiNaK
Wavelength (μm)
A
ε (M⁻¹cm⁻¹)

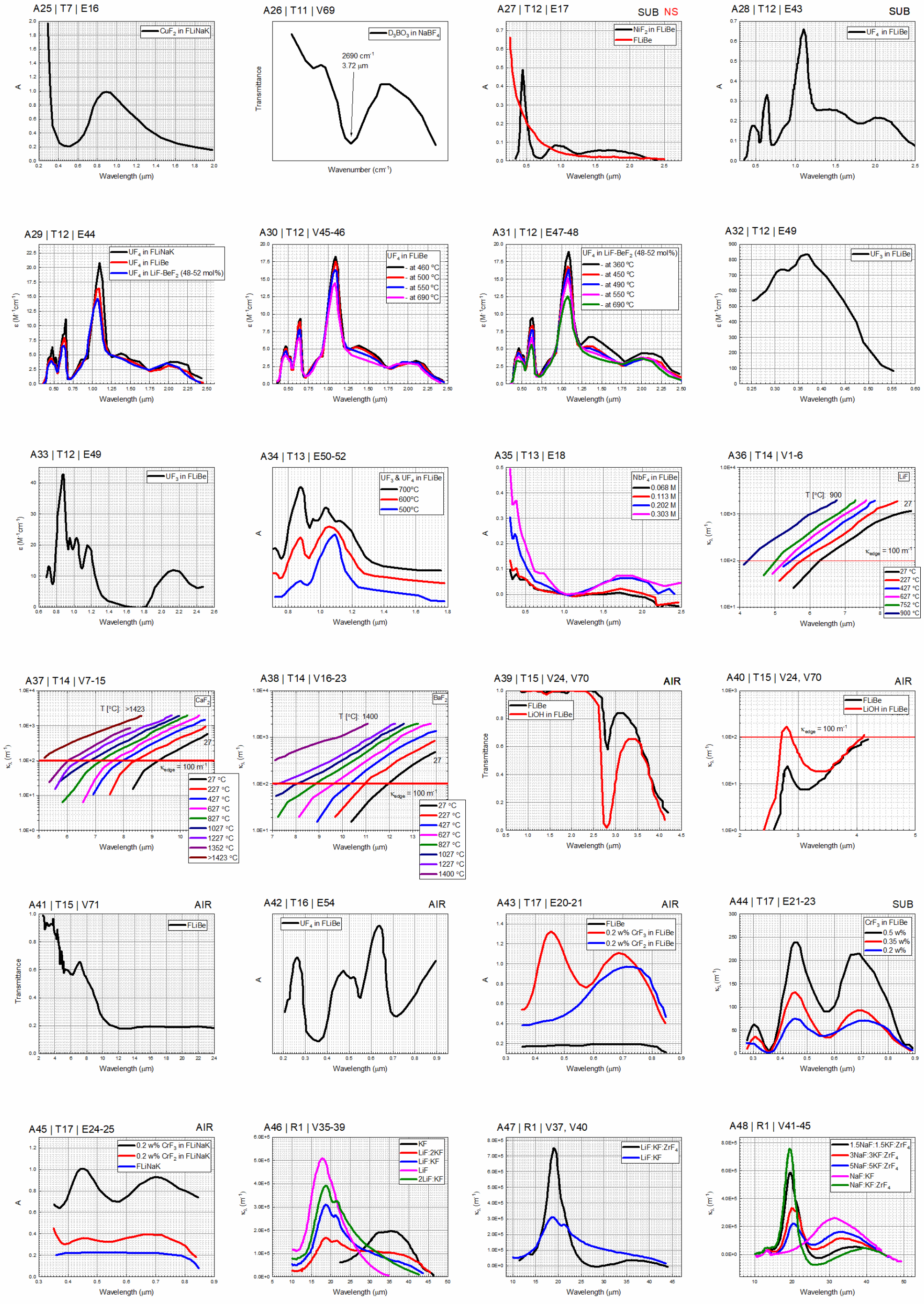

A25 | T7 | E16
CuF2 in FLiNaK
Wavelength (µm)
A26 | T11 | V69
D3BO3 in NaBF4
2690 cm-1
3.72 µm
Transmittance
Wavenumber (cm-1)
A27 | T12 | E17
SUB NS
NiF2 in FLiBe
FLiBe
Wavelength (µm)
A28 | T12 | E43
SUB
UF4 in FLiBe
Wavelength (µm)
A29 | T12 | E44
UF4 in FLiNaK
UF4 in FLiBe
UF4 in LiF-BeF2 (48-52 mol%)
Wavelength (µm)
A30 | T12 | V45-46
UF4 in FLiBe
at 460 °C
at 500 °C
at 550 °C
at 690 °C
Wavelength (µm)
A31 | T12 | E47-48
UF4 in LiF-BeF2 (48-52 mol%)
at 360 °C
at 450 °C
at 490 °C
at 550 °C
at 690 °C
Wavelength (µm)
A32 | T12 | E49
UF3 in FLiBe
Wavelength (µm)
A33 | T12 | E49
UF3 in FLiBe
Wavelength (µm)
A34 | T13 | E50-52
UF3 & UF4 in FLiBe
700°C
600°C
500°C
Wavelength (µm)
A35 | T13 | E18
NbF4 in FLiBe
0.068 M
0.113 M
0.202 M
0.303 M
Wavelength (µm)
A36 | T14 | V1-6
LiF
T [°C]: 900
27
κedge = 100 m-1
27 °C
227 °C
427 °C
627 °C
752 °C
900 °C
Wavelength (µm)
A37 | T14 | V7-15
CaF2
T [°C]: >1423
27
κedge = 100 m-1
27 °C
227 °C
427 °C
627 °C
827 °C
1027 °C
1227 °C
1352 °C
>1423 °C
Wavelength (µm)
A38 | T14 | V16-23
BaF2
T [°C]: 1400
27
κedge = 100 m-1
27 °C
227 °C
427 °C
627 °C
827 °C
1027 °C
1227 °C
1400 °C
Wavelength (µm)
A39 | T15 | V24, V70
AIR
FLiBe
LiOH in FLiBe
Transmittance
Wavelength (µm)
A40 | T15 | V24, V70
AIR
FLiBe
LiOH in FLiBe
κedge = 100 m-1
Wavelength (µm)
A41 | T15 | V71
AIR
FLiBe
Transmittance
Wavelength (µm)
A42 | T16 | E54
AIR
UF4 in FLiBe
Wavelength (µm)
A43 | T17 | E20-21
AIR
FLiBe
0.2 w% CrF3 in FLiBe
0.2 w% CrF2 in FLiBe
Wavelength (µm)
A44 | T17 | E21-23
SUB
CrF3 in FLiBe
0.5 w%
0.35 w%
0.2 w%
Wavelength (µm)
A45 | T17 | E24-25
AIR
0.2 w% CrF3 in FLiNaK
0.2 w% CrF2 in FLiNaK
FLiNaK
Wavelength (µm)
A46 | R1 | V35-39
KF
LiF:2KF
LiF:KF
LiF
2LiF:KF
Wavelength (µm)
A47 | R1 | V37, V40
LiF:KF:ZrF4
LiF:KF
Wavelength (µm)
A48 | R1 | V41-45
1.5NaF:1.5KF:ZrF4
3NaF:3KF:ZrF4
5NaF:5KF:ZrF4
NaF:KF
NaF:KF:ZrF4
Wavelength (µm)

A49 | R2 | V46-48

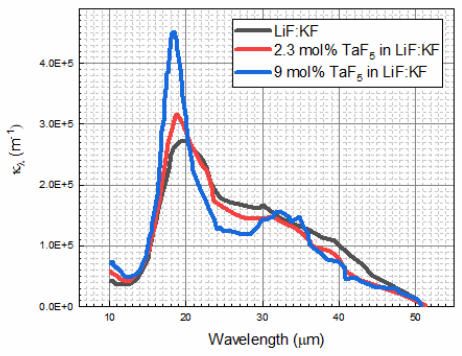

A50 | R2 | V49-50

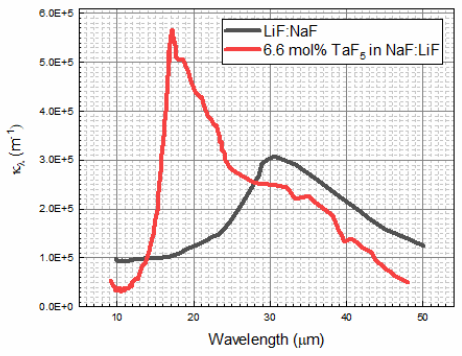

A51 | R2 | V47-48 SUB

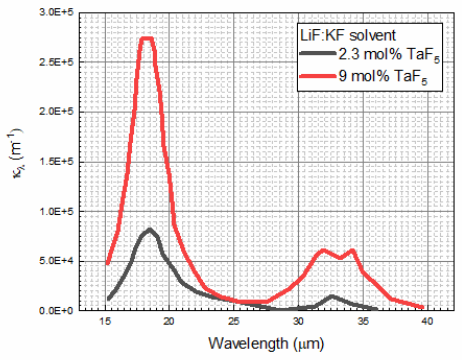

A52 | R2 | V50 SUB

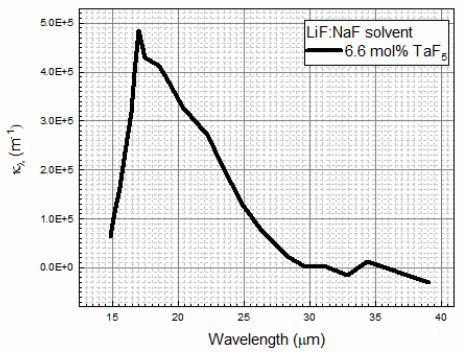

A53 | R2 | V51-52

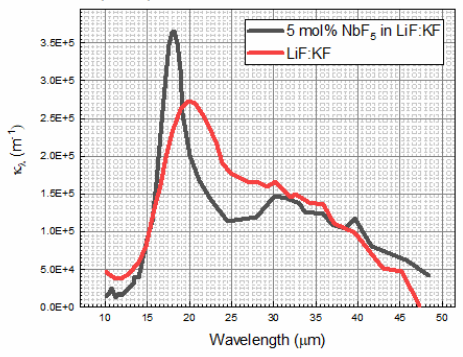

A54 | R2 | V52 SUB

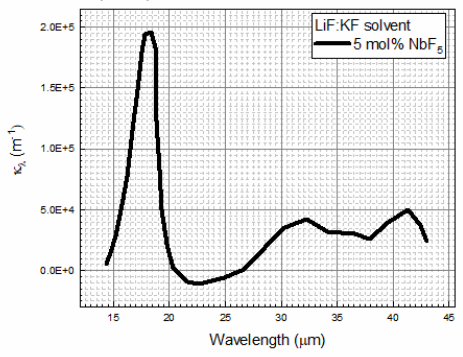

A55 | R2 | V53-54

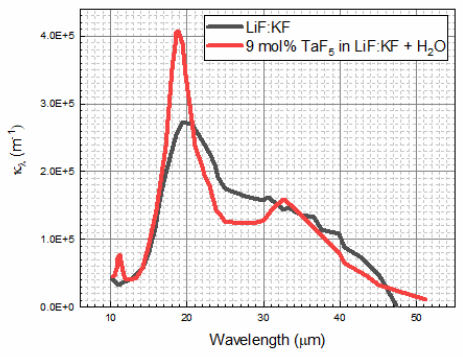

A56 | R2 | V55-57 SUB

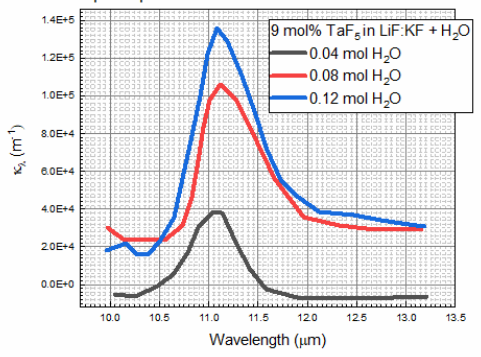

A57 | R3 | V25-29

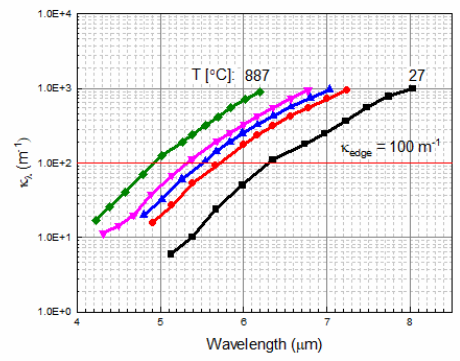

A58 | R4 | V58

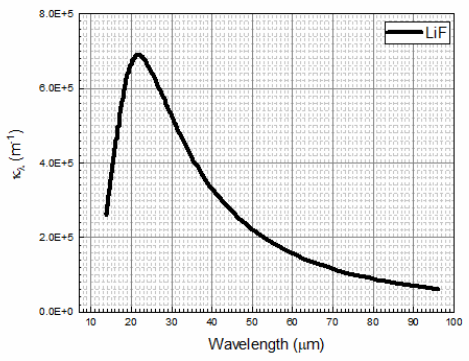

A59 | R4 | V30-33

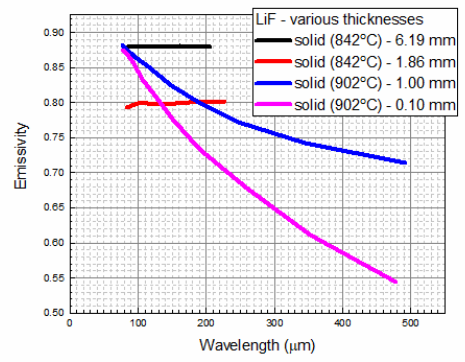

A60 | R4 | V59-61

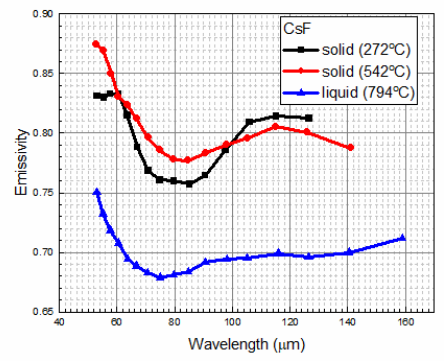

A61 | R5 | V65-68 SUB

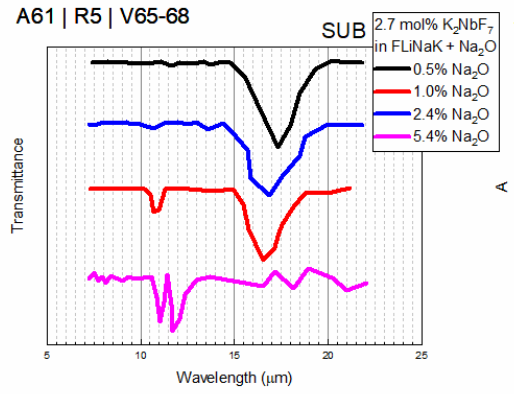

A62 | R7 | E55-56 SUB

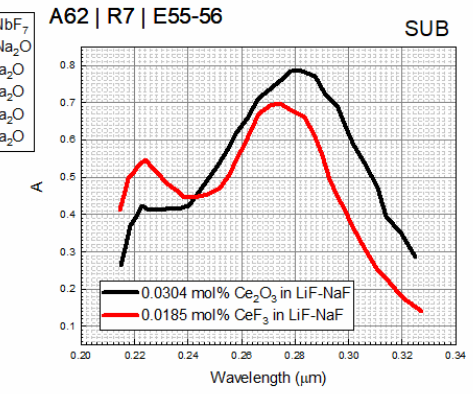

A63 | R7 | E26 SUB

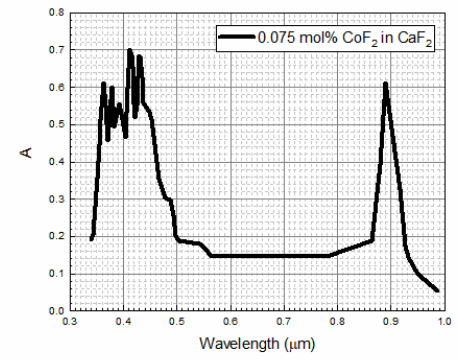

A64 | R7 | E27 SUB

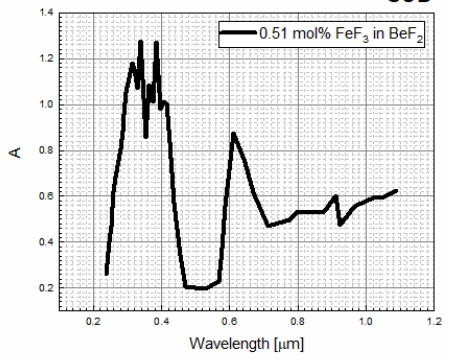

A65 | R7 | E28

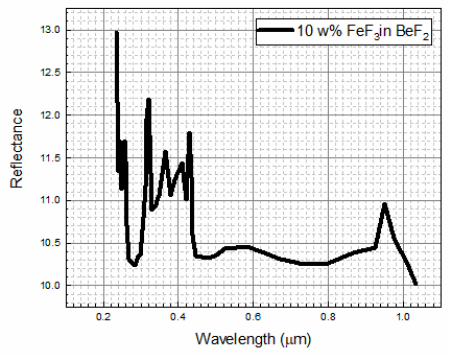

A66 | R7 | E29

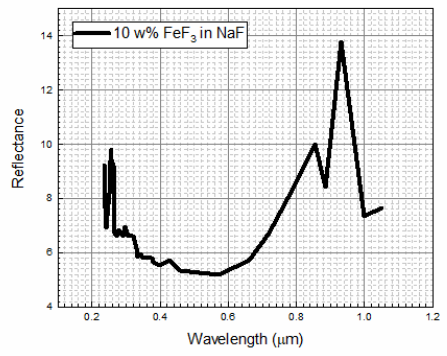

A67 | R7 | E57-60 SUB

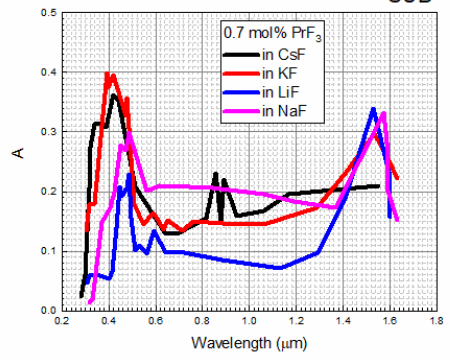

A68 | R7 | E61 SUB

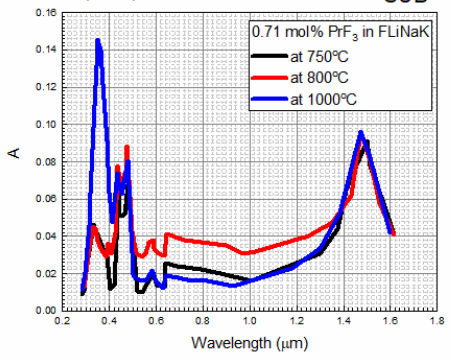

A69 | R7 | E62

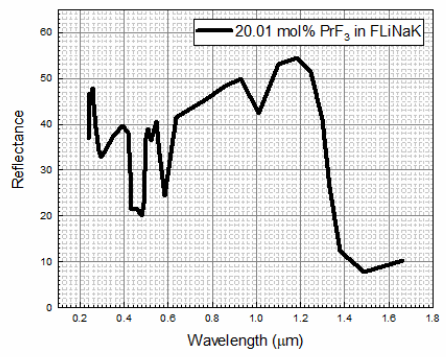

A70 | R7 | E63 SUB

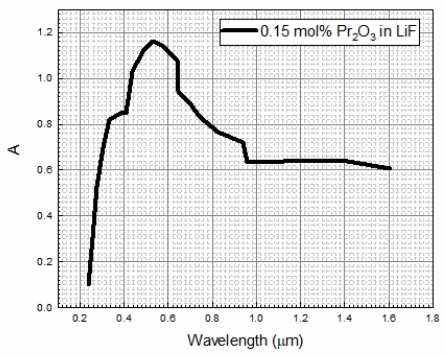

A71 | R7 | E64-69 SUB

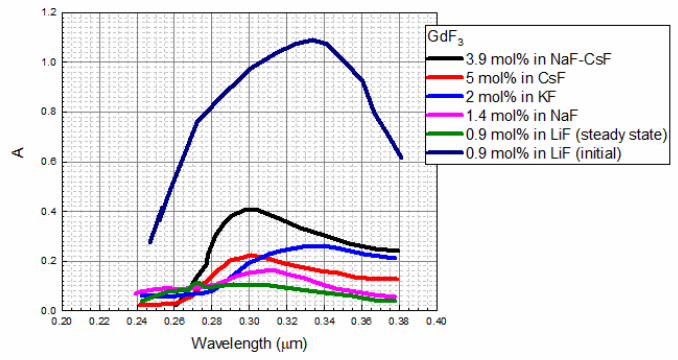

## 4.2. Tabulation of Vibrational Absorption Data

**Table 4. Measurements of infrared absorption edge data in molten fluoride salts. The edge location is defined as the wavelength at which $\kappa_{\lambda_{min}}$ = 100 m$^{-1}$ is reached. Details on the setup used can be found by looking up the setup index (e.g., T#, R#) in Table 1 and Table 3. Plots can be found in Figure 4.**

| Spectrum# | Plot # | Method | Temperature (˚C) | Composition | Path Length (mm) | Edge location (μm) |
|---|---|---|---|---|---|---|
| V1 | A36 | T14<br>Setup and study: Varlamov [49]<br>Wavelength: 4 – 14 μm | 27 | LiF | NR | 6.3 |
| V2 | | | 227 | | | 5.8 |
| V3 | | | 427 | | | 5.5 |
| V4 | | | 627 | | | 5.3 |
| V5 | | | 752 | | | 5.1 |
| V6 | | | 900 (liquid) | | | 4.2 |
| V7 | A37 | | 27 | $CaF_2$ | | 9.2 |
| V8 | | | 227 | | | 8.4 |
| V9 | | | 427 | | | 8 |
| V10 | | | 627 | | | 7.6 |
| V11 | | | 827 | | | 7.2 |
| V12 | | | 1027 | | | 6.8 |
| V13 | | | 1227 | | | 6.4 |
| V14 | | | 1352 | | | 6.1 |
| V15 | | | >1423 (liquid) | | | ~5 |
| V16 | A38 | | 27 | $BaF_2$ | | 11.9 |
| V17 | | | 227 | | | 10.9 |
| V18 | | | 427 | | | 10.3 |
| V19 | | | 627 | | | 9.6 |
| V20 | | | 827 | | | 8.9 |
| V21 | | | 1027 | | | 8.1 |
| V22 | | | 1227 | | | 7.2 |
| V23 | | | 1400 (liquid) | | | <7 |
| V24 | A39<br>A40 | T15<br>Setup and study: S. Liu [50]<br>Wavelength: 0.8 – 4.7 μm | 600 | FLiBe | 10 | 4.2 |
| V25 | A57 | R3<br>Setup: Barker [58]<br>Study: Barker [67]<br>Wavelength: 4 – 8 μm | 27 | LiF | 1-2 | 6.3 |
| V26 | | | 432 | | | 5.7 |
| V27 | | | 562 | | | 5.5 |
| V28 | | | 702 | | | 5.3 |
| V29 | | | 890 (liquid) | | | 4.9 |
| V30 | A59 | R4<br>Setup: Barker [58]<br>Study: Mead [57]<br>Wavelength: 100 – 500 μm | 842 | LiF | 6.19 | NR |
| V31 | | | | | 1.86 | |
| V32 | | | 902 (liquid) | | 1.00 | |
| V33 | | | | | 0.10 | |
| V34 | N/A | Study: Chaleff [9]<br>Wavelength: 4 – 10 μm | 700 | FLiNaK | NR | 7.2 |

NR indicates the parameter was not reported

**Table 5. Measurements of vibrational resonance absorption in molten fluoride solvents and solutes. Details on the setup used can be found by looking up the setup index (e.g., T#, R#) in Table 1 and Table 3. Plots can be found in Figure 4.**

| Spectrum # | Plot # | Method | Temp (˚C) | Composition (mole ratio / mole percent) | | Speciation | Peaks | | | | | |
|---|---|---|---|---|---|---|---|---|---|---|---|---|
| | | | | Solute | Solvent | | NIR/MIR: < 20 µm | | MIR: 20-25 µm | | FIR: >25 µm | |
| | | | | | | | Location (µm) | Intensity ($m^{-1}$) | Location (µm) | Intensity ($m^{-1}$) | Location (µm) | Intensity ($m^{-1}$) |
| V35 | A46 | R1<br>Setup: Wilmshurst [61]<br><br>Study: Wilmshurst [63]<br><br>Wavelength: 10 – 50 | NR | LiF | | NR | 17.75 | 5.1E+05 | $20.25^{sh}$ | $4.0E+05^{sh}$ | | |
| V36 | A46 | | | 2LiF-KF | | | 18.75 | 3.9E+05 | 21.5 | 3.25E+05 | | |
| V37 | A46<br>A47 | | | LiF-KF | | | 18.75 | 3.1E+05 | 21.25 | 2.65E+05 | | |
| V38 | A46 | | | LiF-2KF | | | 18.75 | 1.75E+05 | 21.5 | 1.55E+05 | $31.5^{sh}$ | $1.1E+05^{sh}$ |
| V39 | A46 | | | KF | | | | | | | $33^{sh}$<br>36.75 | $1.9E+05^{sh}$<br>1.95E+05 |
| V40 | A47 | | | LiF-KF-$ZrF_4$ | | | 19 | 7.5E+05 | $21.5^{sh}$ | $3.7E+05^{sh}$ | 36.5 | 3.5E+04 |
| V41 | A48 | | | NaF-KF-$ZrF_4$ | | | 19.2 | 7.6E+05 | | | 40 | 4.0E+04 |
| V42 | A48 | | | 1.5NaF-1.5KF-$ZrF_4$ | | | 19.5 | 5.9E+05 | $21.5^{sh}$ | $3.5E+05^{sh}$ | $35^{sh}$<br>37 | 5.0 E+04<br>6.0E+04 |
| V43 | A48 | | | 3NaF-3KF-$ZrF_4$ | | | | | 20<br>$21.5^{sh}$ | 3.3E+05<br>$2.1E+05^{sh}$ | 33<br>$36^{sh}$ | 1.2E+05<br>1.1E+05 |
| V44 | A48 | | | 5NaF-5KF-$ZrF_4$ | | | | | 20.5 | 2.2E+05 | 33<br>$36^{sh}$ | 1.6E+05<br>1.4E+05 |
| V45 | A48 | | | NaF-KF | | | | | | | 31.5<br>$36^{sh}$ | 2.6E+05<br>2.0E+05 |
| V46 | A49 | R2<br>Setup: Fordyce [64]<br><br>Study: Fordyce [64]<br><br>Wavelength: 10 - 50 | 680 - 750 | LiF-KF | | NR | 19.5 | 2.7E+05 | | | 30.5 | 1.7E+05 |
| V47 | A49 | | | $TaF_5$ 2.3% | LiF-KF | $TaF_7^{2-}$ | 19 | 3..2E+05 | | | 31 | 1.5E+05 |
| | A51 | | | | | | $18.5^{*}$ | $8.20E+04^{*}$ | | | $33^{*}$ | $1.5E+04^{*}$ |
| V48 | A49 | | | $TaF_5$ 9% | LiF-KF | $TaF_7^{2-}$ | 18.5 | 4.5+05 | | | 32 | 1.6E+05 |
| | A51 | | | | | | $18.5^{*}$ | $2.75E+05^{*}$ | | | $32.5^{*}$ | $6.0E+04^{*}$ |
| V49 | A50 | | | NaF-LiF | | NR | | | | | 30.5 | 3.1E+05 |
| V50 | A50 | | | $TaF_5$ 6.6% | NaF-LiF | $TaF_6^{-}$<br>$TaF_7^{2-}$ | 17.5 | 5.65E+05 | | | 30 | 2.5E+05 |
| | A52 | | | | | | $19^{*}$ | $4.8E+05^{*}$ | | | $34.5^{*}$ | $1.0E+04^{*}$ |
| V51 | A53 | R2<br>Setup: Fordyce [64]<br><br>Study: | 700 | LiF-KF | | | 19 | 2.75E+05 | | | 31 | 1.65E+05 |
| V52 | A53 | | | $NbF_5$ 5% | LiF-KF | $NbF_7^{2-}$ | 18.5 | 3.65E+05 | | | 31<br>41 | 1.45E+05<br>1.1E+05 |
| | A54 | Fordyce [65]<br><br>Wavelength: 10 - 50 | | | | | 18.5* | 1.95E+05* | | | 32*<br>42* | 4.0E+04*<br>5.0E+04* |
| V53 | A55 | R2<br>Setup: Fordyce [64]<br><br>Study: Fordyce [66]<br><br>Wavelength: 10 - 50 | 720 | LiF-KF | | NR | 20 | 2.7E+05 | | | $34^{sh}$ | 1.4E+05 |
| V54 | | | | Solute: $TaF_5$ 9% + $H_2O$ | LiF-KF | $TaOF_6^{3-}$ | 19<br>11 | 4.1E+05<br>8.0E+04 | | | 32 | 1.6E+05 |
| V55 | A56 | | | $TaF_5$ 9% + 0.04 mol $H_2O$ | | | 11* | 3.8E+04* | | | | |
| V56 | | | | $TaF_5$ 9% + 0.08 mol $H_2O$ | | | 11* | 1.0E+05* | | | | |
| V57 | | | | $TaF_5$ 9% + 0.12 mol $H_2O$ | | | 11* | 1.4E+05* | | | | |
| V58 | A58 | R4<br>Setup: Barker [58]<br><br>Study: Mead [57]<br><br>Wavelength: 10 – 100 | 902 | LiF (liquid) | | NR | | | 21.5 | 6.9E+05 | | |
| V59 | A60 | R4<br>Setup: Barker [58]<br><br>Study: Mead [68]<br><br>Wavelength: 50 – 160 | 272 | CsF | | NR | | | | | 72 | NR |
| V60 | | | 542 | | | | | | | | | |
| V61 | | | 794 (liquid) | | | | | | | | | |
| V62 | N/A | Setup & Study: | -265 | LiF (solid) | | NR | $20^{sh}$ | 9.0E+05 | 31 | 7.6E+06 | | |
| V63 | N/A | | 147 | | | | $20^{sh}$ | 9.0E+05 | 32 | 2.5E+06 | | |
| V64 | N/A | Jasperse [56]<br><br>Wavelength: 10 – 100 | 787 | | | | $20^{sh}$ | 9.0E+05 | 32 | 1.3E+06 | | |
| V65 | A61 | R5<br>Setup & Study: Andersen [69]<br><br>Wavelength: 7 - 22 | 600 | $K_2NbF_7$ 2.7% + $Na_2O$ 0.5% | FLiNaK | $NbF_7^{2-}$ | 17.25 | NR | | | | |
| V66 | | | | $K_2NbF_7$ 2.7% + $Na_2O$ 1.0% | | $NbOF_5^{2-}$ | 10.75<br>16.75 | NR | | | | |
| V67 | | | | $K_2NbF_7$ 2.7% + $Na_2O$ 2.4% | | $NbOF_5^{2-}$ | 10.75<br>16.75 | NR | | | | |
| V68 | | | | $K_2NbF_7$ 2.7% + $Na_2O$ 5.4% | | $NbO_2F_4^{3-}$ | 11<br>12 | NR | | | | |
| V69 | A26 | T11<br>Furnace: Young [27]<br><br>Cell and Study: Bates [41]<br><br>Wavelength: NR | 425 | $D_3BO_3$ | $NaBF_4$ | $BF_3OD^-$ | 3.72 | NR | | | | |
| V70 | A39<br>A40 | T15<br>Setup & study: Liu [50]<br><br>Wavelength: 0.83 - 25 | 600 | LiOH | FLiBe | $OH^-$ | 2.78 | 170 | | | | |
| V71 | A41 | | | FLiBe | | $BeF_4^{2-}$ | 6.25<br>12.5 | NR | | | | |
| V72 | N/A | Setup & study: Quist [77] | Room temperature | FLiBe (Li-7) (solid) | | $BeF_4^{2-}$ | 11.6<br>12.4<br>12.9 | NR | 20<br>21.6<br>23<br>24.7 | NR | 26.9<br>27.8<br>30 | NR |
| V73 | N/A | | | FLiBe (Li-6) (solid) | | $BeF_4^{2-}$ | 11.6<br>12.4<br>13 | NR | 19.2<br>20.6<br>21.9<br>24 | NR | 26.3<br>27.7<br>29.4 | NR |

[sh]indicates a shoulder rather than a sharp peak

*indicates signal has been subtracted from solvent background

NR indicates the parameter was not reported

## 4.3. Tabulation of Electronic Absorption Data

**Table 6. Measurements of electronic absorption for transition-metal elements and tellurium. Details on the setup used can be found by looking up the Setup Number (e.g., T#, R#) in Table 1 and Table 3. Plots can be found in Figure 4.**

| Spectrum # | Plot # | Method | Temp (˚C) | Composition | | Speciation | Path Length (cm) | Visible Peaks (<0.8 µm) | | Infrared Peaks (>0.8µm) | |
|---|---|---|---|---|---|---|---|---|---|---|---|
| | | | | Solute | Solvent | | | Peak Location (µm) | Peak intensity $\kappa_\lambda$ (m$^{-1}$) or $\varepsilon$ (M$^{-1}$ cm$^{-1}$) | Peak Location (µm) | Peak intensity $\kappa_\lambda$ (m$^{-1}$) or $\varepsilon$ (M$^{-1}$cm$^{-1}$) |
| E1 | A2 | T1<br>Furnace and cell: Young [23]<br><br>Study: Young [24]<br><br>Wavelength: 0.3 – 1.3 µm | 500 | $CoF_2$<br>3 w% | FLiNaK | NR | 1 | 0.515<br>0.575 | NR | NR | NR |
| E2 | A3 | | 650 | $CrF_3$<br>1.5 w% | | NR | 1 | 0.45<br>0.69 | NR | NR | NR |
| E3 | A1 | T2<br>Furnace and cell: Young [23]<br><br>Study: Young [24]<br><br>Wavelength: 0.3 – 1.3 µm | 500 | $NiF_2$<br>1 w% | FLiNaK | NR | 0.5 | 0.35<br>0.44 | NR | NR | NR |
| E4 | A7 | T3<br>Furnace: Young [23]<br><br>Cell: Young [25]<br><br>Study: Young [26]<br><br>Wavelength: 0.3 – 2 µm | 525 | $NiF_2$<br>0.2 – 1.0 w% | FLiNaK | $NiF_6^{4-}$ | NR | 0.43 | $\varepsilon =$<br>12 | 0.9<br>1.5 | $\varepsilon =$<br>2.1<br>0.9 |
| E5 | A13 | T4<br>Furnace: Young [27]<br><br>Cell: Young [25]<br><br>Study: Young [29]<br><br>Wavelength: 0.3 – 2.3 µm | 540 | $FeF_2$ | FLiBe | $FeF_6^{4-}$ | NR | NR | NR | 1.04<br>1.72 | $\varepsilon =$<br>4.8<br>3.0 |
| E6 | A14 | | 550 | $NiF_2$ | | $NiF_6^{4-}$ | | 0.44 | $\varepsilon =$<br>11.2 | 0.92<br>1.62 | $\varepsilon =$<br>1.3<br>0.8 |
| E7 | A15 | | | $CrF_2$ | | $CrF_6^{4-}$ | | 0.71 | $\varepsilon =$<br>5.9 | NR | NR |
| E8 | A16 | | | $CrF_3$ | | $CrF_6^{3-}$ | | 0.3<br>0.45<br>0.69 | $\varepsilon =$<br>8.6<br>10<br>6.9 | NR | NR |
| E9 | A20 | T9<br>Furnace: Young [27]<br><br>Cell and study: Bamberger [38]<br><br>Wavelength: 0.25 – 0.7 µm | 525 | $Te_2$ (g) | FLiBe | NR | 1 | 0.41 | NR | NR | NR |
| E10 | | | 600 | $LiTe_3$<br><br>$Li_2Te$ +oxidants | | $Te_3^-$ | | 0.478 | NR | NR | NR |
| E11 | A21 | T10<br>Furnace:<br>Young [27]<br><br>Cell and study:<br>Whiting [39]<br><br>Wavelength:<br>0.2 – 0.5 μm | 500 | $NaO_2$ | FLiNaK | $O_2^-$ | 1 | 0.254 | NR | NR | NR |
| E12 | A22 | | | $K_2CrO_4$<br>~13 wppm | | $CrO_4^{2-}$<br>($Cr^{6+}$ valence state) | | 0.265<br>0.372 | $\varepsilon =$<br>2940<br>3970 | NR | NR |
| E13 | N/A | | | $KNO_2$ | | $NO_2^-$ | | 0.365 | NR | NR | NR |
| E14 | A23 | T7<br>Furnace:<br>Young [27]<br><br>Cell and study:<br>Whiting [35]<br><br>Wavelength:<br>0.2 – 2 μm | 500 | $MnF_3$ | FLiNaK | $MnF_6^{3-}$ | NR | 0.525 | NR | NR | NR |
| E15 | A24 | | | $MnF_2$<br>~5.6 w% | | $MnF_6^{4-}$ | | 0.408 | $\varepsilon =$<br>0.11 | NR | NR |
| E16 | A25 | | | $CuF_2$<br>~0.5 w% | | $CuF_6^{4-}$ | | NR | NR | 0.88 | $\varepsilon =$<br>11 |
| E17 | A27 | T12<br>Furnace:<br>Young [27]<br><br>Cell and study:<br>Toth [42]<br><br>Wavelength:<br>0.3 – 2.5 μm | 550 | $NiF_2$ | FLiBe | NR | 0.635 | 0.45 | NR | 0.92<br>1.72 | NR<br>NR |
| E18 | A35 | T13<br>Furnace:<br>Toth [44]<br><br>Cell:<br>Toth [42]<br><br>Study:<br>Toth [47]<br><br>Wavelength:<br>0.25 – 3.4 μm | 550 | $NbF_4$<br><br>0.068 M<br>0.113 M<br>0.202 M<br>0.303 M | FLiBe | $NbF_7^{3-}$ | NR | 0.38<br>0.51<br>0.74 | NR | 1.82 | NR |
| E19 | N/A | T13<br>Furnace:<br>Toth [44]<br><br>Cell: Toth [42]<br><br>Study: Toth [48] | 550 – 600 °C | $Li_2Te$<br>+oxidants | FLiBe | $Te^-$<br>(or likely same species as E10) | .635 | 0.478 | $\varepsilon =$<br>3370 | NR | NR |
| E20 | A43 | T17<br>Furnace:<br>H. Liu [51]<br><br>Cell and study:<br>Y. Liu [52]<br><br>Wavelength:<br>0.35 – 0.85 μm | NR | $CrF_2$<br>0.2 w% | FLiBe | NR | 10 | 0.75 | NR | NR | NR |
| E21 | A43<br>A44 | | | $CrF_3$<br>0.2 w% | | NR | | 0.31<br>0.45<br>0.70 | $\kappa_\lambda =$<br>20*<br>75*<br>70* | | |
| E22 | A44 | | | CrF3<br>0.35 w% | | NR | | | $\kappa_\lambda =$<br>35*<br>132*<br>92* | | |
| E23 | | | | $CrF_3$<br>0.5 w% | | NR | | | $\kappa_\lambda$ =<br>62*<br>240*<br>215* | | |
| E24 | A45 | | | $CrF_2$<br>0.2 w% | FLiNaK | NR | | 0.45<br>0.7<br>0.75 | NR | NR | NR |
| E25 | | | | $CrF_3$<br>0.2 w% | | NR | | 0.45<br>0.7 | NR | | |
| E26 | A63 | R7<br>Setup and study: Khokhryakov [70]<br><br>Wavelength:<br>0.23 – 1.03 µm | 1377 | $CoF_2$<br>0.075 mol% | $CaF_2$ | $CoF_6^{3-}$ | NR | 0.36<br>0.38<br>0.39<br>0.41<br>0.43<br>0.49<br>0.53 | NR | 0.89 | NR |
| E27 | A64 | | 1027 | $FeF_3$<br>0.51 mol% | $BeF_2$ | $FeF_6^{3-}$ | | 0.31<br>0.34<br>0.36<br>0.38<br>0.41<br>0.61 | NR | 0.92 | NR |
| E28 | A65 | | 827 | $FeF_3$<br>10 w% | $BeF_2$^R | | | 0.24<br>0.26<br>0.29<br>0.32<br>0.35<br>0.36<br>0.38<br>0.41<br>0.43<br>0.56 | NR | 0.95 | NR |
| E29 | A66 | | 727 | | NaF^R | | | 0.26<br>0.30<br>0.41 | NR | 0.85<br>0.93 | NR |

* the solvent background has been subtracted from the solute
NR indicates the parameter was not reported
^R indicates reflectance was measured and correlated to absorbance

**Table 7. Measurements of electronic absorption for lanthanides and actinides. Details on the setup used can be found by looking up the Setup Number (e.g., T#, R#) in Table 1 and Table 3. Plots can be found in the supplemental material.**

| Spectrum # | Plot # | Method | Temp (˚C) | Composition | | Speciation | Path Length (cm) | Visible Peaks (<0.8 µm) | | Infrared Peaks (>0.8µm) | |
|---|---|---|---|---|---|---|---|---|---|---|---|
| | | | | Solute | Solvent | | | Peak Location (µm) | Peak intensity $\kappa_\lambda$ ($m^{-1}$) or $\varepsilon$ ($M^{-1}cm^{-1}$) | Peak Location (µm) | Peak intensity $\kappa_\lambda$ ($m^{-1}$) or $\varepsilon$ ($M^{-1}cm^{-1}$) |
| E30 | A4 | T1<br>Furnace and cell: Young [23]<br><br>Study:<br>Young [24]<br><br>Wavelength:<br>0.3 – 1.3 µm | 550 | Pr fluoride<br><5 w% | FLiNaK | NR | 0.8 | 0.44<br>0.47<br>0.48<br>0.58 | NR | NR | NR |
| E31 | A5 | | 570 | $UF_4$<br>1.66 w% | | NR | NR | 0.44<br>0.48<br>0.55<br>0.62<br>0.69 | NR | 0.89<br>1.1<br>1.34 | NR |
| E32 | A6 | | 575 | $UO_2F_2$<br><1.7 w% | | NR | NR | 0.43 | NR | NR | NR |
| E33 | A8<br>A9 | T4<br>Furnace and study:<br>Young [27]<br><br>Cell: Young [25]<br><br>Wavelength:<br>0.2 – 1.4 µm | 50 | $UF_4$ | FLiBe | $UF_8^{4-}$<br>$UF_9^{5-}$ | 0.7 | 0.24<br>0.49<br>0.52<br>0.61<br>0.64 | $\varepsilon =$<br>1400<br>3.8<br>3.2$^{sh}$<br>5.2$^{sh}$<br>6.4 | 0.86<br>1.1<br>1.38 | $\varepsilon =$<br>1.8$^{sh}$<br>13<br>0.7 |
| E34 | A10<br>A11 | | | $UF_3$ | | | | 0.36<br>0.73<br>0.82 | $\varepsilon =$<br>430<br>12<br>32$^{sh}$ | 0.87<br>0.96<br>1.03<br>1.16 | $\varepsilon =$<br>42<br>24<br>24<br>21 |
| E35 | | | 525 | $UF_3$ | FLiNaK | | | 0.44<br>0.72 | $\varepsilon =$<br>420<br>32$^{sh}$ | 0.83<br>0.89<br>0.97<br>1.06<br>1.16 | $\varepsilon =$<br>31$^{sh}$<br>43<br>28<br>26<br>25 |
| E36 | A12 | T7<br>Furnace:<br>Young [27]<br><br>Cell:<br>Whiting [35]<br><br>Study:<br>Young [36]<br><br>Wavelength:<br>0.3 – 0.6 µm | 540 | $UF_3$ | LiF-$BeF_2$-$ZrF_4$ (66-29-5 mol%) + 2 w% $UF_4$ | NR | NR | 0.36 | NR | NR | NR |
| E37 | A17<br>A18 | T5<br>Furnace:<br>Young [27] [30]<br><br>Cell:<br>Young [25]<br><br>Study:<br>Young [31]<br><br>Wavelength:<br>0.3 – 2.3 µm | 570 | $PaF_4$ | LiF-$BeF_2$-$ThF_4$ (72-16-12 mol%) | $PaF_7^{4-}$ | NR | 0.26<br>0.33 | $\varepsilon =$<br>694<br>624 | 1.58<br>1.83 | NR |
| E38 | N/A | T8<br>Furnace:<br>Young [27]<br><br>Cell and study:<br>Bamberger [37]<br><br>Wavelength:<br>0.64, 1.09 µm | 498 | $UF_4$ | FLiBe | NR | NR | 0.64 | $\varepsilon =$<br>16.4 | 1.09 | $\varepsilon =$<br>8.0 |
| E39 | | | 560 | | | | | | $\varepsilon =$<br>14.4 | | $\varepsilon =$<br>7.3 |
| E40 | | | 650 | | | | | | $\varepsilon =$<br>13.8 | | $\varepsilon =$<br>6.2 |
| E41 | | | 698 | | | | | | $\varepsilon =$<br>12.9 | | $\varepsilon =$<br>5.7 |
| E42 | A19 | T5<br>Furnace:<br>Young [27] [30]<br><br>Cell:<br>Young [25]<br><br>Study:<br>Bamberger [32]<br><br>Wavelength: | 575 | $PuF_3$<br>0.4 mol% | LiF-$BeF_2$-$ThF_4$ (72-16-12 mol%) | NR | NR | 0.42<br>0.43<br>0.45<br>0.5<br>0.56<br>0.6<br>0.66 | NR | 0.81<br>0.89<br>1.06<br>1.34<br>1.44 | NR |
| | | 0.4 – 1.8 µm | | | | | | | | | |
| E43 | A28 | T12<br>Furnace:<br>Young [27]<br><br>Cell and study:<br>Toth [42]<br><br>Wavelength:<br>0.3 – 2.5 µm | 550 | $UF_4$ | FLiBe | NR | 0.635 | 0.53<br>0.6<br>0.65 | NR | 0.85<br>1.1<br>1.42<br>2.03 | NR |
| E44 | A29 | T12<br>Furnace:<br>Young [27]<br><br>Cell: Toth [42]<br><br>Study: Toth [14]<br><br>Wavelength:<br>0.4 – 2.5 µm | 550 | $UF_4$ | FLiNaK | $UF_8^{4-}$ | 0.63 | 0.43<br>0.44<br>0.47<br>0.53<br>0.61<br>0.66 | $\varepsilon =$<br>$4.4^{sh}$<br>$4.5^{sh}$<br>6.2<br>4.4<br>$8.5^{sh}$<br>11 | 1.1<br>1.41<br>1.61<br>2.13 | $\varepsilon =$<br>20.8<br>5.2<br>$3.8^{sh}$<br>3.8 |
| E45 | A30 | | 460 | $UF_4$ | FLiBe | $UF_8^{4-}$ | | 0.43<br>0.44<br>0.47<br>0.53<br>0.61<br>0.66 | $\varepsilon =$<br>$3.6^{sh}$<br>$4.6^{sh}$<br>5.2<br>3.9<br>$6.2^{sh}$<br>9.2 | 1.1<br>1.41<br>1.61<br>2.13 | $\varepsilon =$<br>18.2<br>5.2<br>$4.2^{sh}$<br>3.4 |
| E46 | | | 690 | $UF_4$ | | $UF_7^{3-}$ | | 0.44<br>0.49<br>0.52<br>0.63 | $\varepsilon =$<br>$3.8^{sh}$<br>4.1<br>$3.2^{sh}$<br>6.6 | 1.08<br>2.02 | $\varepsilon =$<br>$14.4^{+}$<br>$2.8^{+}$ |
| E47 | A31 | | 360 | $UF_4$ | LiF-$BeF_2$ (48-52 mol%) | $UF_8^{4-}$ | | 0.43<br>0.44<br>0.47<br>0.53<br>0.61<br>0.66 | $\varepsilon =$<br>$3.6^{sh}$<br>$4.6^{sh}$<br>5.2<br>4.1<br>$7.2^{sh}$<br>9.5 | 1.1<br>1.41<br>1.61<br>2.13 | $\varepsilon =$<br>18.9<br>6.7<br>$5.0^{sh}$<br>4.4 |
| E48 | | | 690 | $UF_4$ | | $UF_7^{3-}$ | | 0.44<br>0.49<br>0.52<br>0.63 | $\varepsilon =$<br>$3.2^{sh}$<br>3<br>$2.6^{sh}$<br>5.6 | 1.08<br>2.02 | $\varepsilon =$<br>12.4<br>3.7 |
| E49 | A32<br>A33 | T13<br>Furnace and study:<br>Toth [44]<br><br>Cell: Toth [42]<br><br>Wavelength:<br>0.25 – 2.5 µm | NR | $UF_3$ | FLiBe | NR | 0.63 | 0.32<br>0.36<br>0.43<br>0.71 | $\varepsilon =$<br>740<br>840<br>$600^{sh}$<br>13 | 0.87<br>0.96<br>1.04<br>1.14<br>2.14 | $\varepsilon =$<br>43<br>22<br>22<br>20<br>12 |
| E50 | A34 | | 500 | $UF_3$<br>$UF_4$ | | | | NR | NR | 0.88<br>1.09<br>1.6 | NR |
| E51 | | | 600 | | | | | 0.71 | NR | 0.88<br>0.96<br>1.04<br>1.09<br>1.14 | NR |
| E52 | | | 700 | | | | | 0.71 | NR | 0.88<br>0.96<br>1.04<br>1.14 | NR |
| E53 | N/A | T13<br>Furnace: Toth [44]<br><br>Cell: Toth [42]<br><br>Study: Gilpatrick [46] | 500 – 800 | $UF_3$ | FLiBe<br><br>$LiF-BeF_2$ (48-52 mol%)<br><br>$LiF-BeF_2$ ( 57-43 mol%)<br><br>$LiF-BeF_2-ThF_4$ (72-16-12 mol%) | NR | NR | 0.36 | $\varepsilon =$ 11,001 | NR | NR |
| E54 | A42 | T16<br>Furnace and study: H. Liu [51]<br><br>Cell: S. Liu [50]<br><br>Wavelength: 0.2 – 1.0 µm | 600 | $UF_4$<br>0.5 w% | FLiBe | NR | 10 | 0.27<br>0.45<br>0.48<br>0.52<br>0.64 | NR | NR | NR |
| E55 | A62 | R7[a]<br>Setup: Khokhryakov [74][70]<br><br>Study: Khokhryakov [71]<br><br>Wavelength: 0.21 – 0.33 | 667 | $CeF_3$<br>0.0304 mol% | LiF-NaF | $CeF_6^{3-}$ | 1 | 0.23<br>0.26<br>0.29 | NR | NR | NR |
| E56 | | | | $Ce_2O_3$<br>0.0185 mol% | | $CeOF_5^{4-}$ | | 0.23<br>0.25<br>0.29 | NR | NR | NR |
| E57 | A67 | R7<br>Setup: Khokhryakov [70]<br><br>Study: Khokhryakov [72]<br><br>Wavelength: 0.24 – 1.67 µm | 1000 | $PrF_3$<br>0.7 mol% | LiF | $PrF_6^{3-}$ | NR | 0.34<br>0.44[b]<br>0.47<br>0.48<br>0.55<br>0.59 | NR | 1.53 | NR |
| E58 | | | | | NaF | | | 0.38<br>0.45<br>0.48<br>0.59 | NR | 1.5 | NR |
| E59 | | | | | KF | $PrF_6^{3-}$<br>$PrOF_5^{4-}$ | | 0.33<br>0.38<br>0.48<br>0.58 | NR | 1.52 | NR |
| E60 | | | | | CsF | | | 0.35<br>0.41[b] | NR | 0.85<br>0.89<br>1.52 | NR |
| E61 | A68 | | 750<br>800<br>1000 | $PrF_3$<br>0.71<br>mol% | FLiNaK | $PrF_6^{3-}$ | NR | 0.33<br>0.44<br>0.47<br>0.48<br>0.59<br>0.64 | NR | 1.53 | NR |
| E62 | A69 | | 500 | $PrF_3$<br>20.01<br>mol% | FLiNaK<br>R | | NR | 0.27<br>0.40<br>0.50<br>0.54<br>0.64 | NR | 0.95<br>1.18 | NR |
| E63 | A70 | | 1000 | $Pr_2O_3$<br>0.15<br>mol% | LiF | $PrOF_5^{4-}$ | NR | 0.34<br>0.54<br>0.8 | NR | 1.24 | NR |
| E64 | | | | $GdF_3$<br>0.9 mol%<br>initial | LiF | | NR | 0.26<br>0.28<br>0.34 | NR | NR | NR |
| E65 | | | | $GdF_3$<br>0.9 mol%<br>steady | | | NR | 0.27<br>0.3 | NR | NR | NR |
| E66 | A71 | R7<br>Setup:<br>Khokhryako<br>v [70]<br><br>Study:<br>Khokhryako<br>v [73]<br><br>Wavelength:<br>0.24 – 0.38 | 1000 | $GdF_3$<br>1.4 mol% | NaF | $GdF_6^{3-}$ | NR | 0.25<br>0.27<br>0.31 | NR | NR | NR |
| E67 | | | | $GdF_3$<br>2 mol% | KF | | NR | 0.32 | NR | NR | NR |
| E68 | | | | $GdF_3$<br>5 mol% | CsF | | NR | 0.3 | NR | NR | NR |
| E69 | | | | $GdF_3$ 3.9 mol% | NaF-CsF | | NR | 0.3 | NR | NR | NR |

[sh]indicates a shoulder rather than a sharp peak
NR indicates the parameter was not reported
[*] the solvent background has been subtracted from the solute
[R] indicates reflectance was measured and correlated to absorbance

## 5. STRUCTURAL INTERPRETATION

Here we summarize and discuss absorption data from prior optical spectroscopy studies (mostly in the infrared and visible range) on fluoride salts. The data has been divided into two sections based on the physics of the resonance causing absorption to occur at a particular wavelength. First, we summarize and discuss vibrational data, starting with solvents in section 5.1.2 and followed by solutes in 5.1.3. Then, we summarize and discuss electronic data of solutes, starting with lighter elements (mostly transition metals) in section 5.2.2 followed by heavier (lanthanides and actinides) in section 5.2.3.

### 5.1. Vibrational Absorption – NIR/MIR/FIR

#### 5.1.1. Background

Vibrational resonance occur when atoms bound by a restoring force, such as an ionic or covalent bond, are excited by an incoming photon with a frequency that matches the natural vibrational frequencies of the bond. Within a molten fluoride environment, ionic interactions are expected to dominate due to the large electronegativity differences between F and the alkali and alkaline earth metals that typically make up a solvent mixture. These interactions typically occur between a single cation surrounded by multiple anions, a structure referred to as the ion’s first coordination shell. The coordination number (CN) is the number of ions within the first coordination shell, typically defined relative to a specific cation. This structure leads to collective vibration similar to that seen in the solid-state, but of shorter range, leading some authors to refer to the melt structure as a ‘quasilattice’ [57][63].

It is useful to classify these ionic interactions or ‘bonds’ by their lifetime, as the melt structure is clearly dynamic. Alkali fluoride mixtures exhibit bonding at a sub-picosecond time interval (though intermediate range ordering has been observed at slightly longer time-scales) [78], and we refer to them as consisting of dissociated ions.

A longer time scale on the order of 10-100 ps is expected for bonding in alkaline-earth (or other multi-valent) fluorides, as predicted via molecular dynamics (MD) modeling of FLiBe [79][80]. These species will be referred to as associated monomers (e.g., $BeF_4^{2-}$) and oligomers (e.g., $Be_2F_7^{3-}$, etc.). As the concentration of alkaline earth (or other multi-valent ions) ions is increased, formation of F- bridging between two metal cations is increasingly favored, leading to oligomer formation and in some cases large chains or networks of ions [79][81].

### 5.1.2. Solvents

#### 5.1.2.1. Resonance data

We refer to the region of peak absorption as the vibrational absorption resonance region. Vibrational absorption resonance data is available for the mixtures LiF-KF, LiF-NaF, LiF-KF-$ZrF_4$, LiF-NaF-$ZrF_4$, FLiBe, and compounds LiF and CsF. In Table 5 we compile the available data, providing an index (V#) for each spectrum, and reference the plot (A#) and setup indices (T# or R#). For each spectrum, we give the measurement temperature, salt composition, deduced speciation, and list the location and intensity of the various peaks or shoulders.

Dissociated melts

Mead studied the vibrational absorption of LiF and compared with high temperature crystalline spectra [57], shown in Figure 5a. Crystalline LiF displays a relatively sharp, high-intensity peak at 32 µm and a shoulder at 20 µm (Table 5, V64) [56], whereas Mead's data for liquid LiF displays a much broader, lower-intensity peak at 21.5 µm. Using an oscillator model, Mead attributed the peak broadening and intensity reduction upon melting to be due to increased damping caused by loss of long-range order [57]. We note the discrepancy with Wilmshurst's liquid LiF data [63] which shows a sharper, lower-intensity peak at 17.75 µm. Mead thought this to be due to KK integration error or temperature differences, as Wilmshurst's wavelength range was much shorter, and the measurement temperature was not reported.

In addition to LiF, Wilmshurst studied LiF-KF mixtures over various compositions [63], shown in Figure 5b. The spectra of the mixtures are approximately a superposition of the peaks for the pure components; however, the LiF peak shifts slightly from 17.75 µm to 18.75 µm, does not diminish proportionally as the LiF concentration decreases, and a secondary peak appears at about 21.5 µm. Wilmshurst explained the apparent preference for Li-F vibrations as 'typical of the spectra of highly associated liquids exhibiting quasilatticelike behavior', without giving much further discussion, and the 'doublet' nature of the LiF peak was attributed to 'anharmonicity'. Superposition behavior is also seen in Raman spectra LiF-KF and LiF-CsF mixtures [82] and an additional peak was observed and attributed to an intermediate-range $(LiF_x)K$ or $(LiF_x)Cs$ species having short lifetimes (between 0.1 and 1 ps). LiF-KF exhibits a type II Chemla Effect, meaning that the ionic mobilities of both the smaller and larger cations decrease with increasing concentration. Ribiero used rigid-ion MD to investigate this, revealing the presence of intermediate-range Li species [78], and Salanne confirmed the result using polarizable-ion MD [83]. High-temperature pulsed-field-gradient NMR spectroscopy (HT-PFG-NMR) was performed on LiF-KF mixtures, and it was concluded that the melt was completely dissociated and no direct evidence of intermediate structures was found; however, their presence couldn't be ruled out, since the NMR characteristic time is long relative to the short-lived species [84]. It is possible that the behavior observed by Wilmshurst is evidence of intermediate-range Li species, and further experimental and theoretical work in this area would be quite interesting.

FLiNaK or LiF-NaF-KF (46.5-11.5-42 mol%) is an example of a mixture expected to consist of dissociated ions. Its structure has been fairly well documented by MD [85], x-ray diffraction (XRD) [86], and a combined MD and neutron diffraction approach [87]. The XRD-determined cation-anion average bond distances and CN found in molten FLiNaK are near the values obtained for the single-component melts, a result consistent with the vibrational data of alkali fluoride mixtures discussed above. The results of the two computational approaches agree with experimental for average cation-anion bond distances, but the computational-derived CNs are slightly higher than the experimental values. This discrepancy could be due to differences in how CN is calculated, or other methodological differences. Raman measurement of FLiNaK has also been performed and the melt was interpreted as being completely dissociated [52]. There is no infrared vibrational absorption data for FLiNaK, but Wilmshurst's LiF-KF data (Figure 5b) suggests that the ternary mixture FLiNaK would appear as a superposition of the individual constituent peaks, with a weighting towards the LiF peak.

Infrared resonance data for alkali fluoride melts are consistent with results from other techniques, which suggest that absorption is due short-lifetime bonds between cations and anions in the first-coordination shell. Further study would be of interest to understand the potential contribution of intermediate-range Li species.

Associated melts

Wilmshurst also studied mixtures of $ZrF_4$ with LiF-KF (Figure 4, Plot A47, V40) and various compositions with NaF-KF [63], shown in Figure 5c. At the highest $ZrF_4$ concentration (NaF-KF-$ZrF_4$), a sharp, high-intensity peak is displayed at 19.2 µm. As $ZrF_4$ concentration decreases, this peak reduces intensity and shifts slightly to 19-20 µm as a secondary peak or shoulder appears at 21.5 µm. At the lowest $ZrF_4$ concentration (5NaF-5KF-$ZrF_4$), only one peak at 20.5 µm remains. Wilmshurst attributed this peak transition (from about 19 to 21 µm) to the breakdown of a $Zr^{4+}$ polymeric lattice into $Zr^{4+}$ monomers, suggesting either $ZrF_6^{2-}$ or $ZrF_5^-$ monomeric species. Recently, a combined study using MD, high-temperature NMR (HT-NMR), and extended x-ray absorption fine structure (EXAFS) was performed of LiF-$ZrF_4$ mixtures [88]. NMR spectra reveal a relatively sharp transition in the $^{19}F$ chemical environment going from dilute $ZrF_4$ to about 33 mol% followed by a slower transition going to 50 mol%, which correlates well with the MD-determined amount of bridging F. EXAFS and MD-determined CN's are a mixture of 6, 7 and 8. The similarity with Wilmshurst's conclusions regarding degree of bridging (i.e. oligomer formation) suggests that $ZrF_4$ may behave similarly in mixtures of varying alkali fluoride composition. We note that Wilmshurst chose to avoid mixtures of LiF and $ZrF_4$ due to peak overlap.

FLiBe is an example of a melt which consists of dissociated ions in the form of $Li^+$, associated monomers in the form of $BeF_4^{2-}$, and associated oligomers present as chains of tetrahedral $BeF_4^{2-}$. The structure of FLiBe has been studied by several MD approaches [79][80][89] and experimentally via Raman spectroscopy [77][90] and XRD [91]. $BeF_4^{2-}$ and larger (e.g., $Be_2F_7^{3-}$, etc.) structures have been found in broad agreement across the various approaches. There is no quantitative vibrational absorption resonance data for FLiBe, but qualitative transmittance spectra are available of a thin melt [50], shown in Figure 5d. The dominant peak appears at about 12.5 µm, and was attributed to $BeF_4^{2-}$ via quantum-chemical density-functional-theory (QC-DFT). This peak is also observed as an infrared-active Raman band in molten FLiBe and the infrared spectra of crystalline FLiBe powder (Table 5, V72-73) [77]. The crystalline data suggest that another resonance would be found at about 22 µm, corresponding to vibration between $Li^+$ and $BeF_4^{2-}$ ions. Another small peak at 6.25 µm was observed by Liu and attributed as an overtone, but it could be due to larger $Be^{2+}$ oligomers, such as $Be_2F_7^{3-}$, as a similar infrared-active peak was observed in a Raman study of molten FLiBe [90].

Infrared resonance data for $ZrF_4$-containg salt indicates a transition from monomers to an oligomer network as $ZrF_4$ concentration increases, in agreement with other techniques. More infrared resonance data for FLiBe or other associated fluoride melts (e.g., cryolite) would be useful to probe network-forming behavior.

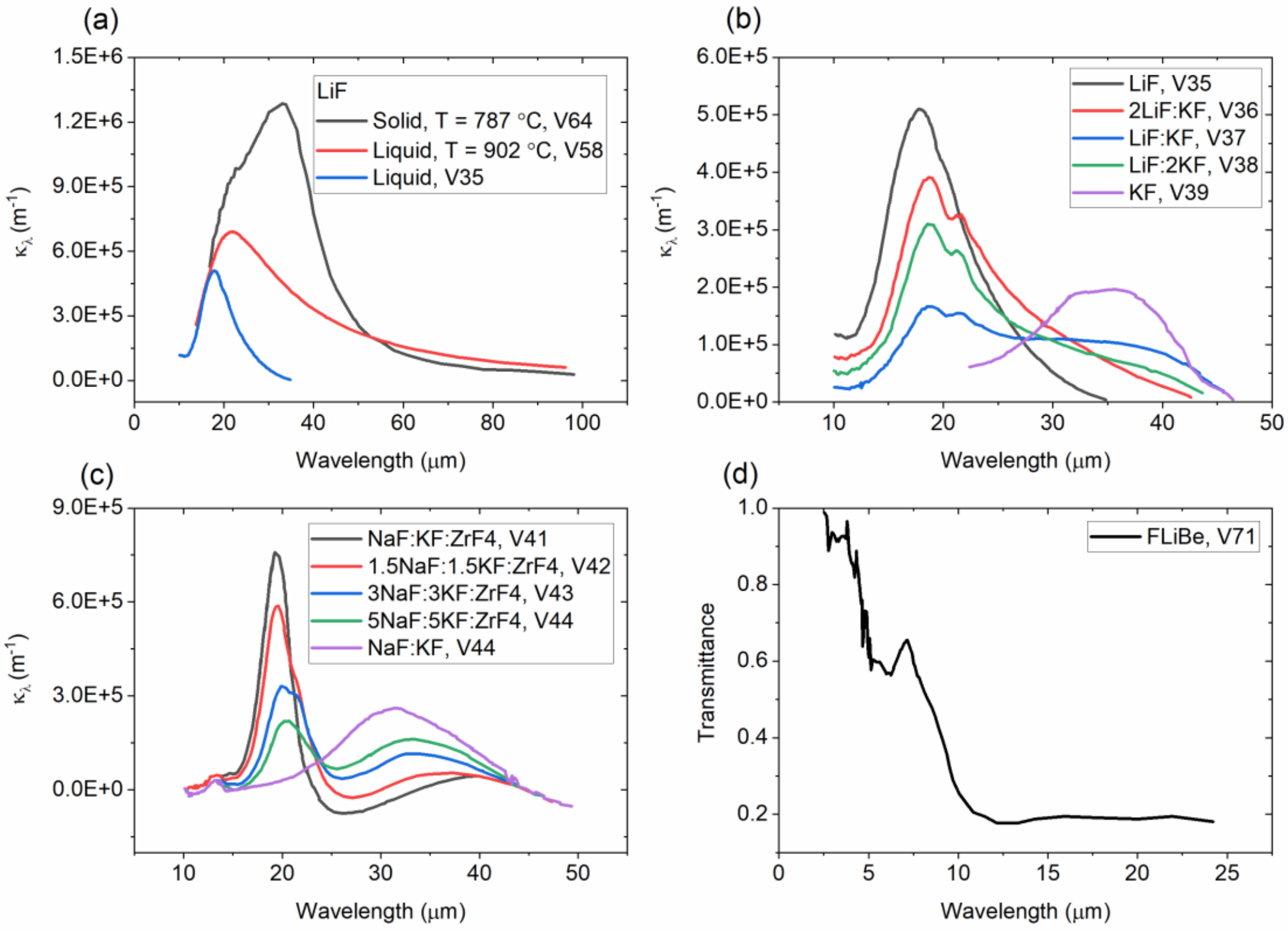

**Figure 5. Vibrational absorption resonance spectra of molten fluoride solvents: (a) LiF, (b) LiF:KF mixtures, (c) NaF:KF:ZrF$_4$ mixtures, (d) FLiBe. We give the spectrum index (V#) for each curve, for reference in Table 5. The negative values seen for $k$ in (c) are due to KK integration errors, which we discuss in section 3.2.1.1.**

### 5.1.2.2. Edge Data

The infrared absorption edge region for a given species occurs as the resonance region is approached and consists of rapidly rising $\kappa_\lambda$ values. NIR edge data is available for LiF, $CaF_2$, $BaF_2$, and FLiBe. FIR edge data is also available for LiF. The NIR edge of FLiNaK has not been measured, but Chaleff predicted $\kappa_\lambda$ (V34) by averaging high-temperature crystalline data for LiF and NaF, neglecting KF [9]. For each of the spectra, we define the edge location as the wavelength at which $\kappa_{edge} = 100\ \mathrm{m}^{-1}$. In Table 4 we compile the available edge data, providing an index (V#) for each spectrum, and reference the plot (A#) and setup indices (T# or R#). For each spectrum, we give the measurement temperature, salt composition, path length, and edge location.

Varlamov noted that the edge of $CaF_2$, $BaF_2$, and LiF shifts to shorter wavelengths with increasing temperature and with the solid-to-liquid phase transition [49], but did not comment further on the behavior. Barker noted a similar shift with LiF, shown in Figure 6a, and interpreted it as peak broadening due to increased lattice disorder with increasing temperature [67]. This NIR shifting is expected to cause the absorption edge of liquid alkali fluorides and alkali fluoride mixtures to approach the region of RHT significance, especially if LiF is a major constituent. More data is needed to characterize the degree to which peak broadening continues as temperature increases further into the liquid state.

Liu measured the NIR edge transmittance of FLiBe (Figure 4, Plot A39, V24) [50], and using the reported path length we present it as $\kappa_\lambda$, shown in Figure 6b. We note that the large peak about 2.8 μm is due to $OH^-$ absorption and discuss this in section 5.1.3. More NIR edge measurements of both FLiBe and FLiNaK over a larger wavelength range are needed to quantify the degree of shifting which occurs with varying temperature in these solvents.

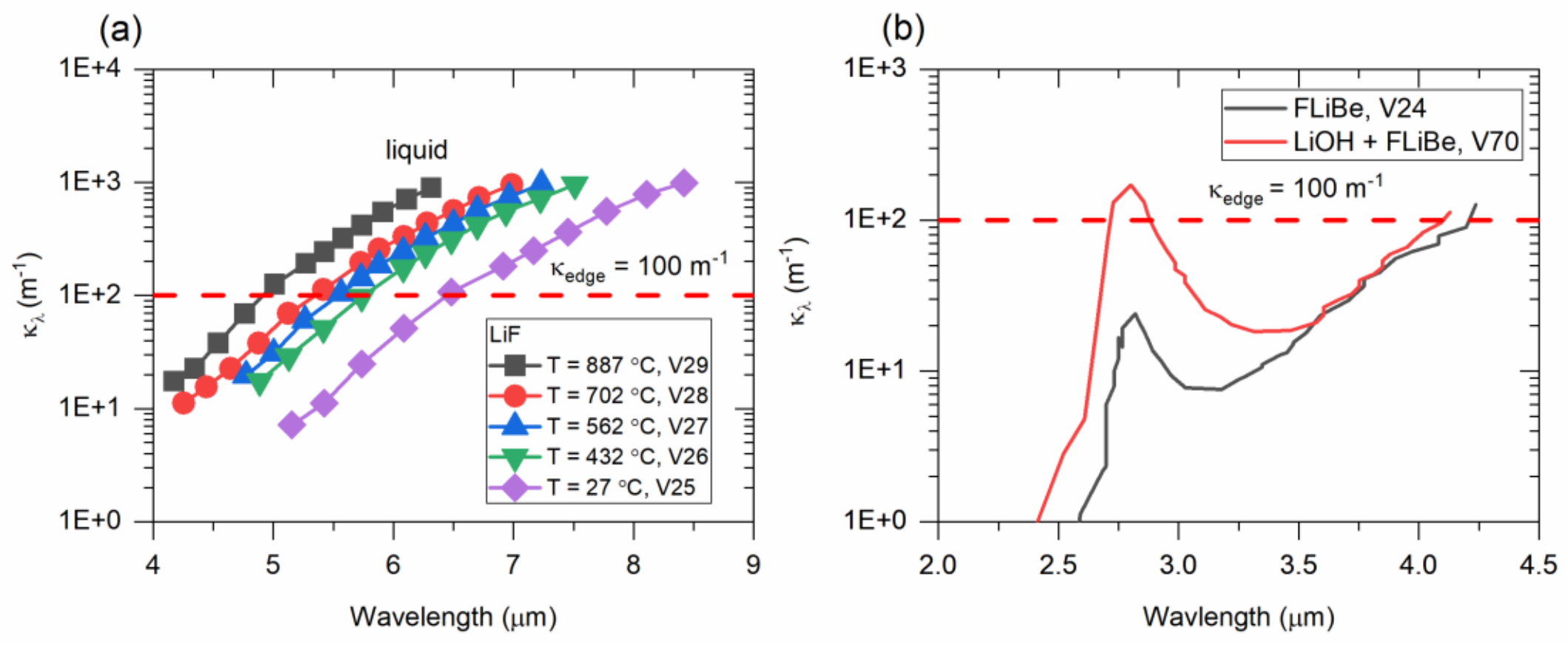

**Figure 6. Vibrational absorption edge spectra of molten fluoride solvents: (a) LiF, (b) FLiBe. We give the spectrum index (V#) for each curve, for reference in Table 4. We note that (b) also contains solute vibrational absorption resonance data for LiOH in FLiBe solvent, which is referenced in Table 5.**

### 5.1.3. Solutes

Vibrational absorption resonance data is available for the solutes $OH^-$ in FLiBe, $OD^-$ in $NaBF_4$, $Ta^{5+}$ in LiF-KF and LiF-NaF, and $Nb^{5+}$ in LiF-KF and FLiNaK. In Table 5 we compile the available data, providing an index (V#) for each spectrum, and reference the plot (A#) and setup indices (T# or R#). For each spectrum, we give the measurement temperature, salt composition, deduced speciation, and list the location and intensity of the various peaks or shoulders.

Addition of LiOH to FLiBe resulted in a large peak at about 2.8 μm (Figure 6b), suggesting the presence of O-H bonding. Similarly, the addition of $D_3BO_3$ to $NaBF_4$ resulted in a peak at about 3.7 μm (Figure 4, Plot A26, V69), indicating the presence of O-D bonding. These results suggest that NIR/MIR measurements could be used to detect $OH^-$ and $OD^-$ contaminants.

Via comparison with crystalline $K_2TaF_7$ and $CsTaF_6$ [64], Fordyce theorized that $Ta^{5+}$ existed as $TaF_7^{2-}$ in LiF-KF, and a mixture of $TaF_6^-$ and $TaF_7^-$ in LiF-NaF [64], shown in Figure 7a. With the addition of $H_2O$ in LiF-KF [66], $Ta^{5+}$ was theorized to exist as $TaOF_6^{3-}$ due to similarity with $TaF_7^{2-}$ and the presence of a peak at about 11 μm attributed to a Ta=O bond [92], also shown in Figure 7a. These interpretations have been partially validated by Raman results [93], in which the authors concluded that $TaF_5$ existed as $TaF_7^{2-}$ in FLiNaK. With some $Na_2O$ addition, $TaOF_5^{2-}$ was proposed to exist; larger $Na_2O$ additions resulted in a new species, presumed to be $TaO_2F_4^{3-}$. Another Raman and QC-DFT study inferred the presence of $TaF_7^{2-}$ and $TaF_8^{3-}$ in FLiNaK, and subsequently $TaOF_5^{2-}$, $TaOF_6^{3-}$, $TaO_2F_2^{-}$, and $TaO_2F_3^{2-}$ as $Li_2O$ was added [94]. More investigation in this area is needed to resolve these discrepancies.

From comparison with crystalline spectra [65] and complementary Raman [95], $NbF_5$ was inferred to exist as $NbF_7^{2-}$ in LiF-KF (Figure 4, Plot A54, V52) [65] and FLiNaK [69], shown in Figure 7b. With the addition of $Na_2O$ to $NbF_5$ in FLiNaK, the species $NbOF_5^{2-}$ and subsequently $NbO_2F_4^{4-}$ were inferred (Figure 4, Plot A67, V66-68). These results are in agreement with Raman measurements of $Nb^{5+}$ in FLiNaK with varying $Na_2O$ [95]. The formation of new species with addition of $O^{2-}$ to the melt indicates that both $NbF_5$ and $TaF_5$ could be used as marker species to detect the presence of $O^{2-}$, as could several lanthanide fluorides (e.g., $CeF_3$, $PrF_3$, etc.).

Infrared resonance data for solutes in fluoride melts have been used to propose structures that are mostly in agreement with other techniques. The data has also shown the ability to distinguish vibrations related to O (e.g., $OH^-$, Ta=O, oxy-fluoride species).

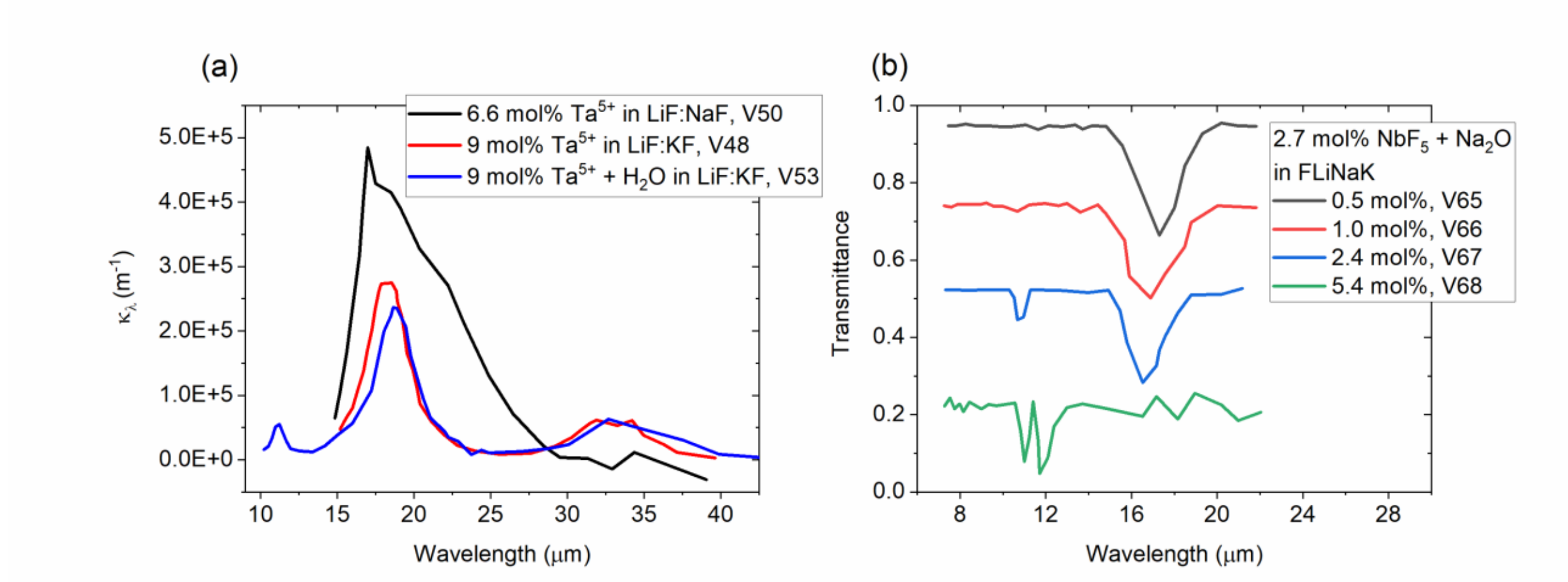

**Figure 7. Vibrational absorption resonance spectra of solutes in molten fluoride solvents: (a) 6.6 mol% $Ta^{5+}$ in LiF:NaF, 9 mol% $Ta^{5+}$ in LiF:KF, and 9 mol% $Ta^{5+}$ + $H_2O$ in LiF:KF (b) 2.7 mol% $Nb^{5+}$ in FLiNaK + varying amounts of $Na_2O$. We give the spectrum index (V#) for each curve, for reference in Table 5. We performed a subtraction procedure (following the method presented by Fordyce** [64]**) to obtain the 9 mol% $Ta^{5+}$ + $H_2O$ in LiF:KF curve in (a), but the data tabulated for entry V53 is un-subtracted. The negative values seen for *k* in (a) are due to KK integration errors, which we discuss in section 3.2.1.1.**

### 5.1.4. Summary

Inferences from solvent vibrational absorption data are consistent with the distinction between melts of dissociated ions (fluorobasic) and associated monomers and oligomers (fluoroacidic). Intermediate structure in Li-containing alkali fluoride mixtures could be an interpretation to ‘doublet’ peak behavior. Solvent and solute vibrational absorption data exhibit strong MIR resonances, resulting in NIR edges of significance to RHT in participating media. Inferences from solute vibrational absorption data are mostly consistent with more recent results and have been useful in identifying vibrations associated with oxygen.

## 5.2. Electronic Absorption – UV-Vis/NIR

### 5.2.1. Background

Molten fluoride salts consist of positively and negatively charged ions, each defined by a unique electronic configuration. Electronic resonance occurs when valence electrons (from the highest occupied energy orbital) are excited by an incoming photon with an energy that is equal to the difference in electronic energy levels between the valence orbital (ground state) and higher energy orbitals (excited states). In general, lighter elements (e.g., $Li^{+}$, $Be^{2+}$; both $1s^2$) have access to s and p orbitals which are separated by relatively large energies, requiring higher energy photons to excite and resulting in higher-energy electronic absorption bands in the UV and transparency in the visible spectrum. For example, Plot A1 shows the visible transparency of FLiNaK (UV cutoff attributed to the MgO cell) and Plots A39 and A43 show the NIR and visible transparency of FLiBe.

On the other hand, heavier elements have access to filled and unfilled d orbitals that are much closer in energy, resulting in lower energy absorption bands in the visible and NIR. For instance, the transition metal $Cr^{3+}$ ($3d^3$) in FLiNaK and FLiBe has three peaks in the range of 0.30 to 0.69 um (Figure 4, Plot A3, E2; Plot A16, E8). As atomic weight increases, f orbitals can be accessed, further increasing the number of peaks, and the wavelength range expands through the NIR. For instance, in $U^{4+}$ ($5f^2$) in FLiBe displays 10 peaks from 0.43 to 2.13 μm (Figure 4, Plot A30, E45-46).

### 5.2.2. Transition metals and Tellurium

Electronic absorption data is available for transition metals $Ni^{2+}$, $Cr^{2+}$, $Cr^{3+}$, $Fe^{2+}$, $Fe^{3+}$, $Co^{2+}$, $Mn^{2+}$, $Mn^{3+}$, $Cu^{2+}$, and $Nb^{4+}$ in molten fluoride media[1]. Using crystal-field theory [96] and comparison with crystalline and aqueous spectra, interpretation of the solvation structure of the tabulated cations in various fluoride solvents was CN = 6 in all cases except for $Nb^{4+}$, which was CN = 7. In Table 6 we compile the available data, providing an index (E#) for each spectrum, and reference the plot (A#) and setup indices (T# or R#). For each spectrum, we give the measurement temperature, salt composition, deduced speciation, path length, and list the location and intensity of the various peaks or shoulders.

No complementary structural data from other experimental or modeling techniques were found,

[1] $Tl^{+}$, $Pb^{2+}$, and $Bi^{3+}$ in FLiNaK were measured in the UV [119]. In this review, the focus has been on compiling data in the visible and infrared ranges due to their increased RHT relevance.

except for the $Cr^{2+}$ and $Cr^{3+}$ solute species which we further discuss in section 5.2.2.1. We then discuss the deduction of anionic Te speciation in section 5.2.2.2, as this is still a debated topic.

5.2.2.1. $Cr^{2+}$/$Cr^{3+}$

Ab initio molecular dynamics (AIMD) simulations have been performed on the $Cr^{2+}$ and $Cr^{3+}$ solutes in molten fluoride solvents [79][85][89]. Nam's results indicated that $Cr^{2+}$ has CN = 5.2 in FLiBe and 4.9 in FLiNaK, and that $Cr^{3+}$ has 5.9 in both solvents [89] Winner's results for $Cr^{2+}$ showed predominantly CN = 4 and some CN = 5 and the results for $Cr^{3+}$ dominantly showed CN = 6; similar coordination numbers were observed in three different solvents: FLiBe, 3LiF-$AlF_3$, and 2KF-NaF [79]

Young measured the electronic absorption spectra of $CrF_2$ in FLiBe (Figure 8a) and $CrF_3$ in FLiBe (Figure 8a) and FLiNaK (Figure 4, Plot A4, E2) [29][24], and compared the various peak locations with crystal-field-theory calculations [97][98]. Due to agreement with the calculated peak locations, which had been obtained under the assumption of octahedral symmetry (i.e., CN = 6), Young concluded that $Cr^{2+}$ and $Cr^{3+}$ had CN = 6 in both FLiBe and FLiNaK [29]. Young also observed that the spectra of $Cr^{2+}$ and $Cr^{3+}$ did not exhibit spectral changes with varying LiF concentration (60-77 mol%) in LiF-$BeF_2$ mixtures. They argued that this reflected the consistency of the Cr coordination environment in solvents of varying fluoroacidity, an observation made by Winner [79] and nearly seen by Nam [89]. Furthermore, Young suggested that in FLiBe, Cr ions would be solvated by $BeF_4^{2-}$ rather than compete with $Be^{2+}$ for $F^-$, a result which was directly predicted by Winner. However, the interpretation of CN = 6 for $Cr^{2+}$ in FLiBe disagrees with the AIMD results of both Nam and Winner. It would be valuable to revisit the interpretation of Young's spectral data using computational chemistry tools that were not available at the time of Young's study (1969). Since optical spectroscopy is an indirect indicator of CN, more direct structural data (e.g., from neutron diffraction, AIMD, etc.) would be useful to gain confidence in the structural interpretation of optical spectra and to address the discrepancies.

Liu performed a study aimed at understanding the differences in stability of $Cr^{2+}$ and $Cr^{3+}$ in FLiNaK and FLiBe, and the effect of $CrF_3$ on the corrosion of Cr metal. First, electronic absorption spectra of $CrF_2$ and $CrF_3$ in FLiBe were measured, shown in Figure 8c, both in agreement with Young's data (Figure 7a) [29]. Next, Cr metal was added to a solution containing FLiBe and $CrF_3$, and a decrease in absorbance of the 450 nm peak and a shift from 700 to 750 nm was observed, implying the conversion of $CrF_3$ to $CrF_2$ via the following reaction.

$$3CrF_2 \leftrightarrow 2CrF_3 + Cr \qquad (24)$$

Following the investigation of FLiBe, the electronic absorption spectra of $CrF_2$ and $CrF_3$ in FLiNaK were measured, shown in Figure 8d. The $CrF_3$ absorbance agreed with Young's data (Figure 4, Plot A3, E2) and looked similar to the FLiBe data; however, the $CrF_2$ absorbance looked markedly different from the FLiBe data, showing a peak at 450 nm in addition to a broad peak from about 600 to 750 nm. This behavior was interpreted as the conversion of $CrF_2$ to $CrF_3$ and Cr metal, further evidenced by the presence of Cr metal particles in the frozen salt after experimentation. Following the investigation of FLiNaK, $CrF_2$ was added to LiF-$BeF_2$ (2.25:1) and a similar reaction occurred. Liu concluded that $CrF_3$ becomes more favored as fluorobasicity, or dissociated $F^-$ concentration, increases.

This concept is further explored computationally for $Cr^0$, $Cr^{2+}$ and $Cr^{3+}$ in [79], which demonstrates that besides dissociated $F^-$ and associated $F^-$, bridging $F^-$ is also of importance, as solutes can be solvated by dissociated $F^-$ as well as by incorporation into solvent oligomers that consist of strongly-associated $F^-$ containing complex ions; it is hypothesized that steric hindrance makes the oligomer-solvation less prevalent for higher CN ions ($Cr^{3+}$ having a higher CN than $Cr^{2+}$) [79]. This could explain the stability of $Cr^{2+}$ over $Cr^{3+}$ for fluoroacidic melts. Further study of the $Cr^{2+}$, $Cr^{3+}$ equilibria are needed to understand more clearly how the equilibrium shifts with changes in temperature, fluoroacidity, and heterogeneous phases (e.g., Cr metal, Cr carbides, etc.). An example of such a study is provided by Toth [44], in which the equilibrium quotient of $U^{4+}$ and $U^{3+}$ with uranium carbides and graphite was measured over varying conditions, and we further discuss in section 5.2.3.2.

5.2.2.2. $Te^-$

Te was found in the MSRE as a fission product and caused embrittlement of structural alloys [99], and could pose issues due to its volatility. As opposed to the cationic transition metal ions, Te was studied in a negative oxidation state via addition of $Li_2Te$. Bamberger [38] found that if purified $Li_2Te$ were added to pre-reduced FLiBe, there would be no absorption. If oxidants (e.g., $Te_2$, $FeF_2$) or the impurity $LiTe_3$ were present, a peak at 0.478 µm would appear, shown in Figure 8b. Measurement of $Te_2$ vapor resulted in a peak at about 0.41 µm, also shown in Figure 8b, ruling it out as the potential species. Moreover, Bamberger showed that the equilibrium concentration of dissolved $Te_2$ in FLiBe was immeasurable, resulting in a solubility estimate of 300 wppm at 525 °C. Thus, Bamberger concluded that $Li_2Te$ would convert to $LiT_3$ (n=1, m=5/2 in eq. 25) in oxidizing conditions.

Toth [48] criticized Bamberger's use of a $SiO_2$ cell (T9), arguing that it could act as an oxidant, and performed similar experiments with the diamond windowed cell (T12). Toth controlled the concentration of $Te_2$ in the melt with a thermal gradient furnace. The equilibrium reaction proposed by Toth is given in equation 25.

$$nLi_2Te + mTe_2 \leftrightarrow 2Te_{\left(\frac{n}{2}\right)+m}{}^{n-} + 2nLi^+ \qquad 25$$

Through additions of different amounts of $Te_2$, Toth proposed that $Te^-$ was present (n = 1, m = ½). Toth reported that this behavior was consistent between fluoride and chloride solutions; however, spectral data was only given for chloride solution. Thus, more work is needed to study the tellurium solvation in fluoride salts and to provide conclusive evidence for the structure of the (-1) oxidation state of Te.

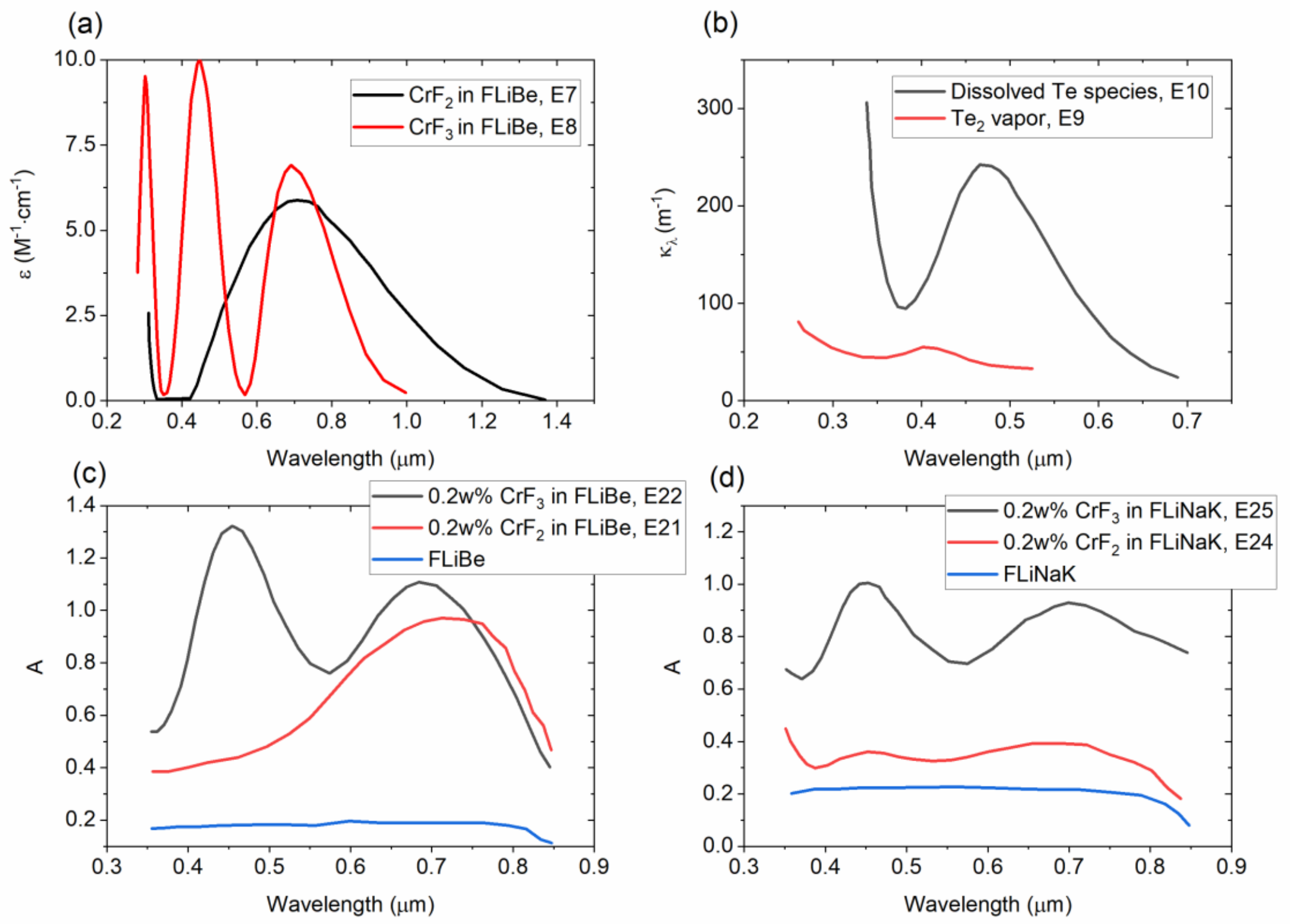

**Figure 8. Electronic absorption spectra of lighter solutes in molten fluoride solvents: (a) molar absorptivity ($\varepsilon$) spectra of $Cr^{2+}$ and $Cr^{3+}$ in FLiBe, (b) linear absorption coefficient ($\kappa_\lambda$) spectra of a dissolved Te species in FLiBe and $Te_2$ vapor, (c) absorbance resulting from additions of $CrF_2$ and $CrF_3$ in FLiBe, and (d) absorbance resulting from additions of $CrF_2$ and $CrF_3$ in FLiNaK . We give the spectrum index (E#) for each curve, for reference in Table 6.**

### 5.2.3. Lanthanides and actinides (f block)

$Ce^{3+}$, $Pr^{3+}$, $Gd^{3+}$ and $Pa^{4+}$, $U^{3+}$, $U^{4+}$, $Pu^{3+}$ constitute the available lanthanide and actinide electronic absorption data in molten fluoride media. In Table 7 we compile the available data, providing an index (E#) for each spectrum, and reference the plot (A#) and setup (T# or R#). For each spectrum, we give the measurement temperature, salt composition, deduced speciation, path length, and list the location and intensity of the various peaks or shoulders.

#### 5.2.3.1. Lanthanides

Using ligand field theory, Khokhryakov interpreted $Ce^{3+}$ (Figure 4, Plot A62, E55-56) [71], $Pr^{3+}$ (Figure 4, Plots A67-70, E57-63) [72], and $Gd^{3+}$ (Figure 4, Plot A71, E64-69) [73] as CN = 6, consisting of either 6 $F^-$ ions, or 5 $F^-$ and 1 $O^{2-}$ ions if $O^{2-}$ were present. This interpretation is supported by Raman results, which have shown that if the lanthanide (La, Ce, Nd, Sm, Dy, Yb, Y) mole fraction is less than 0.25, their structures tend to be undistorted octahedra (CN = 6) [100][101], a result consistent across bromides and chlorides [102][103]. High temperature nuclear magnetic resonance (HT-NMR) measurements of $LaF_3$ mixtures with alkali fluorides have agreed with the result as well [104].

Recently, some work has addressed the structure of oxy-fluoride lanthanide species, which can form when $O^{2-}$ is present. A Raman study was performed of $CeF_3$ in FLiNaK with and without $Li_2O$ addition [105]. Without $Li_2O$, consistent spectra were observed up to 20 mol% $CeF_3$ and were attributed to $CeF_6^{3-}$ (CN = 6) via QC-DFT analysis. Next, 5 mol% $Li_2O$ was added to 5 mol% $CeF_3$ in FLiNaK, and several new peaks appeared, followed by further spectral changes as 10 mol% and 20 mol% $Li_2O$ were added. This behavior was attributed via QC-DFT to formation of $Ce_2O_2F_{10}^{6-}$ (CN = 6) at lower $Li_2O$ concentration and a transition to $Ce_2O_2F_8^{4-}$ (CN = 5) at higher concentration. Khokhryakov theorized the presence of $CeOF_5^{4-}$ (CN = 6) with addition of about 1E-04 mole fraction of $Ce_2O_3$ in LiF-NaF [71], which would be consistent with the results of Cui if $O^{2-}$ bridging does not occur for very low $O^{2-}$ concentrations and CN = 6 is maintained. Raman and QC-DFT of $LuF_3$ in FLiNaK found a similar result in which oligomers formed of two Lu ions bridged by $O^{2-}$, and also found that that CN decreased with increasing $O^{2-}$ content, from 5 to 3 or 4 [106].

5.2.3.2. Actinides

Through comparison with crystalline [107] and aqueous spectra [108][109], Young theorized $UF_3$ to have a CN of 8 or 9 in FLiBe and FLiNaK (Figure 4, Plots A9-10, E34-35) [27] and $PaF_4$ to have a CN of 7 within the LiF-$BeF_2$-$ThF_4$ (72-16-12 mol%) solvent (Figure 4, Plots A17-18, E37) [31] . Bamberger measured absorption spectra of $PuF_3$ in LiF-$BeF_2$-$ThF_4$ (72-16-12 mol%) solvent (Figure 4, Plot A19, E42) [32], but did not hypothesize on the CN. More work is needed to investigate the structure of these species via complementary experimental and modeling methods.

$UF_4$ was of particular interest due to its use as fuel in the MSRE. Toth proposed an equilibrium between the CN = 7 and CN = 8 species as the solvent and temperature were varied [14]. With decreasing LiF concentration (i.e., decreasing dissociated $F^-$) and increasing temperature, CN = 7 was favored. This was inferred through observation of subtle shifts in peak intensity and location and comparison with crystalline spectra of $U^{4+}$-doped $KTh_2F_9$ (CN = 9), $K_7Th_6F_{31}$ (CN = 8), and $Cs_3ThF_7$ (CN = 7) [14]. Figure 9a shows the disappearance of several peaks as solvent is varied from FLiNaK, to FLiBe, to LiF-$BeF_2$ (48-52 mol%), (i.e. increasing fluoroacidity). Plots A30 and A31 in Figure 4 show similar shifts as temperature increases. Two recent studies have used a combined MD and EXAFS approach to investigate alkali fluoride mixtures with $ThF_4$ and $UF_4$ [110][111]. The results show qualitative agreement with Toth, as the average coordination was about 8 for both species.

As part of the redox control for the MSRE fuel, for the purpose of corrosion control, a small amount of $U^{4+}$ would be reduced to $U^{3+}$. There was potential for $U^{3+}$ to react with graphite to form insoluble $UC_2$ via the back-reaction of equation 26 [43][45][44], which would cause similar reactivity concerns as that of $UO_2$ precipitation [4].

$$3UF_4 + UC_2 \leftrightarrow 4UF_3 + C \quad (26)$$

As a result, the equilibrium quotient of the reaction in equation 26 was measured as a function of temperature, varying solvent composition between FLiBe and $LiF\text{-}BeF_2$ (48-52 mol%), and varying carbide stoichiometry between $UC_2$ and $U_2C_3$ [44]. Figure 9b shows a decrease in the $U^{4+}$ peak at 1.1 µm (9147 $cm^{-1}$) and an increase in the $U^{3+}$ peak at 0.88 µm (11,360 $cm^{-1}$), as temperature increases from 500 to 700 °C. Accordingly, the equilibrium quotient increased by a factor of $10^6$ over this temperature range. $U_2C_3$ is less stable than $UC_2$, causing an increase in the equilibrium quotient. The use of $LiF\text{-}BeF_2$ (48-52 mol%) as solvent caused the equilibrium quotient to increase as well. Using the van't Hoff equation, Toth estimated the exothermic heats of solution for $UF_4$ and $UF_3$ in FLiBe, with that of $UF_4$ having a substantially more negative value. Toth suggested that this was because '$UF_4$ forms complexes of greater stability than does $UF_3$' [44]. Moreover, it was suggested that fluoroacidic melts stabilize lower oxidation state cations due to dissociated '$F^-$ deficiency;' conversely, as fluorobasicity increases, the higher oxidation state is favored.

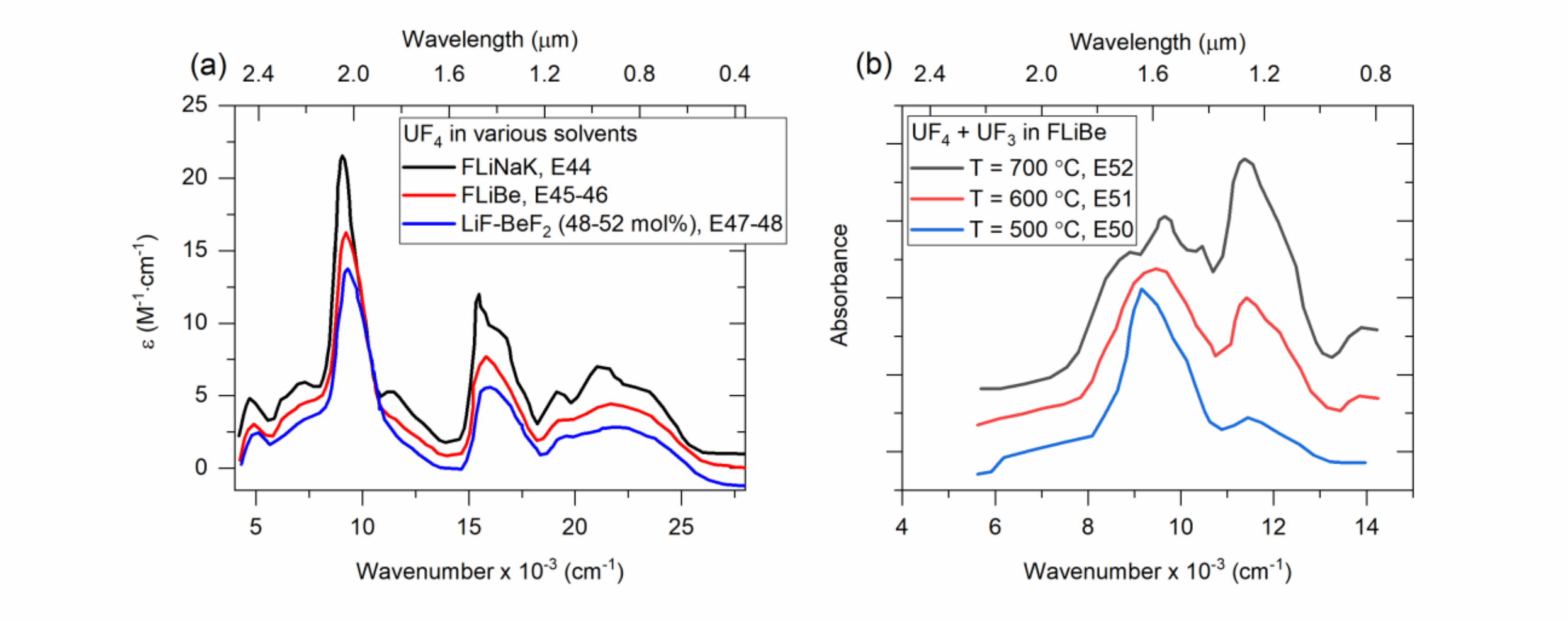

**Figure 9. Electronic absorption spectra of actinide solutes in molten fluoride solvents: (a) Molar absorptivity ($\varepsilon$) spectra of $U^{4+}$ in FLiNaK, FLiBe, and $LiF\text{-}BeF_2$ (48-52 mol%) (b) qualitative absorbance spectra of FLiBe solutions containing $U^{3+}$ and $U^{4+}$ . We give the spectrum index (E#) for each curve, for reference in Table 7. We note that multiple spectra indices are given in (a) because the measurement temperature of 550 °C falls between the two entries in Table 7. We plot in wavenumber (and wavelength) because the spectral changes are more easily visible. Following Toth's convention, we have separated each cruve in (a) by a single molar absorptivity unit.**

### 5.2.4. Summary

Inferences from electronic absorption data are mostly consistent with octahedral symmetry (CN = 6) for transition metal and lanthanide solutes and equilibria between CN = 7, 8, and 9 for actinides, depending on solvent fluoroacidity. Measurements of redox equilibria of $Te^-$, $Cr^{2+}/Cr^{3+}$, and $U^{3+}/U^{4+}$ have been conducted. Study of disproportionation has shown that higher valent ions appear to be more stable in fluorobasic melts.

## 6. Discussion and Summary of Absorption Data

### 6.1. Structural interpretation

#### 6.1.1. Vibrational

Comparison of liquid LiF infrared vibrational edge and resonance data with crystalline data, yielded the view of the melt as a highly disordered lattice, which we consider to be analogous to vibrational resonance among a cation's first coordination shell in a melt of dissociated ions. LiF-KF mixtures were interpreted similarly; however, observed intensity changes were not monotonic as the LiF peak appeared to be favored and a secondary peak was observed. Raman [82] and MD [78][83] results suggest that this behavior could be due to the presence of intermediate range order. There is no infrared vibrational resonance data for FLiNaK. As a melt of dissociated ions, it is predicted to exhibit similar behavior as LiF-KF, consisting of a superposition of pure component peaks weighted towards LiF.

Studies of LiF-KF-$ZrF_4$ and NaF-KF-$ZrF_4$ (varying $ZrF_4$) were interpreted as evidence of associated monomer and oligomer formation by $Zr^{4+}$, in qualitative agreement with combined HT-NMR, EXAFS and MD [88]. Likewise, transmittance measurement of a thin film of FLiBe exhibited a peak at 12.5 μm, in agreement with the QC-DFT-predicted spectra for the associated monomer $BeF_4^{2-}$, as well as with Raman [77] and MD [80]. The spectra exhibited another peak at 6.25 μm that was attributed to an overtone, but we also suggest a vibration of the oligomer $Be_2F_7^{3-}$ (indicated from Raman [90]).

In crystalline alkali and alkaline earth fluorides, increasing temperature and the phase transition from solid to liquid results in peak broadening and shifting of the vibrational absorption edge to the NIR. More data is needed to quantify the degree of shifting that occurs in FLiBe and FLiNaK, and other molten-fluoride solvents.

The infrared resonance spectra of solutes $OH^-$, $OD^-$, $Ta^{5+}$, and $Nb^{5+}$ dissolved in fluoride melts have been measured. $Ta^{5+}$ and $Nb^{5+}$ were hypothesized to exist as $TaF_7^{2-}$ , $TaF_6^-$, and $NbF_7^{2-}$, and were also found to be quite sensitive to the addition of $O^{2-}$, forming several different hypothesized oxy-fluoride species.

In summary, infrared spectral data can be used to probe the structural dynamics of different types of solvents (dissociated vs associated) as well as solutes species. Deduction of the coordination environment from vibrational resonances needs to be validated by other experimental and modeling techniques. Once validated, vibrational data can be conveniently used to probe for changes in structure due to different variables such as temperature and solvation environment. We note that Raman spectroscopy has traditionally been used for this purpose among molten fluorides (e.g., [112]); however, the additional benefits of infrared spectroscopy are to provide complementary data for infrared-active peaks and to give quantitative optical property data.

### 6.1.2. Electronic

Using various analysis techniques, the transition metal ions $Ni^{2+}$, $Cr^{2+}$, $Cr^{3+}$, $Fe^{2+}$, $Fe^{3+}$, $Co^{2+}$, $Mn^{2+}$, $Mn^{3+}$, $Cu^{2+}$, and $Nb^{4+}$ have been interpreted as CN = 6, with the exception of $Nb^{4+}$ as 7. AIMD results predict CN = 5 for $Cr^{2+}$ and CN = 6 for $Cr^{3+}$, and they are the only transition-metal ions that have been subject to extensive structural study (in the form of AIMD [89][85][79]) and optical data and AIMD suggest that these CNs are invariant to solvent fluoroacidity [29][79]. Electronic absorption data of $Cr^{2+}$ and $Cr^{3+}$ in FLiNaK and FLiBe [52] provide evidence that in a more fluorobasic melt (higher activity of dissociated $F^-$), the higher oxidation state ion is favored . More experimental study of the structure of these solutes and AIMD modeling of the electronic spectra is needed to better correlate the optical spectra with the electronic structure and coordination of these ions.

Using ligand field theory, the lanthanide ions $Ce^{3+}$, $Pr^{3+}$, and $Gd^{3+}$ were interpreted as CN = 6, which is consistent with the current understanding from HT-NMR [104] and Raman [100][101][102][103]. From comparison with aqueous and crystalline spectra, $Pa^{4+}$ and $U^{3+}$ were interpreted as CN = 7 and 8 or 9. The electronic absorption spectrum of $Pu^{3+}$ has been measured, but the CN was not theorized. An equilibrium between CN = 7 and 8 for $U^{4+}$ species was proposed, favoring higher coordination in fluorobasic melts (i.e., high concentrations of dissociated $F^-$). These CN's for $U^{4+}$ have been partially validated by recent combined MD and EXAFS [110][111]. In addition to structural studies, the $U^{4+}/U^{3+}$ equilibrium with $UC_2/C$, $U_2C_3/C$, and $H_2/HF$ was studied as a function of temperature in different solvents, finding that the lower valence $U^{3+}$ was increasingly favored at higher temperatures and in more fluoroacidic melts, a finding similar to that of $Cr^{2+}/Cr^{3+}$.

As with vibrational absorption, electronic absorption data provides a unique window into the solute-solvent coordination environment and with information from complementary methods (e.g., NMR, EXAFS, modeling), structural changes can be inferred from spectral changes with temperature and composition.

## 6.2. Radiative heat transfer

Electronic absorption data for some solutes in molten fluoride solvents exhibit peaks into the NIR. As discussed in section 5.2.2.1, Cr species are particularly important due to the prevalence of Cr in structural alloys and the relative stability of Cr species in molten fluoride solvents. In an experiment to measure the thermal conductivity of FLiNaK, Ewing found a decrease in plate-to-plate RHT correlated with increased Cr species concentration in FLiNaK [113]. To further investigate the RHT-significance of Cr species,

Chaleff developed an AIMD model of Cr in FLiNaK [114]. The oxidation state of Cr used in the model was not specified, but the authors validate their results in the visible spectrum by comparison to Young's data on $Cr^{3+}$ in FLiNaK (Figure 4, Plot A4, E2). In addition to the peaks seen in the visible, Chaleff's results predict a wide band that slowly increases from about 2 to 8 µm. The measured $\kappa_\lambda$ values for $CrF_3$ only extend to 1.0 µm and for $CrF_2$ they extend slightly further to about 1.4 µm, and both are decreasing with increasing wavelength at these points (Figure 8a); experimental absorption data over 2 to 8 µm is needed to test Chaleff's prediction of an electronic peak at higher wavelengths. In addition to measurements for the absolute value of $\kappa_\lambda$, it also important to know if there are sharp absorption peaks around the peak of the Planck distribution at the temperatures of interest, in which case the gray $\kappa$ assumption would introduce more distortion than that due to an absorption edge.

Other corrosion products such as $Fe^{2+}$ (Figure 4, Plot A14, E5), and $Ni^{2+}$ (Figure 4, Plot A15, E6), show absorption peaks up to 2 µm and would therefore be good candidates to investigate for potential absorption further into the NIR and MIR. Absorption peaks for lanthanide and actinide elements tend to reach farther into the infrared (between 2-3 µm) than transition metals. Thus, they likely would have a greater impact on RHT; MIR absorption data is needed to experimentally verify this prediction.

In addition to electronic absorption by solute species, vibrational edge absorption by solvent species may be significant. We present the extrapolated $\kappa_\lambda$ of FLiBe and FLiNaK in Figure 10, which show non-negligible overlap with the Planck distribution at 700 ˚C. Chaleff generated the FLiNaK prediction by performing a molar average of tabulated (by Li [115]) high-temperature $\kappa_\lambda$ data for crystalline LiF and NaF [9]; KF was neglected because its NIR edge is expected to be beyond the region of significance for RHT due to its resonance being at much further wavelength than LiF and NaF [116]. We presented Liu's NIR edge data for FLiBe (Figure 6b). Then, we apply Deutch's exponential data fit (equation 27) to Liu and Chaleff's data, assuming a minimum value, $\kappa_{\lambda,min} = 0.7\ m^{-1}$, for both FLiBe and FLiNaK (which is near the $\kappa_{\lambda,min}$ measured for NaCl-KCl (50-50 w%) [117]).

$$\kappa_\lambda = \kappa_{\lambda,min} + A \cdot \exp(-\frac{B}{\lambda}) \quad (27)$$

Figure 10 plots this fit (dashed lines). The extrapolated $\kappa_\lambda$ for FLiBe and FLiNaK in the NIR/MIR show values that fall in the 1 to 6000 $m^{-1}$ range identified by heat transfer modeling to be of engineering relevance [10]. Assuming a 1 cm pipe, and $\kappa$ range of relevance to RHT of 10 – 6000 $m^{-1}$, we calculate gray $\kappa$ (using the Planck-mean absorption coefficient, given in equation 28) at 700 ˚C to be 493 $m^{-1}$ for FLiBe and 148 $m^{-1}$ for FLiNaK.

$$\kappa = \frac{\int_{\lambda_{min}}^{\lambda_{max}} I_{bb}\kappa_\lambda d\lambda}{\int_0^\infty I_{bb} d\lambda} \quad (28)$$

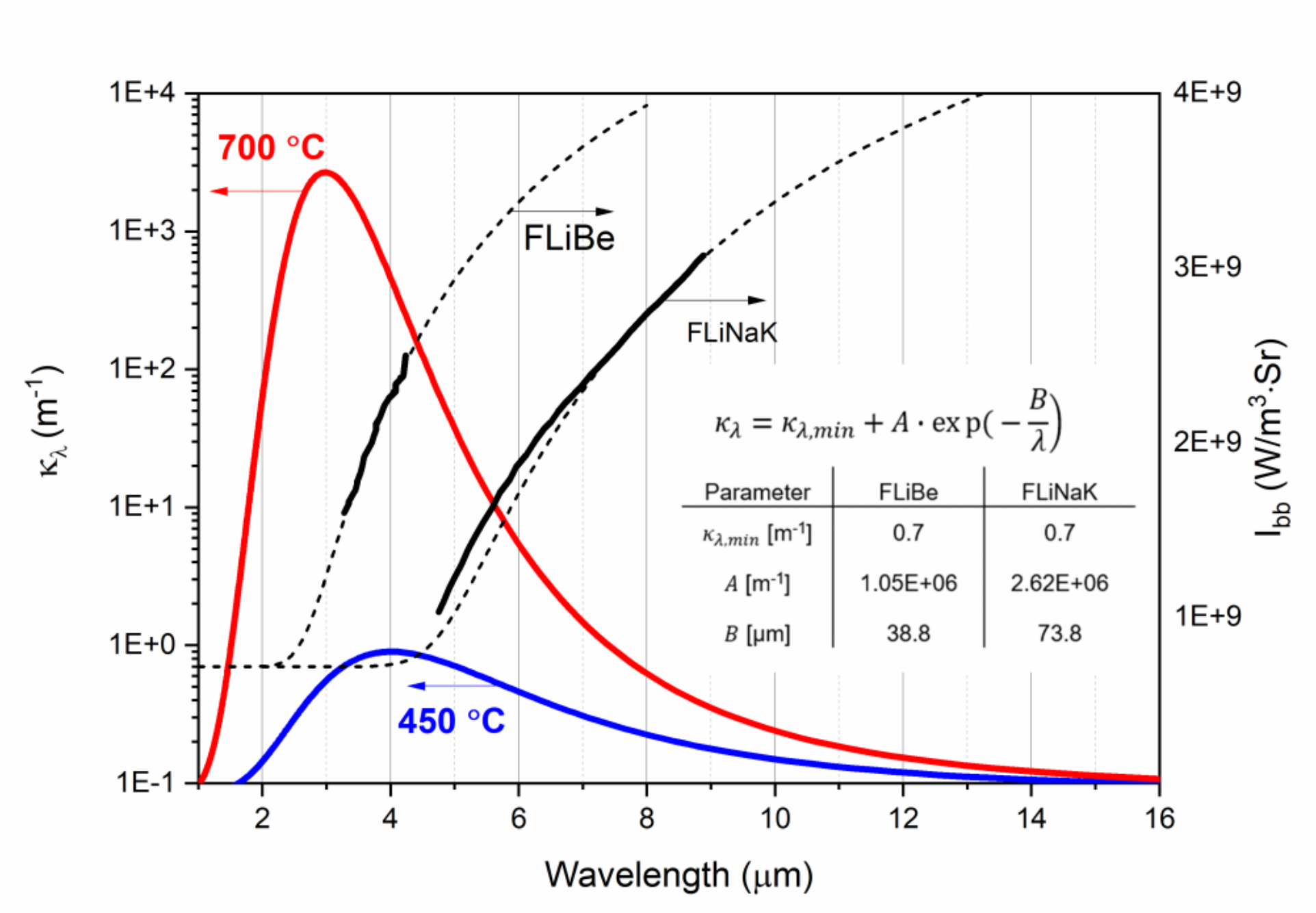

**Figure 10. Extrapolated $\kappa_\lambda$ of FLiBe (V24) and FLiNaK (V34). Dashed line: Deutch's equation (eq. 27) is fitted to Liu's FLiBe data (V24)** [50] **and Chaleff's FLiNaK molar-averaging prediction based on single component data (V34)** [9]**. The Planck distribution is plotted at temperatures of 450 and 700 ˚C, to illustrate the spectral range of interest to RHT at different temperatures. Equation 27 is also inset, along with fitting parameters for FLiBe and FLiNaK.**

## 7. CONCLUSION

Advanced nuclear reactors that utilize molten salts are being designed to operate between 500 and 700 ˚C and reactor designers need to account for RHT in thermal-hydraulic analysis. $\kappa_\lambda$ (i.e. the imaginary part of refractive index) of the molten salts in the NIR/MIR range is a necessary input for RHT modeling. We note that the real part of the refractive index should be known as well because it affects the Planck distribution, though it is not the focus of this review. As shown by the molten-fluoride studies in this review, optical-absorption spectroscopic measurements can provide this data. Furthermore, as the field currently stands, understanding of the molecular and electronic structure of the salt is not complete, and optical spectroscopy can be a convenient way to provide indirect information of this nature.

Given the need for optical absorption data on molten salts, a critical review of methods for visible and infrared spectroscopic absorption of high temperature liquid fluoride salts is provided in an attempt to spur further experimentation. To serve as a reference for researchers investigating the structure of fluoride salts or using optical spectroscopy for other applications such as thermochemistry [44], kinetics [52], online monitoring [118], or radiative heat transfer [13], we selected 37 journal articles and summarized in several tables organizing the methods, available spectra, and structural interpretations. Experimental methods, including transmittance and reflectance, cover operating temperature ranges from 425 to 1527 °C. 17 transmittance setups and 7 reflectance setups were found, and they are summarized in Table 1 (transmittance) and Table 3 (reflectance). Electronic absorption data has been compiled for the transition metal ions $Cr^{2+}$, $Cr^{3+}$, $Mn^{2+}$, $Mn^{3+}$, $Fe^{2+}$, $Fe^{3+}$, $Co^{2+}$, $Ni^{2+}$, $Cu^{2+}$, $Nb^{4+}$, the metalloid ion $Te^{-}$, the lanthanide ions $Pr^{3+}$, $Ce^{3+}$, $Gd^{3+}$, and the actinide ions $Pa^{4+}$, $U^{3+}$, $U^{4+}$, and $Pu^{3+}$. Vibrational absorption data has been compiled for the mixtures LiF-KF, LiF-NaF, LiF-KF-$ZrF_4$, LiF-NaF-$ZrF_4$, FLiBe, compounds LiF, $CaF_2$, $BaF_2$, and CsF and the solute species $OH^{-}$, $OD^{-}$, $Nb^{5+}$, and $Ta^{5+}$.

From the structural perspective, more complementary theoretical and experimental studies are needed for transition metal solutes, as $Cr^{2+}$/$Cr^{3+}$ currently have received sole attention. Electronic absorption data exists for $U^{3+}$, $Pa^{4+}$, and $Pu^{3+}$ but needs comparison with complementary methods as well. Vibrational absorption data in the resonance region is needed for both FLiBe and FLiNaK, which could elucidate bridging behavior and intermediate-range structures, respectively. While UV-vis/NIR electronic and MIR vibrational spectra are an indirect measure of coordination environment and oxidation state, once connected with a structural model they form a powerful way to probe for changes in these properties with respect to temperature or different solvation environments.

From the RHT perspective, it would be useful to investigate the electronic absorption of otherwise well-studied solutes (e.g., $Cr^{2+}$/$Cr^{3+}$, $U^{4+}$/$U^{4+}$) extending from the NIR into the MIR. The measured $\kappa_\lambda$ values for $CrF_3$ only extend to 1.0 µm and for $CrF_2$ they extend slightly further to about 1.4 µm, and both are decreasing with increasing wavelength at these points (Figure 8a); experimental absorption data over 2 to 8 µm is needed to test Chaleff's prediction of an electronic peak at higher wavelengths [114]. In addition to measurements for the absolute value of $\kappa_\lambda$, it also important to know if there are sharp absorption peaks around the peak of the Planck distribution at the temperatures of interest, in which case the gray $\kappa$ assumption would introduce more distortion than that due to an absorption edge. Characterization of the NIR/MIR edge for potential reactor coolant FLiBe and its thermal-hydraulic surrogate FLiNaK and their temperature dependence is also needed for quantifying RHT effects. By fitting to experimental data for FLiBe (Figure 4, Plot A40, V24) and prediction of FLiNaK (Table 4, V35), we extrapolate $\kappa_\lambda$ values over the range of RHT significance. Assuming a 1 cm pipe, and $\kappa$ range of relevance to RHT of 10 – 6000 $m^{-1}$, we calculate gray $\kappa$ (using the Planck-mean absorption coefficient, given in equation) at 700 ˚C to be 493 $m^{-1}$ for FLiBe and 148 $m^{-1}$ for FLiNaK.

## DECLARATION OF COMPETING INTEREST

At the time at which the article is published, some of the authors of this manuscript have interests in or relationships with entities that are commercializing molten salt technology. The content of this manuscript or the direction of the research presented herein was not influenced by these entities, nor by the author's relationships with these entities.

## ACKNOWLEDGEMENTS

The information, data, or work presented herein was funded in part by the Department of Energy Office of Nuclear Energy's Nuclear Energy University Program under Award Number DE-NE0008680 (Project 17-13232) and DE-NE0008651 (Project 17-14541). This work has also been supported through the US Nuclear Regulatory Commission Graduate Student Fellowship program on Grant NRC-HQ-84-15-G-0040. The views and opinions of authors expressed herein do not necessarily state or reflect those of the United States Government or any agency thereof. The authors would also like to acknowledge support by the UC Berkeley SALT Laboratory researchers and by University of Wisconsin Thermal Hydraulics Laboratory and staff. RS and SM would like to acknowledge funding support from the Hellman Foundation. MK was partially supported by the National Science Foundation (1750341). RG would like to acknowledge funding support from U.S. Department of Energy, Office of Nuclear Energy, Nuclear Energy University Program, Grant DE-NE0008985 and INL Laboratory Directed Research & Development (LDRD) Program under DOE Idaho Operations Office Contract DE-AC07-05ID14517.

## CONTRIBUTIONS

Conceptualization: RS, WD, RG, Writing – Original Draft: WD, RG, Writing – Review & Editing: RS, SM, MA, MK, Formal Analysis: WD, Data Curation: WD, RS, Visualization: WD, SM, RS, Resources: RS, Funding Acquisition: RS, MK, MA, Supervision: RS, MA, MK, Project Administration: RS, MK.